\documentclass{aa}  
\usepackage{graphicx}
\usepackage{txfonts}
\usepackage{pifont}
\usepackage{xfrac}    
\usepackage{multirow}
\usepackage{color, colortbl}
\definecolor{Gray}{gray}{0.9}
\newcommand{\cmark}{\ding{51}}
\newcommand{\xmark}{\ding{55}}
\usepackage{accessibility}
\begin{document}

\title{The near-infrared degree of polarization in debris disks.\thanks{Based on observations made with ESO Telescopes at the Paranal Observatory under programs ID 105.20GP.001 and 109.237K.001. The fits files of the observations are available in electronic form at the CDS via anonymous ftp to cdsarc.u-strasbg.fr (130.79.128.5) or via http://cdsweb.u-strasbg.fr/cgi-bin/qcat?J/A+A/}}

   \subtitle{Toward a self-consistent approach to model scattered light observations.}

\author{J. Olofsson\inst{\ref{inst:ESO},\ref{inst:MPIA}}
        \and
        P. Th\'ebault\inst{\ref{inst:LESIA}}
        \and
        A. Bayo\inst{\ref{inst:ESO}}
        \and
        Th. Henning\inst{\ref{inst:MPIA}}
        \and
        J. Milli\inst{\ref{inst:IPAG}}
      }
\institute{
    European Southern Observatory, Karl-Schwarzschild-Strasse 2, 85748 Garching bei M\"unchen, Germany\\\email{johan.olofsson@eso.org}\label{inst:ESO}
    \and
    Max Planck Institut f\"ur Astronomie, K\"onigstuhl 17, 69117 Heidelberg, Germany\label{inst:MPIA}
    \and
    LESIA-Observatoire de Paris, UPMC Univ. Paris 06, Univ. Paris-Diderot, France\label{inst:LESIA}
    \and
    Univ. Grenoble Alpes, CNRS, IPAG, 38000, Grenoble, France\label{inst:IPAG}
}


\abstract{Debris disks give us the unique opportunity to probe the properties of small $\mu$m-sized particles, allowing us to peer into the constituents of their parent bodies, young analogs of comets and asteroids of our solar system.}
{In the past, studies of the total intensity phase function, the brightness of the disk as a function of the scattering angle, have proven powerful to constrain the main characteristics of the dust particles in debris disks. Nonetheless, there can remain some degeneracies in the modeling that can be alleviated when considering polarized intensity observations.}
{We obtained new near-infrared scattered light observations of four young debris disks, and used state-of-the-art algorithms to recover the total intensity and linear polarimetric images of the disks. These images allow us to constrain the degree of linear polarization as a function of the scattering angle.}
{All four debris disks are detected in polarized intensity, and three are also recovered in total intensity. We measured peak degree of polarization of $\lesssim 40$\% for all three disks. For the disk around HD\,129590, we are furthermore able to determine the degree of polarization in the radiation pressure driven halo. To reproduce the observed polarization fractions, we find that the particles must consist of highly refractive and absorbing material. For HD\,129590, by measuring the polarization fraction beyond the birth ring, we constrain the width of the size distribution to be smaller and smaller, compatible with the effect of radiation pressure. We put these findings to the test and present a self-consistent approach to produce synthetic images, assuming different profiles for the radiation pressure strength, and accounting for the presence of unbound grains. We find the contribution of these grains to be especially critical to reproduce the increasing degree of polarization with stellocentric distances.}
{Some of our results (namely a very small blow-out size and very large $(n,k)$ values for the optical constants) required to reproduce the observed degree of polarization might seem difficult to reconcile with our understanding of cosmic dust. Similar results have been obtained for other disks and we discuss the current limitation of available light scattering models as well as possible avenues to alleviate these unfortunate limitations.}


\maketitle
%

\section{Introduction}

Near-infrared (IR) scattered light and millimeter observations of debris disks probe $\mu$m- and mm-sized particles, the small end of the size distribution. The lifetime of these particles is smaller than the age of the central star, as they can be removed either by radiation pressure, Poynting-Robertson drag, or collisions (\citealp{Wyatt2008}). This implies that there must be a constant replenishment, in a collisional cascade initiated from the larger bodies in the disk (\citealp{Krivov2021}). Characterizing the properties of such planetesimals is key to understand how they grew and the process of planet formation in general. Unfortunately, they cannot be observed directly and we therefore need to rely on indirect methods to try to infer some of their main characteristics. One possible avenue to achieve this goal is to better constrain the properties of the small dust particles that are released in destructive collisions and try to infer what the parent bodies might look like. For instance, \citet{Olofsson2022b} showed that the $\mu$m-sized particles in the debris disk around HD\,32297 have optical properties compatible with fluffy, highly porous aggregates. By comparing with solar system objects, this result indicates that the parent bodies in the disk around HD\,32297 are likely pristine cometary like objects which have not collided frequently, in order to avoid compaction of the particles.

A powerful tool to constrain some of the properties of the smallest dust particles in a debris disk is the study of the phase function, the brightness of the disk as a function of the scattering angle\footnote{The complementary angle between the star, the particle, and the observer, between $0^{\circ}$ and $180^{\circ}$.}. By modeling the phase function with light scattering models (e.g., Mie theory, \citealp{Mie1908} for compact spherical grains), one can infer for instance the typical sizes of the grains. The phase functions can be estimated either in total intensity (Stokes $I$) or in linear polarimetry (a combination of Stokes $Q$ and $U$) and the disk appears quite differently in both cases. In total intensity, the front side of the disk, along its minor axis, is usually best revealed, thanks to the strong forward scattering peak. On the other hand, in linear polarimetric observations, the major axis of the disk is brighter, since the polarized phase function usually peaks (or shows a plateau) at scattering angles close to $90^{\circ}$. Taken in isolation, both approaches can reliably provide first order estimates on the typical sizes and porosity of the particles, but there is a strong synergy in combining both total intensity and polarimetry. As outlined in the conclusions of \citet{Min2016}, for aggregates the shape of the phase function in total intensity is mostly governed by the size of the aggregates, while the polarized phase function is more closely related to the properties of the individual monomers constituting the aggregates. Furthermore, the ratio between polarized flux and total intensity provides the degree of polarization (ideally also as a function of the scattering). Since this is a ratio between two images, the dependency on for instance the dust density distribution is naturally removed, allowing us to further characterize the properties of the dust particles.

Furthermore, near-IR observations of debris disks offer the rather unique opportunity to probe different grain sizes, solely depending on the stellocentric distance. Indeed, several drag forces must be accounted for when modeling the dynamics of dust particles, and these forces are size dependent. The main force is radiation pressure, parametrized by the $\beta$ ratio between the radiation pressure and gravitational forces. As summarized in \citet{Krivov2010}, it increases the eccentricity of the smaller particles as soon as they are released, re-distributing them in an extended halo beyond the birth ring (\citealp{Thebault2023}). The eccentricity of the particles depends on $\beta$ (as $e \sim \beta/(1-\beta$), which in turn depends on the particles size $s$ (beyond a few $\mu$m, $\beta$ becomes proportional to $\sim 1/s$). If the particles are released in the birth ring, at a distance $a_0$ to the star, then their semi-major axis $a$ will be such that their apocenters lie at $a_0$, meaning $a(1-e) = a_0$. We can thus derive the distance of their apocenters as $a(1+e) = a_0(1+e)/(1-e) \sim a_0/(1-2\beta)$. This means that as the stellocentric distances increases, the maximum grain size of the size distribution decreases. Ideally, one should therefore expect a gradient in the optical properties as a function of the distance to the star.

Over the past decades, the degree of polarization has been estimated for about a dozen of debris disks. Using Subaru CIAO $K$-band observations, \citet{Tamura2006} estimated a $10$\% degree of polarization for the disk around $\beta$\,Pictoris and it does not appear to vary across the major axis of the disk, between $\sim 2.5\arcsec$ and $6\arcsec$. At $R$-band wavelengths, \citet{Gledhill1991} had previously reported a polarization degree slightly larger, close to $\sim 17$\% at larger separations ($15-30\arcsec$). \citet{Tamura2006} interpreted the low polarization fraction as the presence of particles with sizes comparable to the wavelength of observation (a few $\mu$m). \citet{Graham2007} presented \textit{Hubble Space Telescope} (HST) observations of the disk around the low-mass star AU\,Mic ($V$-band). They found that the degree of linear polarization increases as a function of the projected distance, from $5$\% in the inner regions to $40$\% at $80$\,au. To reproduce the observations, the authors conclude that the dust particles are likely highly porous ($\sim 90$\%) and that ballistic cluster-cluster aggregation is a promising avenue to explain their findings. 

Similar results of the degree of polarization increasing with stellocentric distances were obtained for several other disks (HIP\,79977, \citealp{Thalmann2013}; HD\,111520, \citealp{Draper2016}; HD\,32297, \citealp{Asensio2016}; HD\,15115, \citealp{Engler2019}). Since these disks are highly inclined, if not perfectly edge-on, the polarized fraction is estimated along the projected major axis, and information about the dependency on the scattering angle is lost. These studies all found similar results, with low ($\lesssim 10$\%) degree of polarization, linearly increasing up to $\sim 30-45$\% at larger separations (only reaching $15-20$\% for HD\,15115). The only face-on disk for which the degree of polarization has been estimated is the disk around the low-mass star TWA\,7 (\citealp{Ren2021}). Combining HST and SPHERE observations the authors estimated a high polarization fraction in the main ring and the faint outermost third ring, of $85$ and $75$\%, respectively. Since the disk is close to face-on, this corresponds to scattering angles close to $90^{\circ}$.

More recently, the degree of polarization as a function of the scattering angle could be measured for a handful of disks (HD\,35841, \citealp{Esposito2018}; HD\,191089, \citealp{Ren2019}; HR\,4796, \citealp{Arriaga2020}; HD\,114082, \citealp{Engler2023}). The near-IR observations display a variety in the shape of the polarization fraction, usually peaking at $\sim 90^{\circ}$ (except for HR\,4796 where the peak is closer to $40^{\circ}$), as well as in the maximum peak values, ranging from $\sim 10$\% up to $\sim 50$\%. In all cases, the degree of polarization as a function of the scattering angle could only be determined for the birth ring, and could not be extracted for the extended halos of the disks.

By design, polarimetric observations do not suffer from significant artifacts when removing the contribution of the star to reveal the faint disks. In a nutshell, the stellar photons are expected to be largely unpolarized, while any photon that has been scattered off by some dust particle will show linear polarization in a preferential direction. Subtracting images obtained with orthogonal polarization directions is therefore an efficient way to remove the stellar photons while keeping photons that have been scattered by the disk. This means that the main challenge in measuring the degree of polarization lies in recovering the image in total intensity and minimizing well-known self-subtraction artifacts (e.g., \citealp{Milli2012}). This has been a very active field of research over the past years, and there are several algorithms that aim at tackling this kind of issue, such as DI-sNMF (data imputation using sequential non-negative matrix factorization, \citealp{Ren2020}), \texttt{MAYONNAISE} \citep{Pairet2021}, \texttt{REXPACO} \citep{Flasseur2021}, or \texttt{mustard} \citep{Juillard2023}, among others. On top of these techniques that are focusing on how to best post-process the observations, alternative approaches have focused on building large libraries of reference images that can be used to best reproduce the observations and perform reference star differential imaging \citep[RDI,][]{Xie2022}.

In this paper, we present new observations of four young debris disks, obtained using the Spectro-Polarimetric High-contrast Exoplanet REsearch (SPHERE, \citealp{Beuzit2019}) instrument at the Very Large Telescope. The aim is to measure the degree of polarization of the debris disks and attempt to constrain some of the properties of the dust particles. In Section\,\ref{sec:observations} we present the observations, which are analyzed in Section\,\ref{sec:analysis}. The results are presented in Section\,\ref{sec:results}. In Section\,\ref{sec:unbound} we present a more complex, self-consistent model, aiming at reproducing the observations of HD\,129590, to provide additional context on our results, before discussing our findings in Section\,\ref{sec:discussion}.

\section{Observations and data reduction}\label{sec:observations}

\begin{table}
\caption{Stellar and disk properties.}
\label{tab:star}
\centering
\begin{tabular}{lccc}
\hline\hline
     Star & SpT & $d_\star$\tablefootmark{a} & $f_\mathrm{disk}$\tablefootmark{b} \\
    & & [pc] & [$10^{-3}$] \\
\hline
    HD\,191089 & F5 & $50.11 \pm 0.05$ & $1.5$ \\
    HD\,157587 & F5 & $99.87 \pm 0.23$ & $3.2$ \\
    HD\,115600 & F2 & $109.04 \pm 0.25$ & $2.3$ \\
    HD\,129590 & G3 & $136.32 \pm 0.44$ & $7.0$\\
\hline
\end{tabular}
    \tablefoot{
        \tablefoottext{a}{Distances from \citet{Gaia2016,Gaia2020}}
        \tablefoottext{b}{Values from \citet{Esposito2020}.}
    }
\end{table}

The selection of the four disks was performed with the following criteria: to be around young stars, bright, sufficiently inclined to maximize the chances of a detection in total intensity (but not too inclined either so that the front and near sides of the disk can be distinguished, that is, $60^{\circ} \lesssim i \lesssim 85^{\circ}$), and to have been previously imaged at near-IR wavelengths (to ensure that a detection will be likely). The stars had to be observable during a given ESO period (in that case odd-numbered, P105 and later P109). Since at the time of the P105 deadline the star-hopping mode was new, we settled for a relatively small sample of disks that had not been previously observed with SPHERE in polarimetry. In the end, among all possible candidates, the following science targets were retained: \object{HD\,191089} (\citealp{Ren2019}, \citealp{Esposito2020}), \object{HD\,157587} (\citealp{Millar-Blanchaer2016}, \citealp{Esposito2020}), \object{HD\,115600} (\citealp{Currie2015}, \citealp{Gibbs2019}, \citealp{Esposito2020}, \citealp{Olofsson2022a}), and \object{HD\,129590} (\citealp{Matthews2017}, \citealp{Esposito2020}, \citealp{Olofsson2022a,Olofsson2023}), with the following reference stars: \object{HD\,191131}, \object{HD\,158018}, \object{HD\,117255}, and \object{HD\,129280}. Table\,\ref{tab:star} summarizes some of the stellar and disk properties. 

The disks were observed using the SPHERE/IRDIS instrument (\citealp{Dohlen2008}), making use of the star-hopping technique (\citealp{Wahhaj2021}). This mode allows to hop back and forth between the science target and a reference calibrator (on average $15$\,min per cycle), which has no known companion nor disk. The observations were performed using the dual-beam polarimetric mode (DPI, \citealp{deBoer2020}, \citealp{vanHolstein2020}), in pupil-tracking, with the BB\_H filter. Because the degree of polarization is smaller than unity, there is however the risk that a disk can be detected in total intensity but remains undetected (or poorly recovered) in polarized intensity. To mitigate this risk, we also observed each star individually, in DPI, without observing the reference star (thus allowing to observe the science target longer within the 1-1.5\,hours allocated observing block). 

Table\,\ref{tab:obslog} summarizes the observations used in this study, along with some of the atmospheric conditions during the observations. For the second column of the Table, ``SCI'' and ``CAL'' refer to the science and reference stars of the star-hopping sequence, respectively, while ``DPI'' refers to the stand-alone DPI observations.

\subsection{Linear polarimetric images}

The reduction of the stand-alone DPI observations is done using the IRDAP\footnote{Available at \url{https://github.com/robvanholstein/IRDAP}} package (version 1.3.4, \citealp{vanHolstein2020}), but for the star-hopping sequence, some pre-processing is required. A star-hopping sequence usually consists of a concatenation containing several OBs, alternating between the science and calibrator stars. At the moment, IRDAP only handles files that have the same \texttt{OBJECT} keyword in the headers. For a given concatenation, we therefore grouped the fits files of the same target together (SCI or CAL) and reduced them independently using IRDAP.

The IRDAP package provides several outputs. First, it returns two images, $Q_\phi$ and $U_\phi$, the former revealing the polarized signal from the disk (if there is any), the latter can be used as a proxy for the uncertainties as it contains no astrophysical signal. Second, IRDAP also provides a cube of $N_\mathrm{f}$ frames (the left and right sides of the IRDIS detector are summed together) and a list of parallactic angles for each of the $N_\mathrm{f}$ frames. This cube will be used to derive the total intensity image. Even though the detector integration time (DIT) is the same for all pairs of SCI-CAL, we normalized all the frames by their corresponding DITs. The left column of Figure\,\ref{fig:full} shows the polarimetric $Q_\phi$ images, with a square root scale, where all four disks are detected.

\subsection{Total intensity images}

To retrieve the disk in total intensity, we performed data imputation with sequential non-negative matrix factorization, using the DI-sNMF\footnote{Available at \url{https://github.com/seawander/nmf_imaging}} package (\citealp{Ren2020}). For each science target we only used the star-hopping sequences and did not try to use the DPI stand-alone observations. We first used the cube (produced by IRDAP, see above) of the associated calibrator to build the NMF components (a non-negative, non-orthogonal basis), applying a central numerical mask with a radius of $8$\,pixels. These components are then used to perform RDI on the science frames, applying a mask of $1\arcsec$ in radius ($0.55\arcsec$ for HD\,115600) to make sure the disk signal is not included. For each science frame, a model of the point spread function is constructed from the components, and subtracted to the original image. Afterwards, each frame is de-rotated by its corresponding parallactic angle, and the cube is median-collapsed to produce the final image in total intensity. For all the science targets in our sample, we used $5$ NMF components. The resulting images are shown in the central column of Figure\,\ref{fig:full}, with a square root scaling. The disks around HD\,157587, HD\,11560, and HD\,129590 are well detected in total intensity, while the disk around HD\,191089 cannot be recovered. This disk is the faintest in our sample (Table\,\ref{tab:star}) and from the modeling of the polarimetric data, this is also the disk with the smallest inclination, $i \sim 61^{\circ}$ (see later), which is always more challenging for total intensity observations (\citealp{Milli2012}). 

In Appendix\,\ref{app:data} we further investigate the strength of self-subtracted effects which can severely impair the analysis of total intensity observations. This is done by using the final total intensity image as a ``model'' of the disk and following an approach similar to forward modeling. We find that when using DI-sNMF on a star-hopping sequence, self-subtraction is not significantly impacting neither the total intensity images nor the degree of polarization. 

\subsection{Merging of the different datasets}

As reported in Table\,\ref{tab:obslog}, for each science target, several datasets are available. At the time of the call for proposals the star-hopping mode was a recent addition to the SPHERE instrument, and the performance in total intensity for extended emission remained unknown. We had therefore requested two star-hopping sequences of $1.5$\,hours each for each science target to maximize the on-source integration time, on top of the stand-alone DPI observations. It should also be noted that the star-hopping sequences, being performed in DPI, can also be used to produce the $Q_\phi$ and $U_\phi$ images. 

Figure\,\ref{fig:all_pdi} shows all the reductions from all the different epochs that we obtained. The four leftmost columns show the results in total intensity, while the four rightmost columns show the $Q_\phi$ images. In total intensity we only show the post-processing of the star-hopping sequences, but for the $Q_\phi$ images, we highlight in the lower right corner if the image comes from a star-hopping or stand-alone DPI sequence.

For the DPI images, we median-combined all the different epochs in which the disks were detected (given that all observations were done with the same DITs), and this median-combined image will be used for the rest of this study. Regarding the total intensity images, we tried combining the different datacubes together, along with their respective parallactic angles, but this did not yield significant improvement. For the rest of this study we will therefore use data coming from the best of the star-hopping sequence. The last column of Table\,\ref{tab:obslog} shows which dataset were used. If there is a tick mark for a ``SCI'' object, but not for the corresponding ``CAL'' entry, this means that we only used this dataset for the DPI $Q_\phi$ image.

\subsection{Degree of polarization}

\begin{figure*}
  \centering
  \includegraphics[width=0.95\hsize]{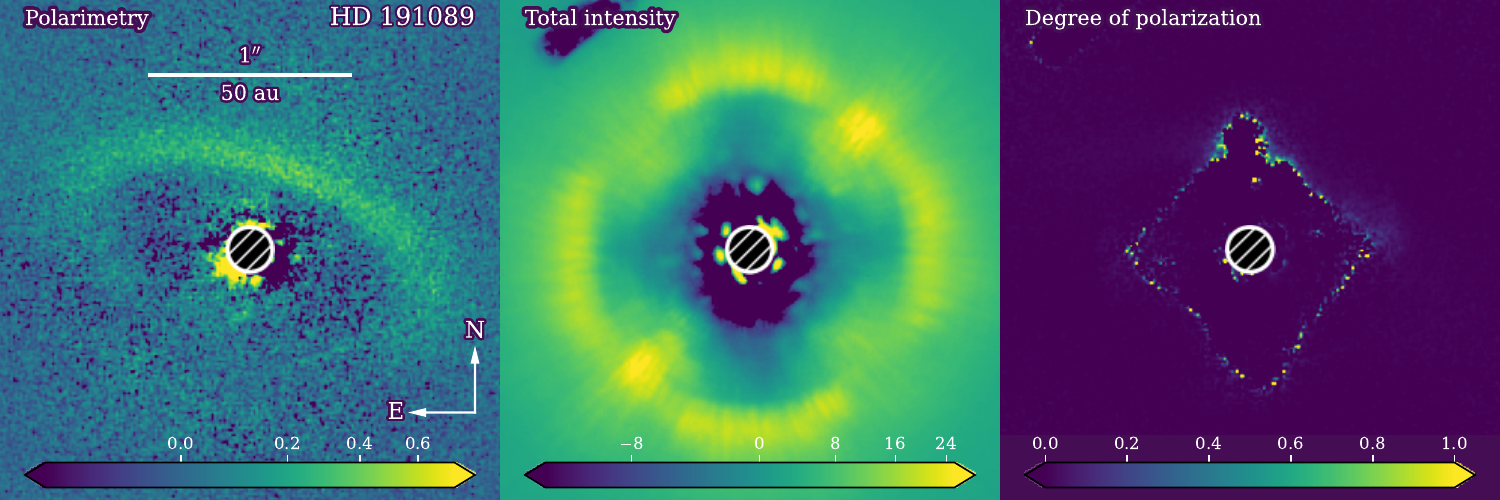}
  \includegraphics[width=0.95\hsize]{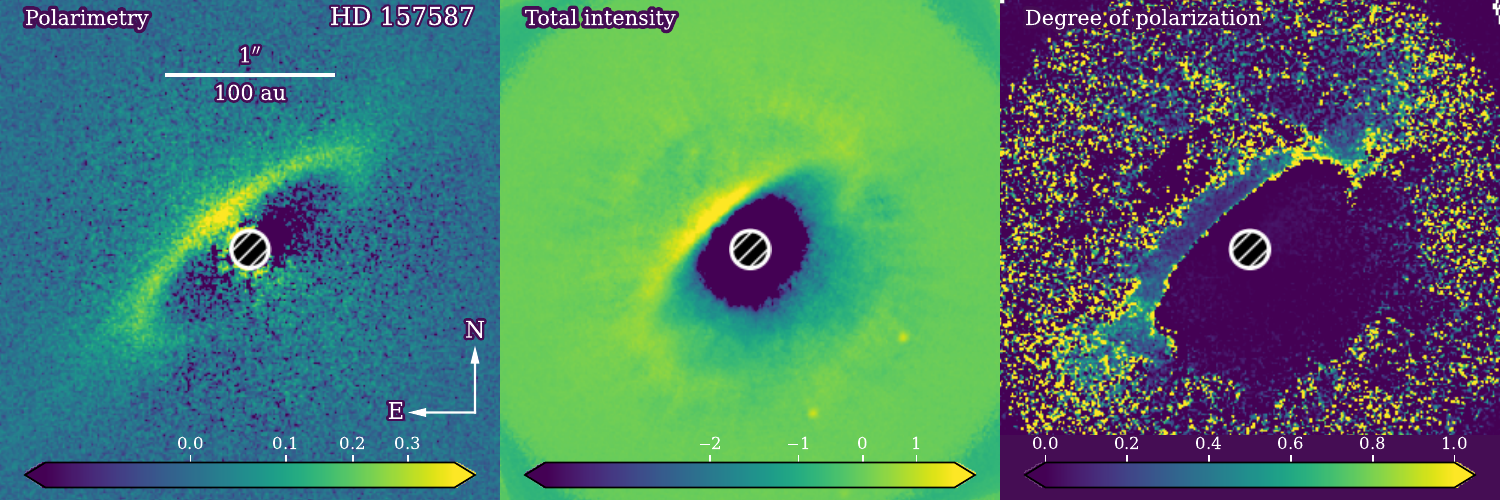}
  \includegraphics[width=0.95\hsize]{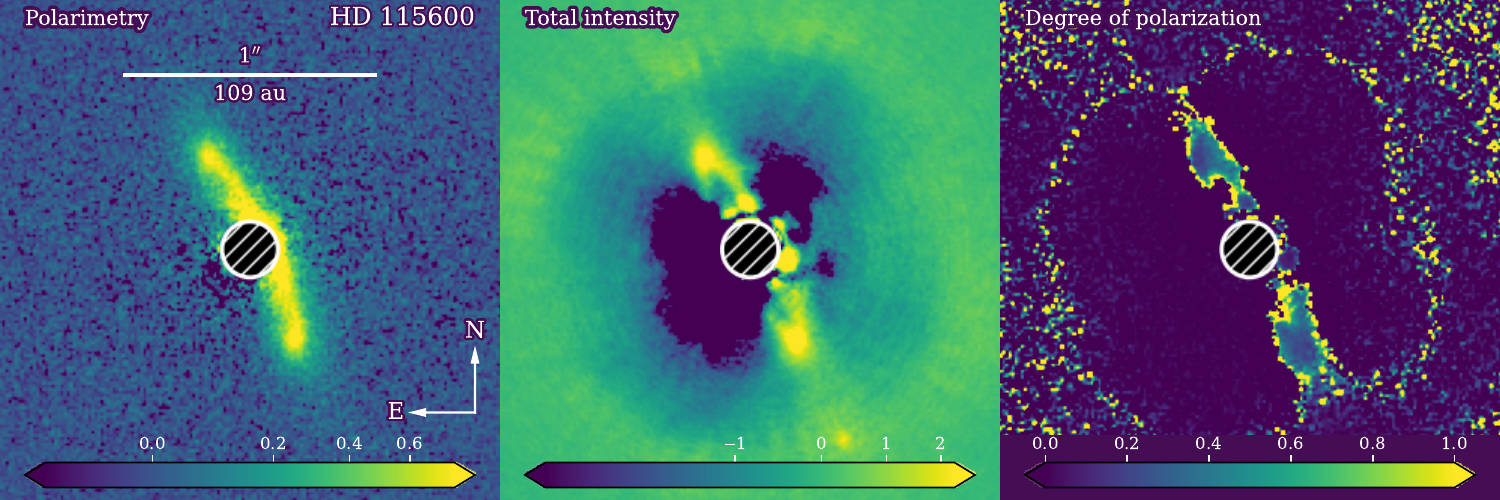}
  \includegraphics[width=0.95\hsize]{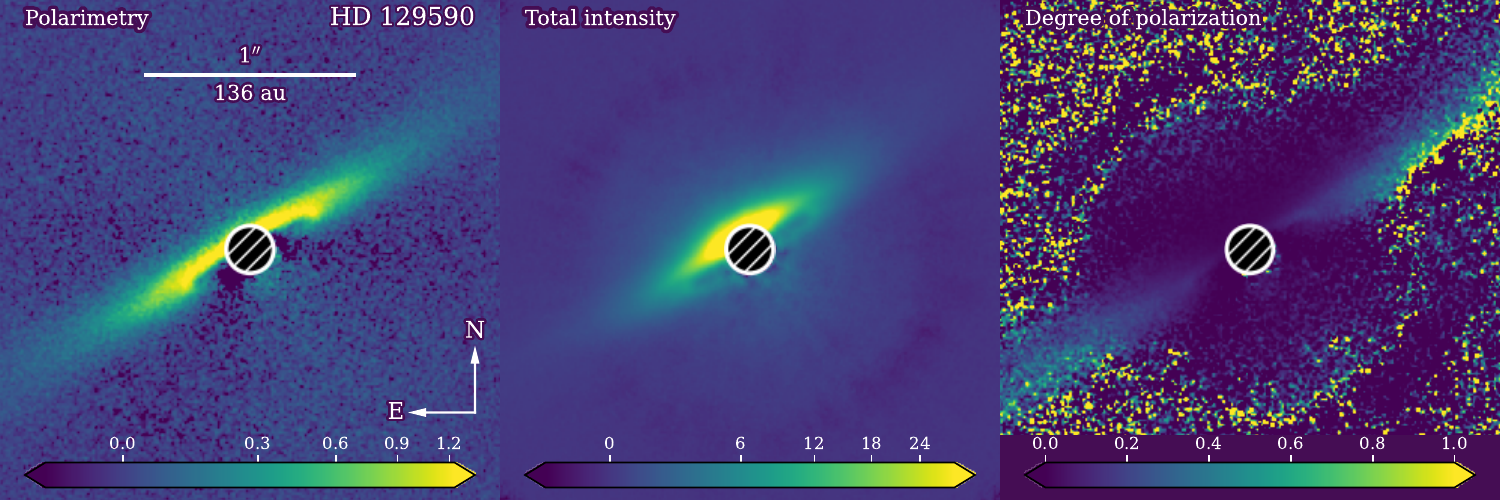}
    \caption{\textit{Top to bottom:} observations of HD\,191089, HD\,157587, HD\,115600, and HD\,129590. \textit{Left:} polarimetry, \textit{center:} total intensity, \textit{right:} degree of polarization (polarimetry over total intensity). The scaling is linear for the right column and between $0$ and $1$, while in square root for the left and center columns. On the left panel, the horizontal bar represents $1\arcsec$ and the distance in au is reported below. North is up, East is right, as indicated by the compass on the leftmost panels.}
  \label{fig:full}
\end{figure*}

From the polarimetric and total intensity images, we are then able to estimate the degree of linear polarization, by simply dividing the former by the latter. The degree of polarization is displayed in the right column of Figure\,\ref{fig:full} using a linear scaling between $0$ and $1$.

Simply dividing the two images one with the other is sufficient since no additional astrophysical effects need to be accounted for. For instance, \citet{Olofsson2020} highlighted the impact that a non negligible vertical scale height can have on the determination of the phase function. For highly inclined and vertically thick disks, there is a column density enhancement along the semi-major axis of the disk, solely due to projection effects. Since this depends on the dust density distribution, this effect affects the polarimetric and total intensity images the same way, and therefore cancels out when computing the degree of polarization.

As mentioned previously, the disk around HD\,191089 is not detected in total intensity, therefore for the rest of this study, we will focus on the remaining three targets (we refer the reader interested in the disk around HD\,191089 to \citealp{Ren2019} and \citealp{Esposito2020}). For HD\,157587, we successfully recover the degree of polarization along the projected semi-minor axis of the disk, in the north east direction. For HD\,115600, on the other hand, the minor axis of the disk is hidden behind the coronagraph, and the degree of polarization can only be measured along the major axis of the disk. For both these disks, we do not recover strong signal beyond the birth ring, and cannot estimate the degree of polarization in the extended halo. This is only possible for the fourth object, HD\,129590, for which we can recover the degree of polarization along the minor and major axis, as well as in the halo beyond the birth ring. On the lower right panel of Fig.\,\ref{fig:full}, the fact that the degree of polarization along the minor axis is close to $0$ does not mean it is unconstrained. At this location, the disk is well detected both in polarimetry and total intensity, meaning that the degree of polarization can reliably be estimated to a few percents (see Section\,\ref{sec:HD129590} for a further analysis). Interestingly, along the major axis, the degree of polarization is increasing with the stellocentric distance.

\begin{figure*}
  \centering
  \includegraphics[width=\hsize]{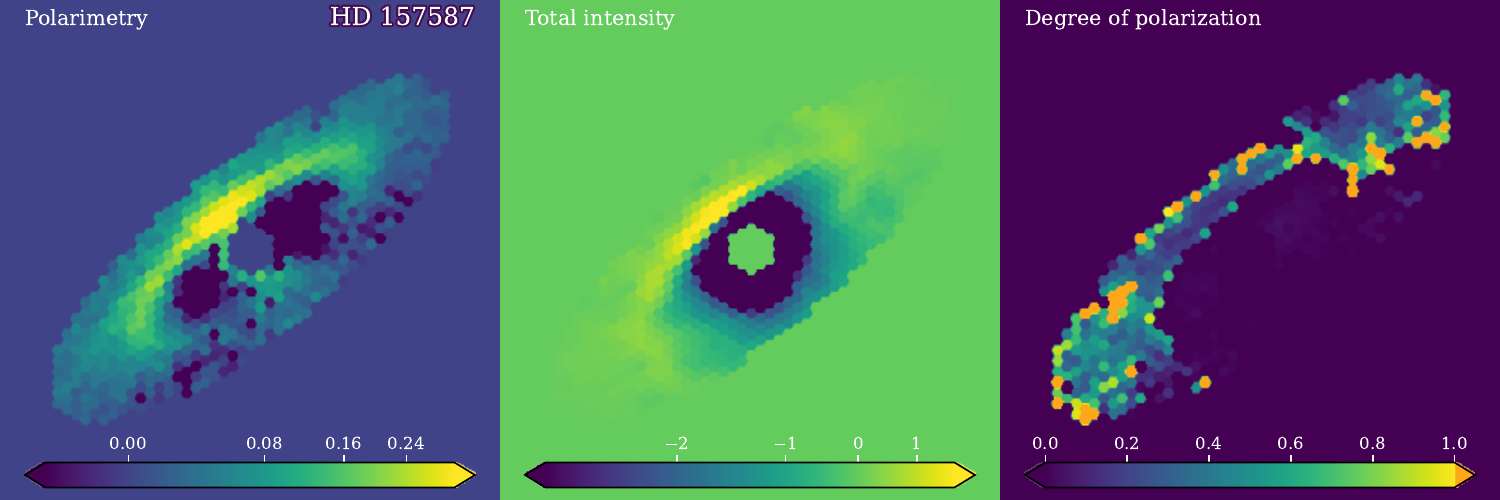}
  \includegraphics[width=\hsize]{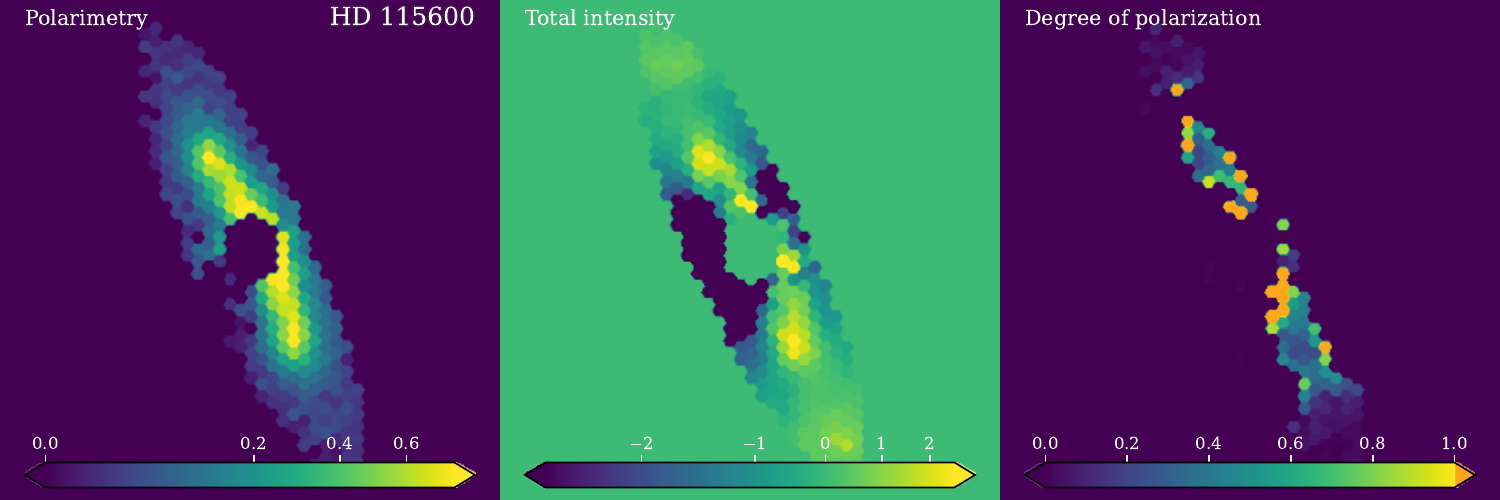}
  \includegraphics[width=\hsize]{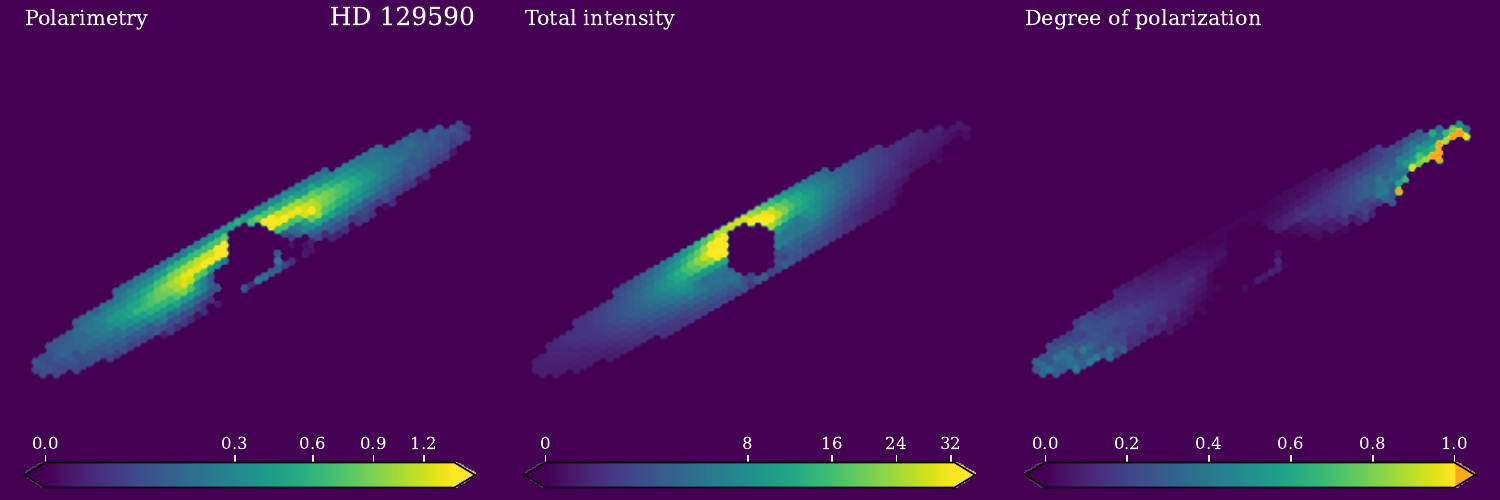}
    \caption{Same as Figure\,\ref{fig:full}, but the images were binned onto an hexagonal grid instead of having the native SPHERE resolution ($12.26$\,mas). On the right column, pixels saturating above unity are represented in orange.}
  \label{fig:binned}
\end{figure*}

Even though the disk around HD\,129590 is very bright both in total intensity and polarimetry, the other two disks around HD\,157857 and HD\,115600 are fainter. Therefore, to increase the signal to noise in the images, we binned the data on an hexagonal grid (and not a squared grid to avoid possible aliasing). For each star, each image (total intensity and polarimetry), and each hexagon, we compute the median value of the flux (as well as the median absolute deviation for the uncertainties), distance to the star, and scattering angle. The latter two quantities are computed in the midplane of the disk, assuming a given inclination and position angle for the disk (see Fig.\,\ref{fig:scatdist}, the scattering angle being the arccosine of the dot product between the line of sight and the coordinates at the disk midplane). These angles are obtained from modeling the polarimetric data, which is described in App.\,\ref{sec:ddit} and the results are presented in Table\,\ref{tab:ddit}. The binned images are shown in Figure\,\ref{fig:binned}, with a similar structure as Fig.\,\ref{fig:full}. The phase functions and degree of polarization as a function of the scattering presented in the rest of this study are computed from these binned images.

Overall, we measure peak degrees of polarization that are relatively low, below $\sim 50$\%, comparable to values reported in the literature for other debris disks. As further discussed in Section\,\ref{sec:results}, for HD\,157857, the maximum degree of polarization is in the range $40-50$\%. For HD\,115600, the peak value is most likely below $40$\%. Finally, for HD\,129590, the degree of polarization is very low in the birth ring, below $20$\%, but increases to values closer to $50$\% as the stellocentric distance increases.

It should be noted that the three disks for which we determined the degree of polarization were shown to display some degree of asymmetry by \citet{Crotts2024}. They reported that though the disk around HD\,115600 appears axisymmetric and does not show color or brightness asymmetries, the analysis of the vertical profile tentatively suggests a possible warp. When modeling the SPHERE observations of this disk, we also find a large opening angle ($\psi \sim 0.13$, Table\,\ref{tab:ddit}), suggesting that the vertical structure of this disk is unusual. However, this should not affect the inclination and position angle determination (crucial for the scattering angle calculation). For HD\,129590, \citet{Crotts2024} noted a brightness asymmetry that is only detected in their $K1$ band data but not in the $H$ band observations. However, the disk does not appear to have strong color asymmetries and our model, which assumes a circular disk (App.\,\ref{sec:ddit}), can account for most of the signal in the SPHERE $H$ band observations. We therefore cannot further comment on the origin of the asymmetry detected in the $K1$ GPI observations. Lastly, \citet{Crotts2024} found a significant color asymmetry for the disk around HD\,157587, the East side of the disk being relatively bluer than the West side. A possible explanation being that the distribution of small dust particles is not uniform throughout the disk. This will be further discussed in Section\,\ref{sec:results}.

\section{Analysis}\label{sec:analysis}

The main objective is to constrain the properties of the dust particles in the three debris disks for which the degree of polarization can be computed. Because we targeted disks with relatively large inclinations, combined with the angular resolution of the SPHERE instrument, we can measure the degree of polarization for a wide range of scattering angles.

\subsection{Modeling approach}

For a given dust model (see next subsections), we can compute the goodness of fit for the degree of polarization phase function, but as discussed in \citet{Milli2024} this is not necessarily the complete picture. For instance, it may well be that the total intensity and polarized intensity phase functions are both completely off with respect to the observations, but that their ratio is still a good match to the data. Therefore, there is non-redundant information in the total intensity phase function that can be included in the modeling approach.

A first approach to find the best fitting model would be to simply sum the $\chi^2$ values calculated for the total intensity phase function and for the degree of polarization as a function of the scattering angle. Nonetheless, there is the risk that one may dominate the other, yielding for instance a good match to the total intensity data and a bad fit to the degree of polarization. For this reason, we chose to normalize the $\chi^2$ values coming from the two observables and computed the ratio between both $\chi^2$ values when the models are set to null. This ratio is then used to weight down one of the $\chi^2$ value. In practice, the ``null'' $\chi^2$ for the total intensity was always larger than for the degree of polarization, and the former was weighted down.

Additionally, the dust density distribution, the flux normalization of the observations, the magnitude of the star, all these parameters do not affect the degree of polarization since it is a ratio between two quantities that are affected the same way by the aforementioned parameters. This is however not the case for the total intensity phase function taken alone (on top of the geometric effects discussed in \citealp{Olofsson2020}). For a given dust model, comparing the total intensity phase function directly to the observations is not straightforward. Therefore, prior to computing the total intensity $\chi^2$, we first find the scaling factor that minimizes the residuals between the model and the observations (see Eqn.\,7 of \citealp{Olofsson2016}). By doing so, we are effectively losing information related to the scattering efficiencies of the particles. Given that this part of the modeling is less ``complete'' compared to the modeling of the degree of polarization, the associated $\chi^2$ is weighted down by a factor \sfrac{1}{2}.

In the following, we first describe some of our attempts to model the observations and how they failed to reproduce the data before presenting the approach that we retained for the rest of this study.

\subsection{Preliminary attempts}\label{subsec:preliminar}

To analyze the variation of the degree of polarization as a function of the scattering angle, we first attempted to use the \texttt{AggScatVIR}\footnote{Available at \url{https://github.com/rtazaki1205/AggScatVIR}} library (\citealp{Tazaki2022,Tazaki2023}). It provides the phase functions in total and polarized intensity, at different wavelengths (in our case $1.63$\,$\mu$m), for different particle shapes. There are two main families for the shape of the particles, namely, aggregates or irregular grains, and both can have different sizes. For the irregular grains, the sizes, or rather, volume-equivalent radius, range between $0.2$ and $1.6$\,$\mu$m, and there are two compositions to choose from. For the aggregates family, several aggregation models are available (from compact to fractal aggregates), with different total number $N$ of spherical monomers (effectively increasing the particle volume). The monomers themselves can have different sizes ($s_\mathrm{mon} = 100$, $200$, and $400$\,nm) and compositions. There are however some limitations, for instance, due to the significant computational cost, optical properties for very large aggregates (e.g., $N = 1024$ or $2048$) are not always available, and this is especially true for highly fractal particles (e.g., FA1.1, see Fig.\,2 of \citealp{Tazaki2023}).

Given that the models are pre-computed and made available in the library, the comparison between the models and the observations is extremely fast. Unfortunately, there are no combinations of particle shapes, sizes, or compositions that can adequately reproduce the measured degree of polarization. This is especially the case in the birth rings of the three disks, where the maximum degree of polarization is $\lesssim 40-50$\%. One possible explanation is that the birth ring should host a wide particle size distribution all the way up to mm-sized grains, which are not included in the \texttt{AggScatVIR} library (the largest particle size being of a few $\mu$m).

The second approach to model the observations is to use optical constants of known material with various compositions, use the Mie theory or the Distribution of Hollow Sphere (DHS, \citealp{Min2005}) model, and compute the phase function over a size distribution, varying the minimum and maximum grain sizes ($s_\mathrm{min}$ and $s_\mathrm{max}$, respectively), the slope of the size distribution, and the porosity of the grains. To compute the total and polarized intensity phase functions, we used the \texttt{optool}\footnote{Available at \url{https://github.com/cdominik/optool}} package (\citealp{optool}). We tried different dust compositions, mixing pyroxene (\citealp{Dorschner1995}), amorphous carbon (\citealp{Zubko1996}), and crystalline water ice (\citealp{Warren2008}) in different proportions. Unfortunately, this approach did not yield any promising results, as it was especially challenging to reproduce low degrees of polarization.

\subsection{Make up your dust}

In the end, we settled for an approach similar to the one described in \citet[][see also \citealp{Milli2024}]{Arriaga2020}. Instead of relying on measured optical constants, they (the real and imaginary parts) become free parameters ($n$ and $\mathrm{log}_{10}(k)$). We used the DHS model, assuming the maximum filling factor $f_\mathrm{max} = 0.8$. There are two additional free parameters related to the size distribution, which are the minimum and maximum grain sizes ($s_\mathrm{min}$ and $s_\mathrm{max}$, respectively). Preliminary tests suggested that the slope of the size distribution and porosity remain largely unconstrained, and we therefore left these values fixed to $-3.5$ and $25$\%. It should be noted that we here assume that the minimum grain size (one of the free parameter) corresponds to the radiation pressure blow-out size $s_\mathrm{blow-out}$. Otherwise, one would need to account for a break in the size distribution, of a given amplitude, for a given size ($s_\mathrm{min} \leq s_\mathrm{blow-out} \leq s_\mathrm{max}$). This is beyond the scope of the exercise and the implications of this assumption are further explored and discussed in Sections\,\ref{sec:unbound} and \ref{sec:discussion}. To identify the most probable solution, we use the \texttt{MultiNest} algorithm (\citealp{Feroz2009}) and the \texttt{Python} package \texttt{PyMultiNest} (\citealp{Buchner2014}). 

\section{Results}\label{sec:results}

Table\,\ref{tab:dust} shows the best fitting parameters for the three debris disks studied here. Results for each individual star are discussed in the rest of this Section.

\begin{table*}
	\centering
    \caption{Best-fitting parameters to model the total intensity phase function and degree of polarization as a function of the scattering angle.}
	\label{tab:dust}
	\begin{tabular}{lccccc}
		\hline\hline
        Star & Region & $s_\mathrm{min}$ & $s_\mathrm{max}$ & $n$ & $\mathrm{log}_{10}(k)$ \\
             &         &  [$\mu$m] & [$\mu$m] & & \\
		\hline
HD\,157587 & $0.70\arcsec \leq r < 1.10\arcsec$ & $0.03_{-0.02}^{+0.06}$ & $97_{-84}^{+375}$ & $3.1_{-0.8}^{+0.5}$ & $0.69_{-0.10}^{+0.07}$ \\
HD\,115600 & $0.35\arcsec \leq r < 0.50\arcsec$ & $0.11_{-0.07}^{+0.06}$ & $0.49_{-0.13}^{+0.14}$ & $2.0_{-0.7}^{+1.1}$ & $1.6_{-0.3}^{+0.2}$ \\
HD\,129590 & $0.30\arcsec \leq r < 0.50\arcsec$ & $0.36_{-0.01}^{+0.01}$ & $110_{-52}^{+75}$ & $4.3_{-0.0}^{+0.1}$ & $-2.69_{-2.28}^{+0.40}$ \\
HD\,129590 & $0.50\arcsec \leq r < 0.70\arcsec$ & $0.16_{-0.03}^{+0.02}$ & $1.9_{-0.3}^{+0.4}$ & $3.4_{-0.1}^{+0.1}$ & $-0.04_{-0.03}^{+0.03}$ \\
HD\,129590 & $0.70\arcsec \leq r < 0.90\arcsec$ & $0.48_{-0.01}^{+0.01}$ & $0.96_{-0.03}^{+0.03}$ & $3.2_{-0.1}^{+0.1}$ & $0.03_{-0.02}^{+0.02}$ \\
HD\,129590 & $0.90\arcsec \leq r < 1.10\arcsec$ & $0.25_{-0.04}^{+0.21}$ & $1.2_{-0.4}^{+0.2}$ & $2.4_{-0.1}^{+0.1}$ & $-0.11_{-0.04}^{+0.07}$ \\
		\hline
	\end{tabular}
\tablefoot{The second column shows the results for different annuli for HD\,129590 (see text for details).}
\end{table*}

\subsection{HD 157587}

\begin{figure}
  \centering
    \includegraphics[width=\hsize]{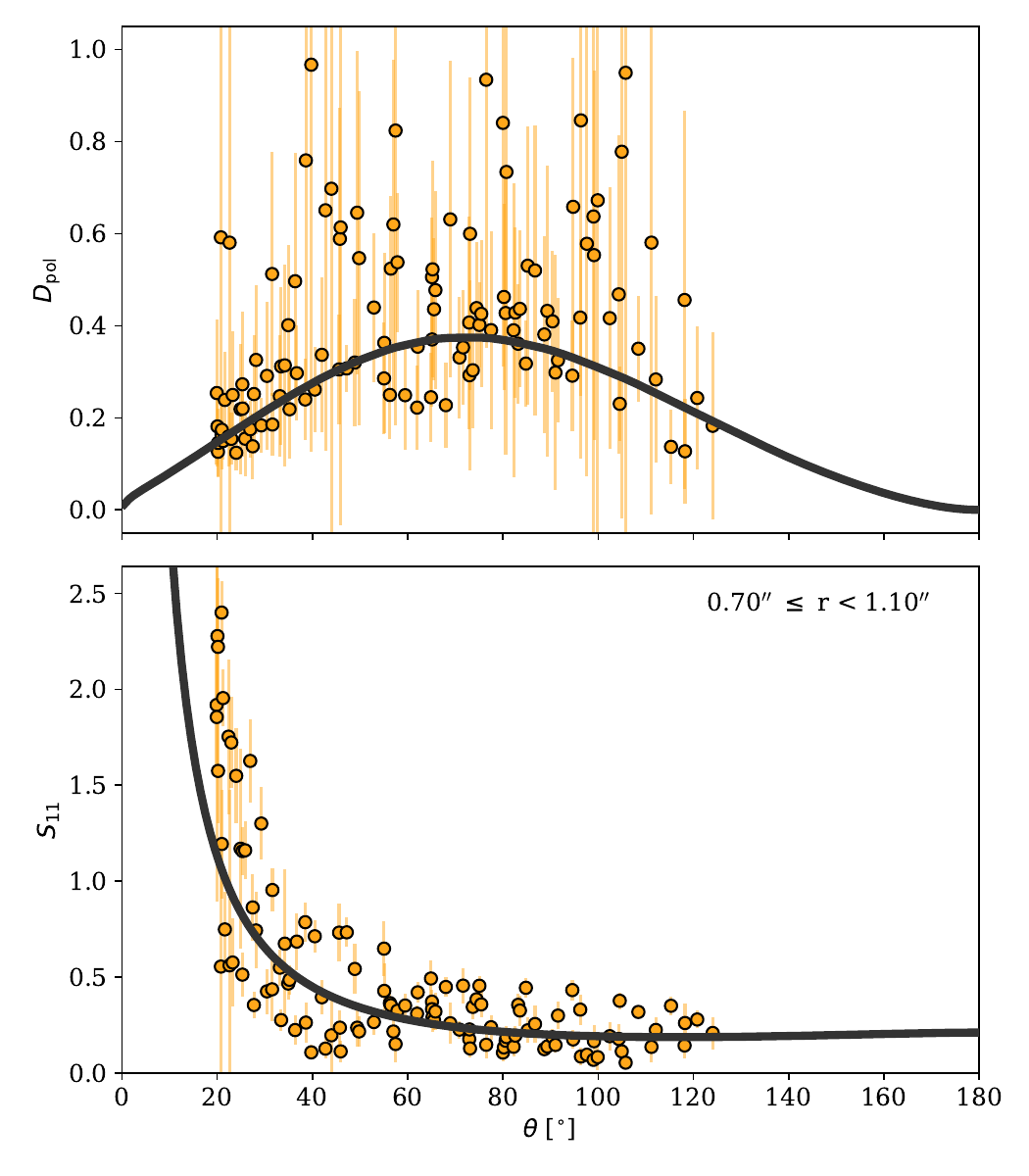}
    \caption{Observations and best fit model for HD\,157587 (orange circles and solid black line, respectively). \textit{Top:} degree of polarization. \textit{Bottom:} total intensity phase function. The inner and outer radii for the stellocentric distances are indicated in the upper right of the bottom panel.}
  \label{fig:HD157}
\end{figure}

Figure\,\ref{fig:HD157} shows the observations and best fit models for the degree of polarization (top) and total intensity (bottom), in an annulus between projected separations of $0.7\arcsec \leq r < 1.1\arcsec$ (corresponding to $70 < r < 110$\,au,tracing the birth ring of the disk). There is a significant dispersion for the degree of polarization, with large uncertainties, especially for scattering angles larger than $40^{\circ}$. This is most likely caused by the overall faintness of the debris disk, especially in total intensity, and most likely not related to the observing conditions. The star was observed with an average seeing of $\sim 0.55\arcsec$, a coherence time of $5.5-6$\,ms, and the on-sky rotation was not negligible ($\sim 50^{\circ}$). The disk around HD\,157857 is the one with the second smallest inclination of the sample ($i \sim 70^{\circ}$), while the disk with the smallest inclination (HD\,191089, $i \sim 61^{\circ}$) is not recovered in total intensity. This might be indicative that recovering the total intensity images of optically thin debris disks remains a challenge if the inclination is smaller than $\sim 75^{\circ}$, even when using state-of-the-art observing and post-processing techniques. Alternatively, the dispersion in the degree of polarization for scattering angles larger than $\sim 40^{\circ}$ could also be the consequence of different size distributions on either sides of the disk. As mentioned previously, \citet{Crotts2024} reported a color asymmetry between the east and west sides of the disk. Since Fig.\,\ref{fig:HD157} shows the degree of polarization of both the east and west sides, an over-abundance of small grains on one side could result in a different (most likely larger) degree of polarization. This would translate in a larger dispersion close to $90^{\circ}$ scattering angle. However, given the low S/N of the detection in both polarized and total intensity and that the brightness asymmetry is only seen in the $J$ band GPI data (\citealp{Crotts2024}), we opt not to further investigate the origin of this large dispersion.

Still, the best fit model can successfully reproduce the overall shape of both the degree of polarization and total intensity as a function of the scattering angle, even though the strong forward scattering peak is slightly under-estimated in the bottom panel. We find that the minimum grain size has to be very small, but with significant uncertainties, suggesting $s_\mathrm{min} \lesssim 0.1$\,$\mu$m, and that the maximum grain size is most likely larger than $\sim 20$\,$\mu$m. Since the spatial region in which the modeling is performed encompasses the birth ring of the disk ($0.7\arcsec \leq r < 1.1\arcsec$), it is expected that we need a wide range of grain sizes, all the way up to at least a few tens of $\mu$m. Regarding the real and imaginary parts of refraction, we obtain values of $n \sim 3.1$ and $k\sim 4.9$, rather large values (see Section\,\ref{sec:discussion}).

\subsection{HD 115600}

\begin{figure}
  \centering
    \includegraphics[width=\hsize]{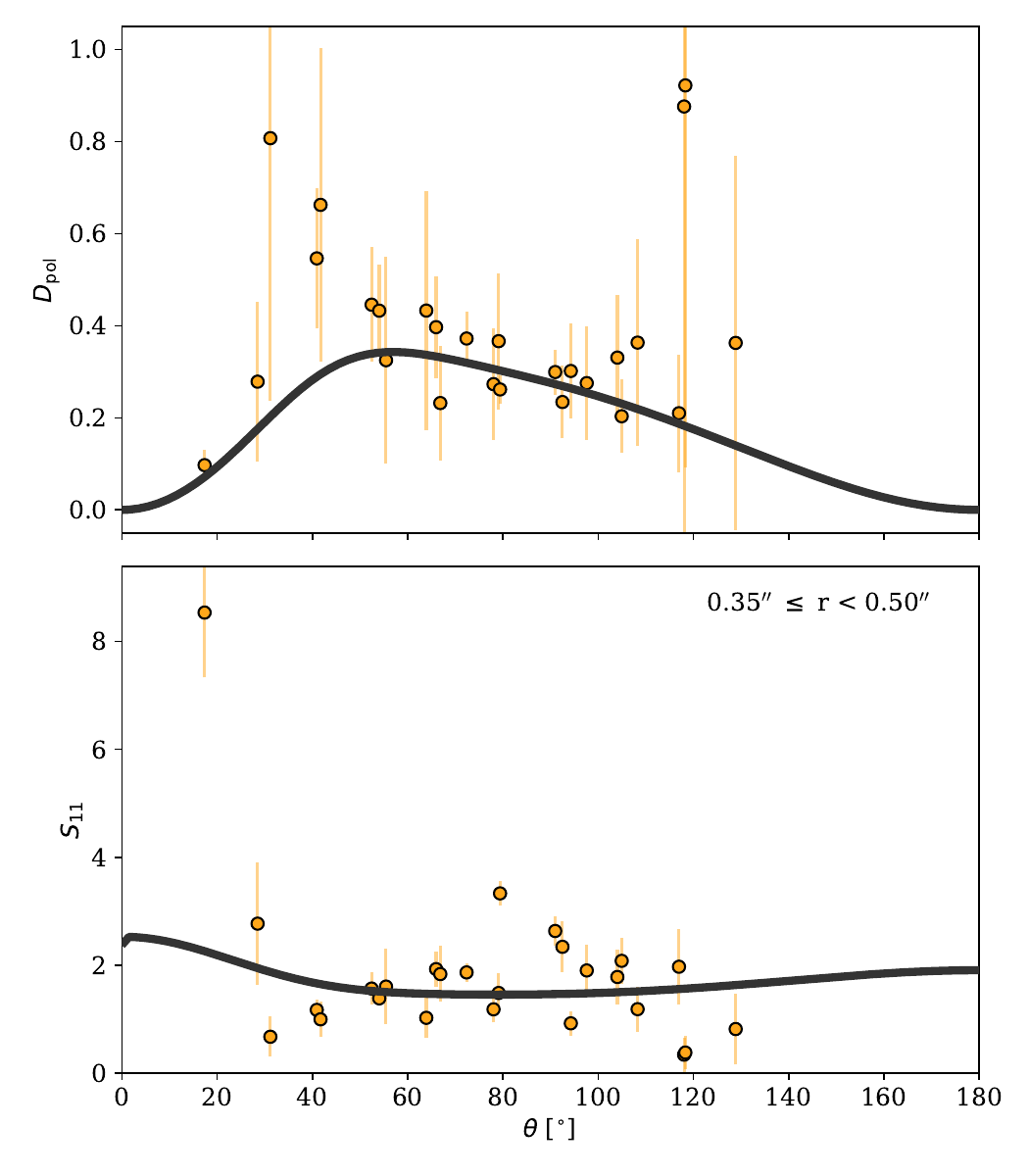}
    \caption{Same as Figure\,\ref{fig:HD157} for HD\,115600.}
  \label{fig:HD1156}
\end{figure}

Figure\,\ref{fig:HD1156} shows the observations and best fitting model for the disk around HD\,115600, where the degree of polarization is estimated between $0.35\arcsec < r < 0.5\arcsec$ (corresponding to $38 < r < 54.5$\,au). Overall, the results for this target are unfortunately not very reliable. The disk is the second faintest of our sample, it is more compact that the one around HD\,157587, and its inclination is larger. Combined with a smaller on-sky rotation during the star-hopping sequence, we are not able to accurately recover the projected semi-minor axis of the disk, especially in total intensity. As shown in the bottom panel of Fig.\,\ref{fig:HD1156}, it is very likely that we are missing the forward scattering peak, casting doubts on the robustness of the results. It seems unlikely that we are probing a very narrow size distribution ($s_\mathrm{min} \sim 0.1$\,$\mu$m and $s_\mathrm{max} \sim 0.5$\,$\mu$m) in the birth ring of the disk, on top of the unrealistic imaginary part for the optical constants ($k \sim 36$). Nonetheless, we can still reliably constrain the degree of polarization to be $\lesssim 40$\% along the major axis of the disk.

\subsection{HD 129590}\label{sec:HD129590}

\begin{figure*}
  \centering
    \includegraphics[width=\hsize]{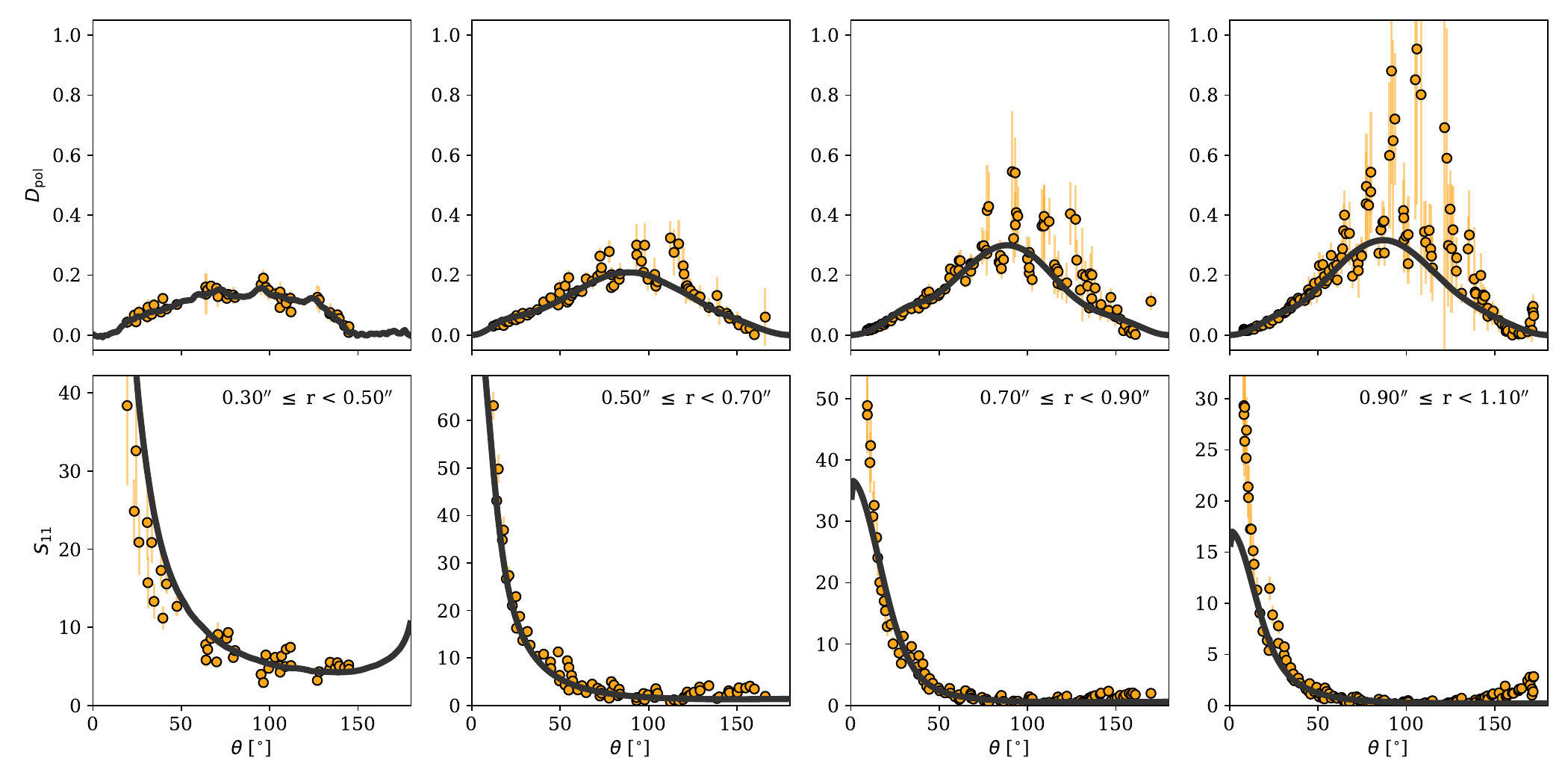}
    \caption{Same as Figure\,\ref{fig:HD157} for HD\,129590, but for four concentric annuli (see text and annotations in the upper right corner of the bottom panels).}
  \label{fig:HD129}
\end{figure*}

The disk around HD\,129590 is better recovered in total intensity compared to HD\,115600. According to Table\,\ref{tab:obslog}, the seeing and coherence time were not too dissimilar for the two datasets, and in fact, the observing conditions were on average better for HD\,115600 compared to HD\,129590. Besides the fact that the disk around HD\,129590 is brighter than the one around HD\,115600 (Table\,\ref{tab:star}), another main difference is the on-sky rotation during the observations, which is a factor two larger for HD\,129590\footnote{It should be noted that the declination of $-59^{\circ}$ of HD\,115600 makes it a challenge to increase significantly the parallactic rotation with typical $1-1.5$\,hours long observing blocks.}, significantly helping to recover the disk in total intensity. The halo beyond the birth ring is well detected both in total and polarized intensity, and as mentioned before, this is the only disk in our sample for which we are able to constrain the degree of polarization in the halo.

Since radiation pressure and gas drag (cold gas has been detected around HD\,129590, \citealp{Kral2020}) are size-sorting processes (see \citealp{Olofsson2022a} for the latter), there should be a size segregation at different stellocentric distances (\citealp{Thebault2014}). We can therefore attempt to estimate the degree of polarization as a function of the scattering angle for different regions beyond the birth ring. We proceed the same way as for HD\,157587 and HD\,115600, but instead of focusing only on the birth ring, we compute the degree of polarization in four concentric rings: $[0.3\arcsec, 0.5\arcsec]$ (encompassing the birth ring, $41 < r < 68$\,au), $[0.5\arcsec, 0.7\arcsec]$, $[0.7\arcsec,0.9\arcsec]$, and $[0.9\arcsec, 1.1\arcsec]$, corresponding to $[68, 95]$\,au, $[95, 123]$\,au, and $[123, 150]$\,au, respectively.

Figure\,\ref{fig:HD129} shows the results for the four regions. The top panels show the best fit models to the degree of polarization as a function of the scattering angle, and the bottom panels show the total intensity phase functions for the four different regions. According to the results reported in Table\,\ref{tab:dust}, all four regions require a minimum grain size that is below $0.5$\,$\mu$m (between $0.16$ and $0.48$\,$\mu$m). Regarding the optical constants, the value of $n$ and $k$ are all fairly comparable to each other outside of the birth ring, even if the different regions are modeled independently from each other ($2.4 < n < 3.3$ and $0.77 < k < 1.1$). In the birth ring, the best fit model requires a larger $n$ value and smaller $k$ imaginary part to reproduce the low degree of polarization. Interestingly, we see that the width of the size distribution is overall decreasing was we venture outside the birth ring. In the birth ring we find $s_\mathrm{max} \sim 110$\,$\mu$m, and the maximum grain sizes in the next three rings are $1.9$, $1.0$, and $1.2$\,$\mu$m, from the inner to the outer regions, respectively. Figure\,4 of \citet{Olofsson2023} shows the radiation pressure ratio $\beta$ as a function of the particle size $s$ for HD\,129590, and they found that the blow-out size for this solar-type star would be close to $1.2$\,$\mu$m. The comparison is not one-to-one though because they assumed a different dust composition (a mix of pyroxene and carbon). Nonetheless, the differences between the small minimum grain size derived here and the blow-out size reported in \citet{Olofsson2023} will be further discussed in Section\,\ref{sec:discussion}. It should however be noted that with the approach that we followed here, we cannot constrain the density of the dust particles, and therefore cannot compute the variation of $\beta$ as a function of $s$. Still, it seems that we are probing a wide range of sizes in the birth ring, but as we probe farther and farther out in the extended halo, the width of the size distribution decreases while the minimum size remains (relatively) constant, the expected behavior of radiation pressure.

\section{Toward a self-consistent model}\label{sec:unbound}

To provide additional context to our findings, we describe here an attempt at a more rigorous modeling approach. Since we are able to measure the degree of polarization in the extended halo for the disk around HD\,129590, we use this star as a reference. \citet{Thebault2019} showed that the contribution of grains that are on unbound orbits ($e \geq 1$ because of radiation pressure) can be significant. Since our results put very stringent constraints on the minimum and maximum grain sizes, our goal here is to assess if and how much these unbound grains can affect the degree of polarization. In the following, we first give a brief description of how synthetic images are computed, before describing the starting hypotheses for four fiducial cases.

\subsection{Model description}

The approach follows the methodology of the DyCoSS code (see \citealp{Thebault2012}), which is an iterative one. In a nutshell, we first draw $n_\mathrm{part}$ particles, following a size distribution $\mathrm{d}n(s) \propto s^{-3.5} \mathrm{d}s$, between $0.01$\,$\mu$m and $1$\,mm. As will be further discussed in Section\,\ref{sec:fiducial}, each particle is assigned a $\beta$ value. The central star has a mass $M_\star = 1.3$\,$M_\odot$ (as reported in \citealp{Kral2020} for HD\,129590, the star chosen as an example in this Section), and each particle feels a central mass of $M_\star (1-\beta)$. The $n_\mathrm{part}$ are released all at once, with their semi-major axis drawn from a normal distribution, centered at $48$\,au with a standard deviation of $1.5$\,au (at a distance of $136.32$\,pc this is meant to mimic the birth ring of the disk around HD\,129590). The opening angle of the disk is set to be constant with $h/r = 0.03$, and the mean anomaly is uniformly drawn between $-\pi$ and $\pi$. We then run a first N-body simulation, with a fine timestep (to ensure that we actually register the unbound grains), and at each timestep we save the position of all particles. We let the simulation evolve for $n_\mathrm{iter}$ timesteps and eventually stop the simulation. From the positions of the particles, and their sizes, we can compute an optical depth radial profile, which maximum\footnote{A more exact approach would be to compute the resulting spectral energy distribution and normalize the optical depth over the range of distances that contribute most to it, but the approach used here is sufficient for first-order estimates.} is normalized to $7 \times 10^{-3}$ (the fractional luminosity of the disk, \citealp{Esposito2020}). This first simulation cannot be used as is, since the result will depend on the total duration of the N-body integration. Therefore, we then iterate and run another simulation, with the same timestep, and saving the positions of the particles at each timestep. The main difference being that collisions are now accounted for (making use of the optical depth profile from the previous simulation), and the simulation will last until $99.9$\% of the initial $n_\mathrm{part}$ have been destroyed (more details can be found in \citealp{Olofsson2022a}). Since the unbound grains will never come back inside the birth ring, we set another condition for them to be destroyed. If they reach distances larger than $5\,000$\,au, they are removed from the simulation. Once $99.9$\% of the grains have been destroyed, we re-evaluate the optical depth profile, normalize it, and we can iterate once more, until two consecutive simulations have converged (in that case we check that the optical depth profiles are similar). The implicit assumption for this approach to be correct is that the disk is at steady-state, meaning that it will not become brighter or fainter over time.

Once we have iterated a sufficient number of times, we can produce images in scattered light. With the $(x,y,z)$ positions of the particles, saved at all timesteps, we can rotate them to account for the inclination and position angle of the disk ($i = 82^{\circ}$ and $\phi = -60.6^{\circ}$, see Table\,\ref{tab:ddit}), and calculate the value of the scattering angle. We used \texttt{optool} to compute the polarized and total intensity phase functions, at the wavelength of the observations, and assumed $n = 4.25$ and $k = 2 \times 10^{-3}$ (as derived for the birth ring of HD\,129590, Table\,\ref{tab:dust}, with a porosity of $25$\% and using $f_\mathrm{max} = 0.8$). The contribution of each particle is multiplied by its cross-section $s^2$, the value of the phase function (either in polarized or total intensity), and divided by the squared distance to the central star to account for illumination effects. The two images in polarized and total intensity are then convolved with the telescope point-spread function from the observations, and are then used to compute the degree of polarization image.

\subsection{Fiducial cases}\label{sec:fiducial}

\begin{figure}
  \centering
  \includegraphics[width=0.95\hsize]{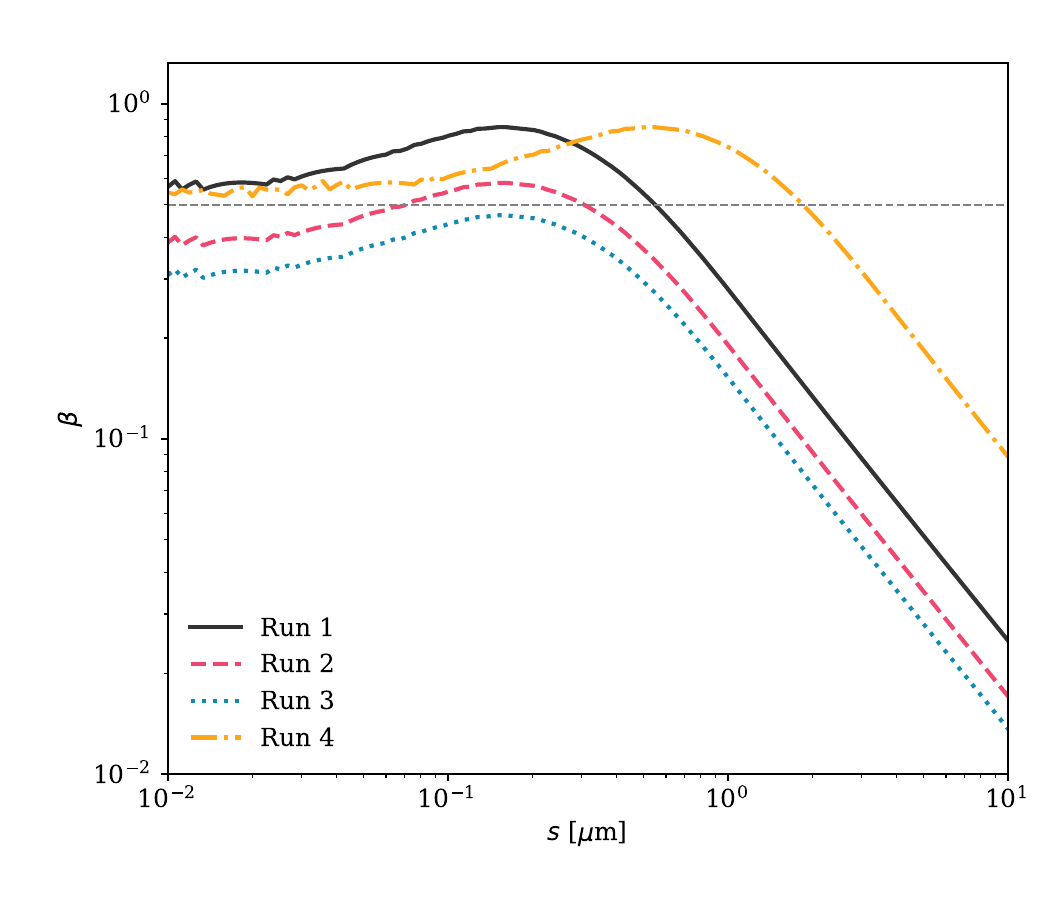}
    \caption{Radiation pressure $\beta$ ratio as a function of the grain size $s$ for three different simulations.}
  \label{fig:beta}
\end{figure}

It should first be noted that running one simulation can take up to several hours, producing very large files (tens of GB, depending on the integration timestep). This approach is therefore (currently) not very suitable to directly fit observations, hence the choice of very specific cases in this Section.

To better illustrate the impact of $\beta$ as a function of the size $s$, we investigate four fiducial cases using the different $\beta(s)$ curves shown in Figure\,\ref{fig:beta} (all the other paramters remaining the same otherwise). The motivation is to use some of the results we presented in Section\,\ref{sec:results} as inputs for the models, compute synthetic images, and qualitatively compare them to the original observations. For Run\,1, there is a clear cut-off at the size $s_\mathrm{blow-out}$ ($\lesssim 0.5$\,$\mu$m, see Section\,\ref{subsec:blow-out} for further discussion) and all the grains smaller than this size are set on unbound orbits. This blow-out size was chosen to be quite close to the minimum grain size derived for the birth ring of HD\,129590 ($0.35$\,$\mu$m) For Run\,2, $\beta(s)$ crosses the threshold $\beta = 0.5$ twice and therefore the smallest grains are bound to the star, but there is a small interval of sizes for which the particles will be unbound. For Run\,3, none of the grains are unbound, but $\beta$ reaches a maximum value of $0.47$, and finally, Run\,4 is quite similar to Run\,1, except that the blow-out size is larger, with $s_\mathrm{blow-out} \sim 1.85$\,$\mu$m.
\subsection{Diagnostics and run comparison}

\begin{figure*}
  \centering
  \includegraphics[width=0.95\hsize]{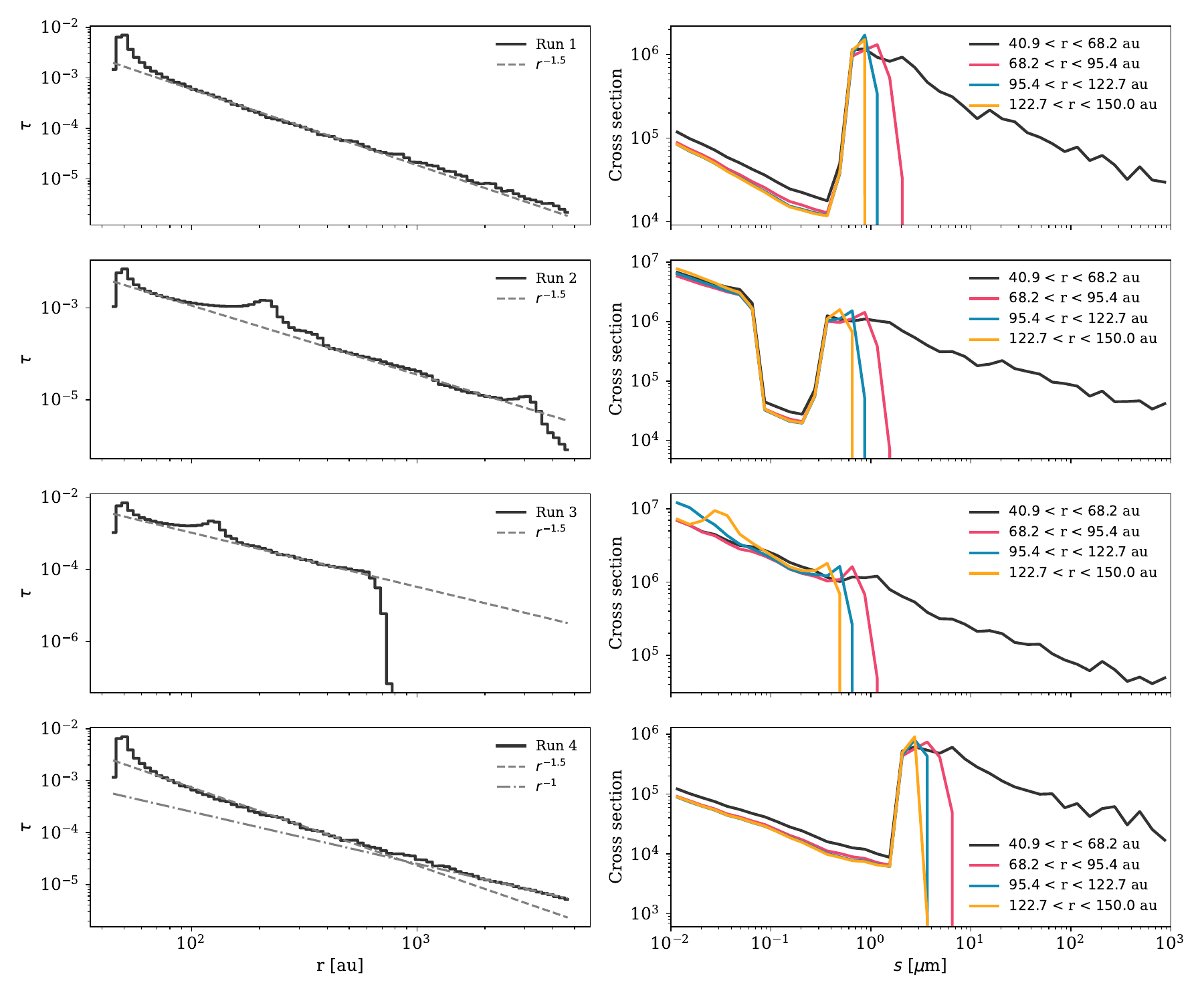}
    \caption{\textit{Top to bottom:} diagnostics for Runs\,1 to 4, respectively. \textit{Left:} optical depth profile as a function of the stellocentric distance. The dashed line shows a profile in $r^{-1.5}$. For the bottom panel, the dashed-dotted line follows a profile in $r^{-1}$. \textit{Right:} cross-section as a function of the grain size $s$. The different colors show the cross-section measured in several concentric rings.}
  \label{fig:diag}
\end{figure*}

It is first interesting to check a few diagnostics plots to better understand the differences between the three runs. The left panels of Figure\,\ref{fig:diag} shows the final optical depth profiles as a function of the stellocentric distances, while the right panels show the cross-section as a function of the grain size $s$, estimated in four concentric annuli (not normalized with respect to each other). The regions are at the same stellocentric distances as the regions used for the analysis of HD\,129590, assuming a distance of $136.32$\,pc. 

Starting with Run\,1, the optical depth beyond the main ring (centered at 48\,au) follows a power-law in $-1.5$, which is expected for the steady-state evolution of a debris disk (\citealp{Thebault2008}). Regarding the contribution of different grain sizes to the geometrical cross-section for Run\,1, the blow-out size is clearly identifiable by the sharp drop at around $0.4$\,$\mu$m, visible for all four regions. For sizes that are smaller than this critical size, the cross-section increases again for smaller $s$, as the size distribution of the released particles in $\propto s^{-3.5}$ is very top-heavy. It is interesting to note that the farther the region considered is, the narrower the distribution is. This is the expected behavior for radiation pressure; particles with a given $\beta$ value, released from a distance $a_0$ can only reach separations up to $\sim a_0/(1-2\beta)$.

For Run\,2, for which there is only a narrow range of sizes with $\beta > 0.5$, the optical depth profile differs from the one of Run\,1. There is a significant ``bump'' at $\sim 200$\,au. This bump is caused by the small-end part of the size distribution. Grains with $s \lesssim 0.03$\,$\mu$m all have similar $\beta$ values close to $0.4$ (the plateau on the left side of Fig.\,\ref{fig:beta}). The bump shows the location of the apocenters of these, very numerous and long-lived (\citealp{Thebault2008}) particles that all have similar $\beta$ value ($a_0/(1-2 \beta) \sim 48/(1-2 \times 0.4) = 240$\,au). Despite this local increase in optical depth, the profile otherwise follows a power-law compatible with $r^{-1.5}$. Looking at the size-dependent cross-section plot, the range of sizes for which the grains are unbound is easily identified by the strong dip between $0.1$ and $0.3$\,$\mu$m. The decrease of the maximum size as a function of the stellocentric distance is quite comparable to the first Run.

For Run\,3, there are no unbound grains, and for the bound grains, the $\beta$ value reaches a maximum of $0.47$. This explains the sharp drop in optical depth at $\sim 800$\,au (indeed $a_0/(1-2 \beta) = 48/(1-2 \times 0.47) = 800$\,au). The ``bump'' at $\sim 120$\,au is caused by the plateau of $\beta(s)$ for very small particle sizes. The bump is closer to the star, because the $\beta$ values flatten at $\beta \sim 0.3$ in this case (compared to $0.4$ for Run\,2). For the cross-section, since there are no unbound particles, there is no discontinuity on the small end of the size distribution. We still see the expected behavior of radiation pressure, as we probe farther and farther out of the birth ring, the maximum grain size decreases.

Run\,4 being in nature very similar to Run\,1, the differences between the two runs are quite minute. The most notable difference, besides the size for which the cross-section is dropping, concerns the optical depth profile at large separation from the star. We can see that the slope deviates from the $r^{-1.5}$ profile and flattens. This is because the unbound grains are much more numerous in this example ($s_\mathrm{min}$ remaining the same but $s_\mathrm{blow-out}$ being larger), and their contribution rather follows a slope in $r^{-1}$, meaning that their relative contribution to the total flux (compared to bound grains) increases with the distance to the star (e.g., Fig.\,2 of \citealp{Thebault2023}). This can best be seen on the lower left panel of Figure\,\ref{fig:diag}, showing that the contribution of the unbound grains takes over the one from bound grains at a separation of $600-700$\,au.

\subsection{Synthetic images}

\begin{figure*}
  \centering
  \includegraphics[width=0.95\hsize]{HD129590_full.pdf}
  \includegraphics[width=0.95\hsize]{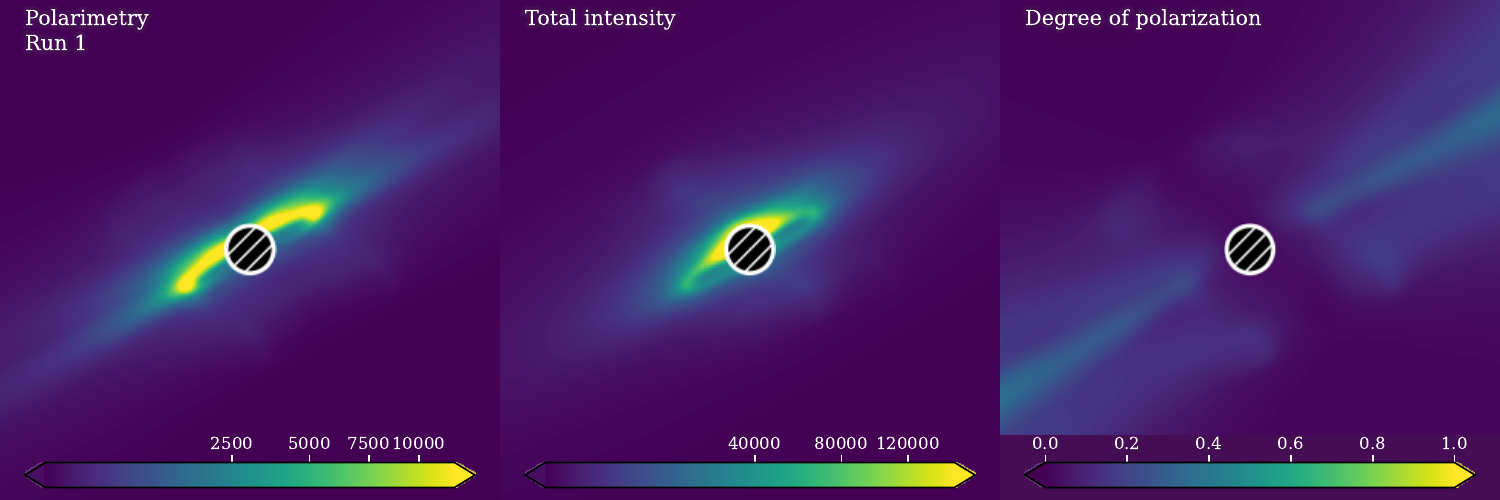}
  \includegraphics[width=0.95\hsize]{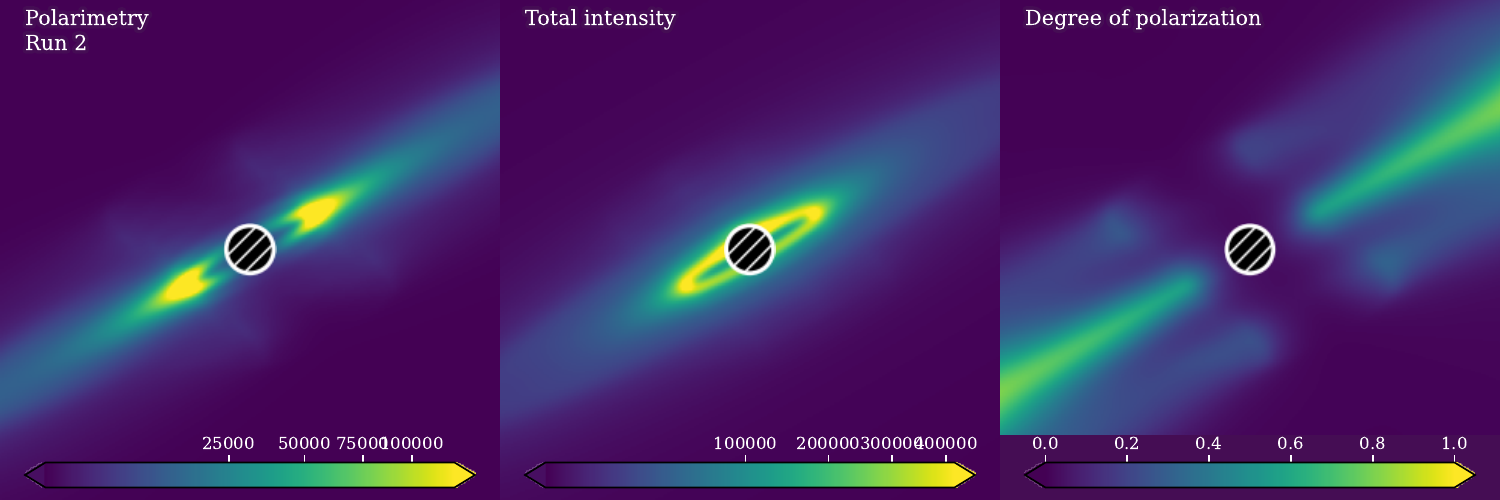}
  \includegraphics[width=0.95\hsize]{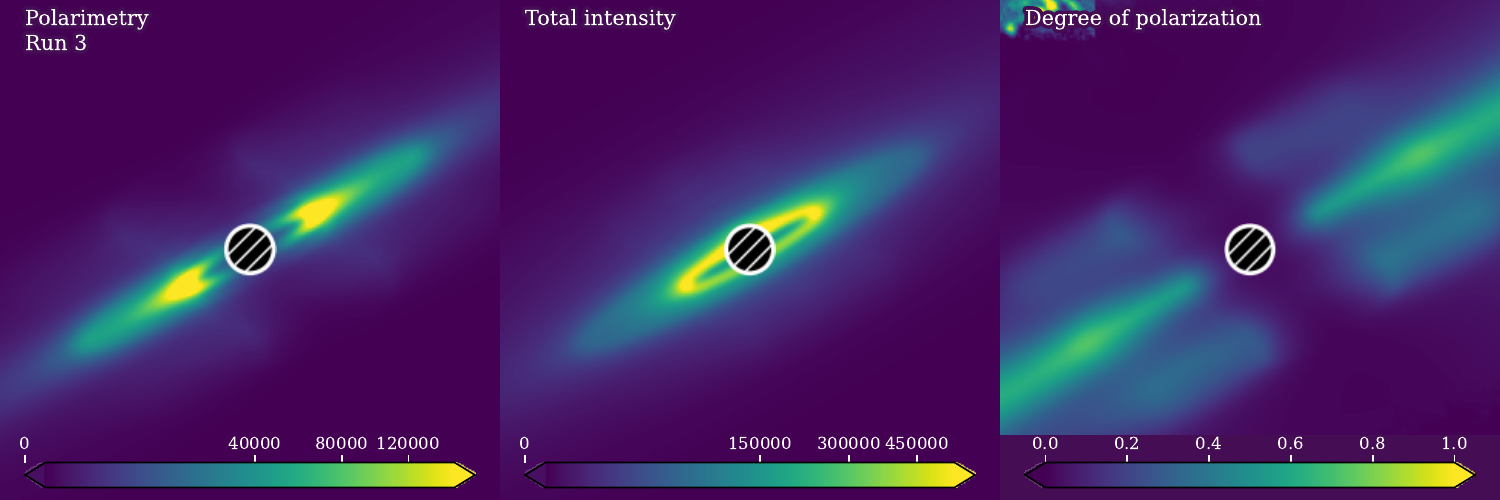}
    \caption{Comparison of the observations (top panels) with the first three models that include the contribution of unbound grains. Left and middle panels show images in polarized and total intensity, respectively, while the right panels show the degree of polarization.}
  \label{fig:unbound}
\end{figure*}

Figure\,\ref{fig:unbound} shows the observations for HD\,129590 at the top, and the images for Runs\,1 to 3. From left to right, we show the polarized intensity, total intensity, and degree of polarization. We can likely rule out the presence of bound very small particles in the disk around HD\,129590. Indeed for Runs\,2 and 3, the major axis of the disk is too bright in polarimetry, and the back side of the disk is too bright in total intensity. The degree of polarization also seems to be too large compared to the observations. This is because very small dust particles have a polarized phase function peaking at $90^{\circ}$ scattering angles, an almost isotropic total intensity phase function, and in general a large degree of polarization.

Conversely, the results of Run\,1 are a validation of our modeling approach presented in Section\,\ref{sec:analysis}. By using the results from the modeling of the birth ring of HD\,129590 (optical constants $n$ and $k$) and using a $\beta(s)$ function that yields a sub-$\mu$m blow-out size below which all grains are unbound, we obtain a good match to the observations not only in the birth ring but also in the extended halo. Indeed, in the polarimetric images, the back side of the disk is equally faint as in the observations, while the front side appears with similar brightness. In total intensity we are able to reproduce the strong forward scattering along the minor axis, as well as the apparent bulge associated with it. The most notable difference between the observations and the model being that the halo might be slightly fainter in the model compared to the observations, and the arc reported in \citet{Olofsson2023} is not as visible in total intensity. This could be due to our parametrization of the birth ring, using a normal distribution. Using an asymmetric distribution that extends farther out might help with the transition between the birth ring and the halo. Even though we are not fitting the observations, the degree of polarization seems to agree quite well with the observations. Its value increases as a function of stellocentric distances, though not as much as in the observations, only from $\sim 20$\% in the birth ring, up to $\sim 30$\% in the outer regions. A possible explanation could be that the blow-out size for Run\,1 is close to $0.5$\,$\mu$m (Fig.\,\ref{fig:diag}), while we found a minimum grain size of $0.35$\,$\mu$m when fitting the observations (Table\,\ref{tab:dust}).

\subsection{Unbound grains, minimum grain size and blow-out size}

\begin{figure*}
  \centering
  \includegraphics[width=0.95\hsize]{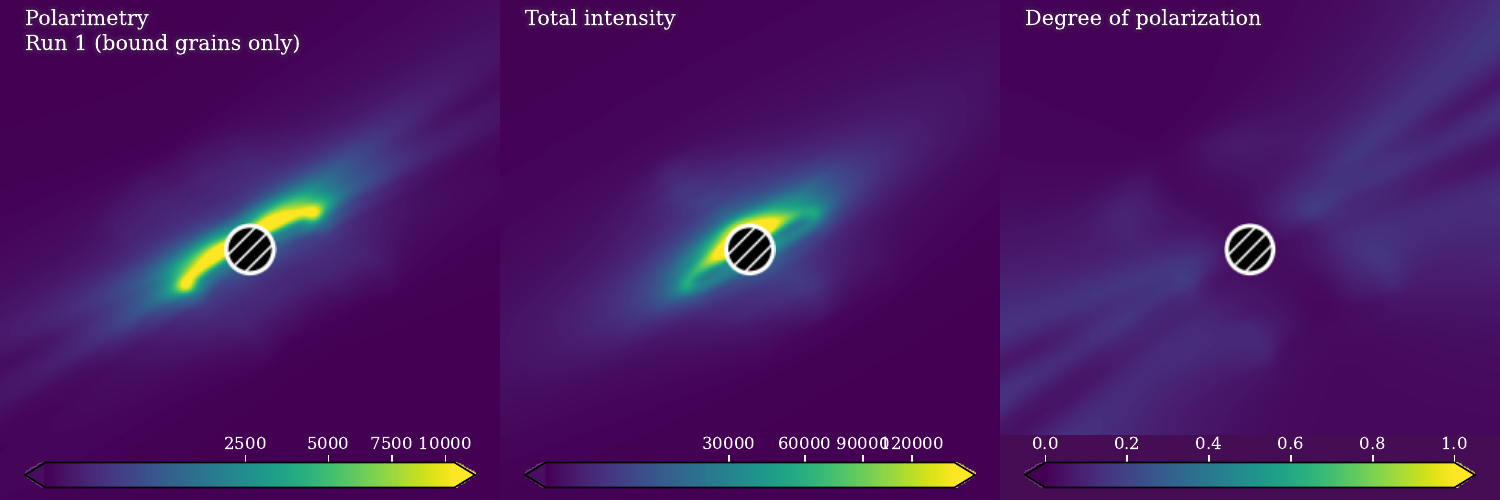}
  \includegraphics[width=0.95\hsize]{HD129_1_images.pdf}
  \includegraphics[width=0.95\hsize]{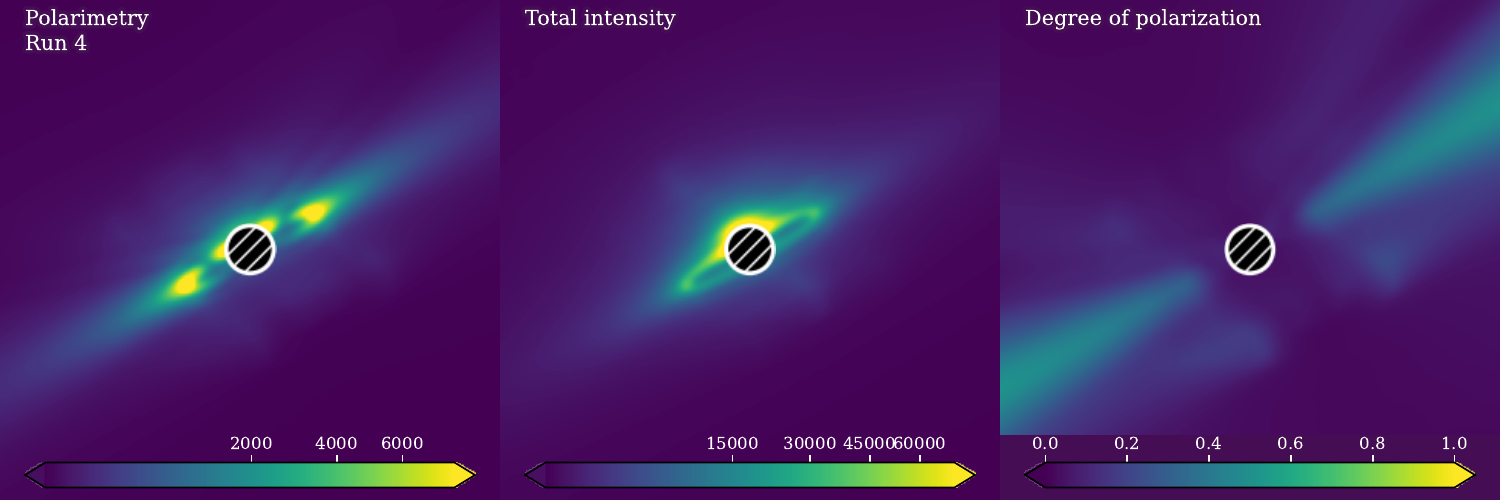}
    \caption{Images for Run\,1 without (\textit{top}) or with (\textit{middle}) the contribution of unbound grains. The \textit{bottom} images show the results for Run\,4 for which the blow-out size is larger (the contribution of unbound grains is included).}
  \label{fig:only_bound}
\end{figure*}

Until now we only investigated the effect the shape of the $\beta(s)$ curve can have on the final images in polarized and total intensity as well as for the degree of polarization. If one wished to ignore the contribution of unbound grains, there are simpler approaches to compute images (e.g., \citealp{Lee2016}, \citealp{Olofsson2022b}). It is therefore worth investigating how much unbound grains impact the final scattered light images and degree of polarization (see also \citealp{Thebault2019} for a more in-depth analysis). For Run\,1 we re-computed scattered light images, but this time ignoring any particle with $\beta \geq 0.5$. The resulting images are shown in the top panels of Figure\,\ref{fig:only_bound}, along side the previous images for Run\,1 (including the contribution of unbound grains) in the middle panels.

At first glance, the differences seem to be quite minute, but upon closer inspection we can see a negative branch in the polarimetric image, and that the front side of the disk appears bulkier compared to the original Run\,1. The biggest difference lies in the degree of polarization. It appears more structured, because of the negative polarization branch, but most importantly, the maximum degree of polarization does not vary with increasing stellocentric distances and remains overall quite low ($\sim 15$\%, see later). 

\begin{figure*}
  \centering
  \includegraphics[width=0.95\hsize]{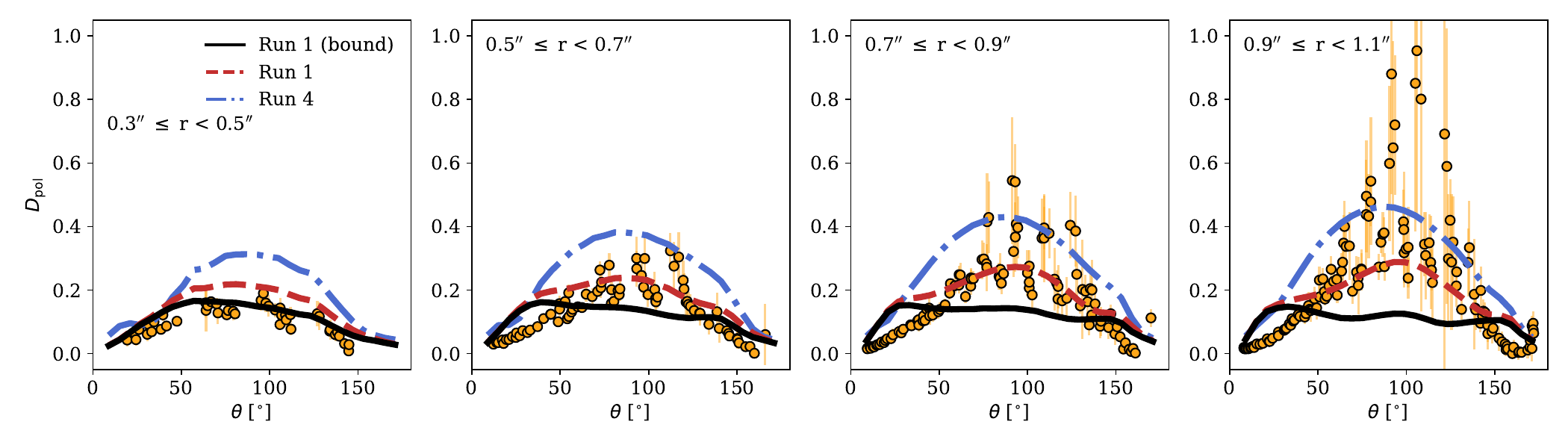}
    \caption{Degree of polarization as a function of the scattering angle for four concentric regions. Observations for HD\,129590 are shown in orange circles. The different lines correspond to the degree of polarization measured from the modeled images for Run\,1 (with and without the contribution of unbound grains) and Run\,4.}
  \label{fig:model_dpol}
\end{figure*}

We can test the hypothesis mentioned in Section\,\ref{sec:analysis}, where we equated the minimum size to the blow-out size. We just discussed the importance of unbound grains and it poses the question of whether their contribution might ``contaminate'' our estimation of $s_\mathrm{min}$ when modeling the observations as we did previously. Run\,4 was computed to begin addressing this point, as the blow-out size is larger ($1.85$\,$\mu$m versus $0.5$\,$\mu$m), but the contribution of unbound grains is included. The bottom panels of Figure\,\ref{fig:only_bound} show that the total intensity image is quite comparable to the one of Run\,1 but the degree of polarization shows slightly larger values ($\sim 30$\% in the birth ring). A positive note is that in this case as well the degree of polarization is increasing with the stellocentric distance, in line with the observations. But the polarimetric image appears quite different than the observed one: there is a strong negative polarization branch for scattering angles close to $\sim 50^{\circ}$. Since the image shows the absolute value of the pixels, this causes this apparent ``null branch'' along the minor axis of the disk, which is not seen in the observations.

This is further quantified in Figure\,\ref{fig:model_dpol}, showing the degree of polarization as a function of the scattering angle, in a similar way as it was measured for the observations, in four different concentric regions. The black dotted line shows the results for Run\,1 without the contribution of unbound grains, the red solid line corresponds to Run\,1 with the unbound grains, and the dashed blue line is for Run\,4, including the unbound grains as well. For the birth ring (leftmost panel), the best agreement is found for Run\,1 without the unbound particles. This is because this is the closest match to the fitting results obtained in Section\,\ref{sec:results}. But as soon as we start venturing outside of the birth ring, the maximum degree of polarization remains constant and fails to reproduce the observations. When the contribution of unbound grains is included, we see that the degree of polarization is slightly too large in the birth ring (though not in total disagreement). However, as we go farther and farther, the degree of polarization continuously increases with separation. This is even more so the case for Run\,4, but in all four regions, the degree of polarization is too large compared to the observations. Overall, this underlines the importance of including unbound grains to best reproduce the observations.

To summarize the main findings of this Section, first, this greatly highlights the importance of not only obtaining observations in polarized and total intensity but also determining the degree of polarization in debris disks, since it can help disentangling otherwise relatively similar images. Second, this also shows that despite their smaller contribution to the cross section (right side of Fig.\,\ref{fig:diag}), the contribution of unbound grains does help reproducing the observations and should therefore not be disregarded, thus strengthening the results of \citet{Thebault2019}. As discussed in \citet{Thebault2023} the relative contribution of unbound grains increases with the stellocentric distance, and we hypothesize that this is the main driver for the degree of polarization to increase with stellocentric distance. As we probe farther and farther out, the average grain size is effectively decreasing. Indeed, small unbound grains are contributing more and more, while the width of the size distribution becomes narrower due to radiation pressure. Since smaller particles are expected to have larger degrees of polarization this provides a natural explanation for the increase in the degree of polarization with stellocentric distance. Nonetheless, considering only the increase of the degree of polarization is not sufficient to fully explain the observations if its absolute value is not correct (e.g., $\sim 15$\% versus $\sim30$\% in the birth ring). This exercise shows that indeed the contribution of unbound grains does contaminate slightly the fitting approach described in Section\,\ref{sec:results} (Run\,1 with or without unbound grains, leftmost panel of Fig.\,\ref{fig:model_dpol}), but that \textit{(i)} this contamination is relatively marginal in the birth ring and \textit{(ii)} that unbound grains must be included to reproduce the increasing degree of polarization in the extended halo. Overall, the minimum grain size that we infer when modeling the observations has to be close to the blow-out size.

\subsection{Future perspectives}

In its current implementation, the approach described in this Section cannot easily be used to perform complex modeling given the time it takes to obtain the final images. That being said, given its overall simplicity, there are reasons to be optimistic, especially when considering additional observations that could help decrease the number of free parameters. There are already several ALMA datasets at high angular resolution (e.g., \citealp{Marino2016} for HD\,181327; \citealp{Kennedy2018} for HR\,4796; and \citealp{Daley2019} and \citealp{Vizgan2022} for AU\,Mic), and more are coming with the ALMA ARKS large program. From these high angular resolution, one can place stringent constraints on some geometric parameters (inclination, position angle, and quite possibly the opening angle of the disk) as well as on the surface density profile of the birth ring. The remaining free parameters should be related to the dust optical properties, either a composition (or mix of compositions), or in the ``worst case scenario'', values for $n$, $k$, and a curve to relate $s$ to $\beta$. This could open the possibility of performing at least a coarse exploration of the parameter space and provide novel constraints on the dynamics of the dust particles.

\section{Discussion}\label{sec:discussion}

\subsection{Radiation pressure and blow-out size}\label{subsec:blow-out}

Even though the results of Run\,1 seem to agree with the observations of HD\,129590, we must acknowledge possible pitfalls or limitations of the approach. Some of the modeling results obtained in Section\,\ref{sec:results} might seem challenging or unrealistic, especially when it comes to the inferred blow-out size (the optical constants will be discussed in Section\,\ref{subsec:optical}). Leaving aside the results for HD\,115600 (since the projected minor axis of the disk is not well detected), we have one F- and one G-type star for which we derive $s_\mathrm{min} \leq 0.1$\,$\mu$m and $s_\mathrm{min} = 0.36$\,$\mu$m, respectively. These minimum sizes appear to be very small and, if representative of the blow-out sizes, would severely challenge our understanding of the strength of radiation pressure.

We here argue that this apparent challenge is not a real issue and that it is most likely due to limitations of available light scattering models that can properly describe the morphology of the dust particles. We want to emphasize that from a dynamical point of view what really matters is the distribution of $\beta$ values. This is the quantity that governs the final dust density distribution as a function of the stellocentric distance. Associating a given size $s$ to a given $\beta$ value is, in the end, nothing more than a ``necessary evil'' to account for the phase function (even the cross-section could in the end be approximated by $1/\beta^2$ for sizes large enough).

For the remainder of this discussion we will focus on the case of HD\,129590, since we are able to measure the degree of polarization in the halo, and were able to propose a self-consistent model that explains the observations reasonably well. The lesson learned from the exercise presented in Section\,\ref{sec:unbound} is that if we have a good description of the dust properties in the birth ring, then we are able to reproduce the observations quite well (even beyond the birth ring). The only caveat being that we need a $\beta(s)$ curve in agreement with the inferred dust properties; in other words, a blow-out size close to the minimum size derived from fitting the phase function and degree or polarization (as discussed previously in Section\,\ref{sec:unbound}).

The apparent challenge of finding very small minimum grain sizes therefore all boils down to the choice of the light scattering model used to fit the observed profiles as a function of the scattering angles. As discussed in Section\,\ref{subsec:preliminar}, we tried several approaches, either trying mixtures of known materials, or pre-computed optical properties of complex geometries. We could only obtain a good match to the observations using the DHS model, combined with rather unusual optical constants (see Section\,\ref{subsec:optical}). We were not able to use the library of phase functions provided in the \texttt{AggScatVIR} library mostly because they failed to reproduce the low degree of polarization observed in the birth rings of the disks. We hypothesized that this is because the largest particle size in the library only goes as far as a few $\mu$m while we should expect to probe a much wider size distribution in the birth ring. It would be very informative to test how the optical properties integrated over a wider size distribution, including aggregates at least for the small-end of the distribution, would look like and what the inferred minimum grain size would be in such case. Unfortunately, this cannot be tested at the moment. 

Still, aggregate grains remain an interesting avenue to further explore. For instance, \citet{Arnold2022} obtained promising results when modeling observations of AU\,Mic using agglomerated debris particles. Despite the challenging edge-on configuration of the debris disk, they show that using aggregates does help reproducing the degree of polarization along the major axis of the disk. Furthermore, \citet{Tazaki2022} showed for instance that some aggregates can yield maximum degree of polarization as low as $\sim 40$\% at a wavelength of $1.6$\,$\mu$m. Porous aggregates, even with an equivalent radius of $2$\,$\mu$m can have relatively low maximum degree of polarization if the size of the individual monomer is sufficiently large ($0.4$\,$\mu$m). Interestingly, this is quite close to the minimum grain size that we infer for the birth ring of HD\,129590, and as the authors mention, the polarization properties of the aggregate as a whole are more correlated to the properties of the monomers themselves, because of the additional internal scattering events that can happen inside the aggregate itself. This was also discussed in \citet{Min2016}, who argued that polarization depends on the size of the monomers, while total intensity observations are more sensitive to the size of the aggregates themselves.

We therefore suggest that our observations could be mostly sensitive to the size of the individual monomers. Because we are using the DHS model, we cannot at the same time have the optical properties of a much larger aggregate particle, leading us to actually under-estimate the minimum grain size in the distribution. To some extent, this is further supported by the fits to the total intensity phase functions, where the forward scattering peak is in some cases under-estimated. Larger aggregates should in principle lead to stronger forward scattering peaks, while preserving the polarized flux, in better agreement with the observations. If correct, this means that the minimum grain size we obtain is in fact a lower limit for the sizes of the particles.

\subsection{Optical constants}\label{subsec:optical}

Our modeling results of the degree of polarization and total intensity phase function from Section\,\ref{sec:results} suggest relatively large $n$ and very large $k$ values for the particles optical constants. Other studies have already reported similar issues in the past, for instance \citet{Duchene2020} for the disk around HD\,32297 ($n \sim 3.8$, $\mathrm{log_{10}}k \sim -1.4$), \citet{Arriaga2020} for HR\,4796 ($n = 3.4$ and $k = 3.7$ when fitting jointly polarized and total intensity observations), or \citet{Milli2024} for HD\,181327 ($n = 3.4$, $k = 1$), all of these studies having used DHS (and also compared with the Mie theory as well). All three studies summarized their findings by comparing the derived optical constants with those from known materials (e.g., astronomical silicates, iron, carbon, among others), highlighting that the inferred real and imaginary parts are quite unusual, and we refer the interested reader to these works for further information.

Instead, we focus here on the possible limitations of the light scattering theory that we used to model our observations, to further pursue the discussion initiated in Section\,\ref{subsec:blow-out}. \citet{Munoz2021} derived the total intensity phase function and degree of polarization for several samples of forsterite particles, for various narrow size distributions. Their sample ``XS'' has a size distribution between $0.1$ and $1.0$\,$\mu$m, while the ``XL'' sample has a size distribution between $\sim 20$ and $100$\,$\mu$m. They also present the results for three other samples that have intermediate distributions between these two (``S'', ``M'', and ``L'', respectively). They derive maximum degree of polarization smaller than $20$\% for all five samples, but also see a strong negative branch for the samples ``XS'' and ``S'' that is not seen for the other three samples. The total intensity phase functions show strong forward scattering peaks, for all five samples, as well as some backward scattering. Overall, their results for samples ``M'', ``L'', or even ``XL'' might compare well (visually at least) with our results obtained for HD\,129590 (for which we have the most reliable measurements and the widest range of scattering angles), these samples having size distributions in the range $\sim 1-10$ or $20-100$\,$\mu$m.

Of greater interest for this discussion, \citet{Munoz2021} also investigated the dust properties that would be inferred by modeling their laboratory measurements with the Mie theory, in other words, performing the same exercise we did in Section\,\ref{sec:results}, but with full knowledge of the ground truth. They show that to be able to reproduce the total intensity phase functions of their low-absorbing, irregular forsterite grains ($k \sim 10^{-5}$), they need to artificially increase the absorptivity of the spherical grains and require $k$ values of $10^{-2}-10^{-1}$. They performed a similar exercise on the degree of polarization as a function of the scattering angle. Only changing the imaginary part $k$ was not sufficient to match the observations, and they therefore used a size distribution of particles. Their results indicate that the inferred grain sizes are severely under-estimated when assuming spherical grains. They also find that the imaginary part of the optical constants are over-estimated, as well as the real part $n$. Overall, even though they used the Mie theory, implying a spherical geometry for the particles, while we use the DHS model more akin to irregular particles, this echoes resoundingly our findings of (very) large $(n,k)$ values, and small minimum sizes. Furthermore, \citealp{Munoz2021} did not attempt to simultaneously fit the degree and total intensity phase functions as we attempted here. 

\subsection{Future perspectives}

The bottleneck clearly seems to be related to how the optical properties of dust particles are computed (the phase functions, degree of polarization but also the variation of $\beta$ as a function of $s$). Despite the slightly pessimistic tone of the discussion in Sections\,\ref{subsec:blow-out} and \ref{subsec:optical}, we show in Section\,\ref{sec:unbound} that if we can find a reasonable match to the optical properties in the birth ring, regardless how ``unrealistic'' they might seem, then the rest of the pieces fall in place together. With the adequate $\beta(s)$ function, and accounting for the contribution of the unbound particles, we are able to reproduce the observations on a wide range of spatial scales. Even if the absolute values remain highly uncertain, the relative values for $n$ and $k$ for instance can still be used to better understand the kind of cosmic dust debris disks are made of. It is indeed quite interesting that we find similar values as for the disk around HD\,32297, HR\,4796, and HD\,181327, most of these studies being performed using different models, assumptions, and approaches. Even though we are clearly in the small statistics regime, regardless of the exact composition of the dust particles, it seems to be relatively similar around different stars, of different spectral types. Constraining the degree of polarization (and not only the polarized or total intensity phase functions) for a larger sample of young debris disks is therefore crucial to better understand the building blocks of planetesimals.

Additionally, one could try to step away from depending on light scattering models when trying to reproduce observations. Recently, both \citet{Lawson2021} and \citet{Hom2024} took steps in this direction, as both studies fixed the total intensity phase function, based on measurements of solar system bodies or other debris disks. However, the phase function they used still remains independent of the grain size in both studies and it would be the same in the birth ring and halo of the disk. Nonetheless, we can still think of other approaches to depend less on the most commonly used light scattering models. The first one would be to pursue the work on optical properties of aggregates to increase the maximum size of the particles, possibly setting some limits on the fractal dimension or ``fluffiness'' of the largest grains. Another approach would be to directly make use of laboratory measurements such as the ones presented in \citet{Munoz2021}. Granted, there is ``only'' one composition analyzed, but there are several samples, of a wide range of sizes, that are almost fully characterized, the main ingredient missing being the radiation pressure efficiency in order to compute $\beta(s)$. Otherwise, one could interpolate the phase function and degree of polarization over the different sizes, and such a model would provide stringent constraint on the strength of radiation pressure, $\beta(s)$ being one of the few free parameters of the model (assuming the composition is indeed representative of dust in debris disks).

Another approach, which needs maturing, would be to rely mostly on the observations. With high angular resolution ALMA observations, the geometric parameters of the birth ring can be derived quite accurately. Using this prior knowledge it might be possible to work our way back to the optical properties, in a data driven way. First the degree of polarization and polarimetric (or total intensity) images are highly complementary. Indeed, the former does not depend on the dust density distribution, meaning that we have direct access to some of the optical properties, which are not entirely disconnected from what is measured in the polarimetric (or total intensity) images. Furthermore, using cross-section plots such as the one presented in the right panels of Fig.\,\ref{fig:diag}, it might be possible to retrieve the dust properties outside-in. Indeed, the outermost regions probe a relatively narrow range of grain sizes. As we probe closer in, the maximum grain size of the distribution will increase. In \citet{Olofsson2020} we presented another data driven approach to model polarimetric observations in which the polarized phase function is an output of the modeling, and it might be possible to expand on a similar approach, informed by the dynamics imposed by radiation pressure. 

\section{Summary}

In this paper, we presented new SPHERE observations of four debris disks, with the aim of measuring the degree of polarization. We show that this remains a challenge, despite significant progress on the instrumentation side as well as post-processing techniques; HD\,191089 is not detected in total intensity, the results for HD\,115600 cannot be exploited to fully constrain the dependence with the scattering angle (since the minor axis is not well recovered). Nonetheless, we could measure the degree of polarization for the birth ring of the disk around HD\,157857, and more spectacularly, we measured the degree of polarization in the birth ring as well as in the extended halo of the disk surrounding HD\,129590.

For the three disks for which we could determine the degree of polarization in the birth ring, we find small maximum degree of polarization $\lesssim 40$\% (with some dispersion for HD\,157857), which remains challenging to model assuming compact spherical grains. Our attempt to reproduce the observations using a pre-computed library of optical properties of aggregates also revealed to be challenging, which we interpreted as the contribution of larger particles, as expected in the birth rings of debris disks. We were only able to find satisfactory match to the observations by ``making up'' our own dust, leaving the optical constants $(n,k)$ as free parameters. Nonetheless, for the disk around HD\,129590, we are also able to constrain the degree of polarization in the halo, and fitting the data independently, we do find that the results are quite consistent with each other (except for the maximum grain size), which brings confidence in our approach. For this disk, for the regions outside of the birth ring, we find that the maximum grain size is much smaller compared to the inner regions, which is naturally explained by the stellar radiation pressure; only small particles are set on eccentric orbits and can venture outside where they were released.

We vetted around findings by using them as inputs of a self-consistent model, including the contribution of unbound particles, and presented our results in Section\,\ref{sec:unbound}. The main conclusion of this exercise is that we can reproduce the observations fairly well only if unbound grains are included in the final images and if the blow-out size is close to the minimum grain size inferred from the modeling. The driving argument to include the contribution of unbound particles is to reproduce the increase of the degree of polarization with stellocentric distances, which we argue is the consequence of a decreasing average effective size with increasing separation.

Our modeling results presented in Section\,\ref{sec:results} might appear unusual on several aspects, namely the inferred minimum grain size, as well as the optical constants. We discussed at length in Section\,\ref{sec:discussion} that the root cause for these (apparent) inconsistencies is most likely related to the light scattering model that we used (though the only one that could provide a good fit to the data). In spite of the challenges underlined in Section\,\ref{sec:discussion}, we discussed possible avenues to circumvent some of these issues and further our understanding of cosmic dust in young debris disks. The new approach presented in Section\,\ref{sec:unbound} to produce synthetic near-IR scattered light images requires very few free parameters, as the dust density distribution in the halo is solely governed by the effect of radiation pressure. We showed that if we have a suitable solution for the dust properties in the birth ring, we are then able to (at least qualitatively) reproduce the extended halo beyond the birth ring. Such an approach will greatly benefit from high angular resolution ALMA observations in the near future.

\begin{acknowledgements}
    We thank the anonymous referee for a constructive, timely, and helpful report. The comments helped strengthen the results of this paper, especially (but not only) with respect to self-subtraction effects, as well as improving the overall clarity of the paper.
We are thankful to Ryo Tazaki for pointing out interesting references that helped shape the discussion of this paper.
J.\,M. acknowledges funding from the European Research Council (ERC) under the European Union’s Horizon Europe research and innovation program (grant agreement No. 101053020, project Dust2Planets, PI F.\,M\'enard).
SPHERE is an instrument designed and built by a consortium consisting of IPAG (Grenoble, France), MPIA (Heidelberg, Germany), LAM (Marseille, France), LESIA (Paris, France), Laboratoire Lagrange (Nice, France), INAF–Osservatorio di Padova (Italy), Observatoire de Gen\`eve (Switzerland), ETH Zurich (Switzerland), NOVA (Netherlands), ONERA (France) and ASTRON (Netherlands) in collaboration with ESO. SPHERE was funded by ESO, with additional contributions from CNRS (France), MPIA (Germany), INAF (Italy), FINES (Switzerland) and NOVA (Netherlands).  SPHERE also received funding from the European Commission Sixth and Seventh Framework Programmes as part of the Optical Infrared Coordination Network for Astronomy (OPTICON) under grant number RII3-Ct-2004-001566 for FP6 (2004–2008), grant number 226604 for FP7 (2009–2012) and grant number 312430 for FP7 (2013–2016). We also acknowledge financial support from the Programme National de Plan\'etologie (PNP) and the Programme National de Physique Stellaire (PNPS) of CNRS-INSU in France. This work has also been supported by a grant from the French Labex OSUG@2020 (Investissements d'avenir – ANR10 LABX56). The project is supported by CNRS, by the Agence Nationale de la Recherche (ANR-14-CE33-0018). It has also been carried out within the frame of the National Centre for Competence in Research PlanetS supported by the Swiss National Science Foundation (SNSF). MRM, HMS, and SD are pleased to acknowledge this financial support of the SNSF. 
This work has made use of data from the European Space Agency (ESA) mission {\it Gaia} (\url{https://www.cosmos.esa.int/gaia}), processed by the {\it Gaia} Data Processing and Analysis Consortium (DPAC, \url{https://www.cosmos.esa.int/web/gaia/dpac/consortium}). Funding for the DPAC has been provided by national institutions, in particular the institutions participating in the {\it Gaia} Multilateral Agreement.
This research made use of Astropy,\footnote{\url{http://www.astropy.org}} a community-developed core Python package for Astronomy \citep{astropy:2013, astropy:2018}, Numpy (\citealp{numpy}), Matplotlib (\citealp{matplotlib}), Scipy (\citealp{scipy}), and Numba (\citealp{numba}).
\end{acknowledgements}

\bibliographystyle{aa}

\begin{thebibliography}{73}
\expandafter\ifx\csname natexlab\endcsname\relax\def\natexlab#1{#1}\fi

\bibitem[{{Arnold} {et~al.}(2022){Arnold}, {Weinberger}, {Videen}, \&
  {Zubko}}]{Arnold2022}
{Arnold}, J.~A., {Weinberger}, A.~J., {Videen}, G., \& {Zubko}, E.~S. 2022,
  \apj, 930, 123

\bibitem[{{Arriaga} {et~al.}(2020){Arriaga}, {Fitzgerald}, {Duch{\^e}ne},
  {Kalas}, {Millar-Blanchaer}, {Perrin}, {Chen}, {Mazoyer}, {Ammons}, {Bailey},
  {Barman}, {Bulger}, {Chilcote}, {Cotten}, {De Rosa}, {Doyon}, {Esposito},
  {Follette}, {Gerard}, {Goodsell}, {Graham}, {Greenbaum}, {Hibon}, {Hom},
  {Hung}, {Ingraham}, {Konopacky}, {Macintosh}, {Maire}, {Marchis}, {Marley},
  {Marois}, {Metchev}, {Nielsen}, {Oppenheimer}, {Palmer}, {Patience},
  {Poyneer}, {Pueyo}, {Rajan}, {Rameau}, {Rantakyr{\"o}}, {Ruffio},
  {Savransky}, {Schneider}, {Sivaramakrishnan}, {Song}, {Soummer}, {Thomas},
  {Wang}, {Ward-Duong}, \& {Wolff}}]{Arriaga2020}
{Arriaga}, P., {Fitzgerald}, M.~P., {Duch{\^e}ne}, G., {et~al.} 2020, \aj, 160,
  79

\bibitem[{{Asensio-Torres} {et~al.}(2016){Asensio-Torres}, {Janson},
  {Hashimoto}, {Thalmann}, {Currie}, {Buenzli}, {Kudo}, {Kuzuhara}, {Kusakabe},
  {Abe}, {Akiyama}, {Brandner}, {Brandt}, {Carson}, {Egner}, {Feldt}, {Goto},
  {Grady}, {Guyon}, {Hayano}, {Hayashi}, {Hayashi}, {Henning}, {Hodapp},
  {Ishii}, {Iye}, {Kandori}, {Knapp}, {Kwon}, {Matsuo}, {McElwain}, {Mayama},
  {Miyama}, {Morino}, {Moro-Martin}, {Nishimura}, {Pyo}, {Serabyn}, {Suenaga},
  {Suto}, {Suzuki}, {Takahashi}, {Takami}, {Takato}, {Terada}, {Turner},
  {Watanabe}, {Wisniewski}, {Yamada}, {Takami}, {Usuda}, \&
  {Tamura}}]{Asensio2016}
{Asensio-Torres}, R., {Janson}, M., {Hashimoto}, J., {et~al.} 2016, \aap, 593,
  A73

\bibitem[{{Astropy Collaboration} {et~al.}(2018){Astropy Collaboration},
  {Price-Whelan}, {Sip{\H{o}}cz}, {G{\"u}nther}, {Lim}, {Crawford}, {Conseil},
  {Shupe}, {Craig}, {Dencheva}, {Ginsburg}, {Vand erPlas}, {Bradley},
  {P{\'e}rez-Su{\'a}rez}, {de Val-Borro}, {Aldcroft}, {Cruz}, {Robitaille},
  {Tollerud}, {Ardelean}, {Babej}, {Bach}, {Bachetti}, {Bakanov}, {Bamford},
  {Barentsen}, {Barmby}, {Baumbach}, {Berry}, {Biscani}, {Boquien}, {Bostroem},
  {Bouma}, {Brammer}, {Bray}, {Breytenbach}, {Buddelmeijer}, {Burke},
  {Calderone}, {Cano Rodr{\'\i}guez}, {Cara}, {Cardoso}, {Cheedella}, {Copin},
  {Corrales}, {Crichton}, {D'Avella}, {Deil}, {Depagne}, {Dietrich}, {Donath},
  {Droettboom}, {Earl}, {Erben}, {Fabbro}, {Ferreira}, {Finethy}, {Fox},
  {Garrison}, {Gibbons}, {Goldstein}, {Gommers}, {Greco}, {Greenfield},
  {Groener}, {Grollier}, {Hagen}, {Hirst}, {Homeier}, {Horton}, {Hosseinzadeh},
  {Hu}, {Hunkeler}, {Ivezi{\'c}}, {Jain}, {Jenness}, {Kanarek}, {Kendrew},
  {Kern}, {Kerzendorf}, {Khvalko}, {King}, {Kirkby}, {Kulkarni}, {Kumar},
  {Lee}, {Lenz}, {Littlefair}, {Ma}, {Macleod}, {Mastropietro}, {McCully},
  {Montagnac}, {Morris}, {Mueller}, {Mumford}, {Muna}, {Murphy}, {Nelson},
  {Nguyen}, {Ninan}, {N{\"o}the}, {Ogaz}, {Oh}, {Parejko}, {Parley}, {Pascual},
  {Patil}, {Patil}, {Plunkett}, {Prochaska}, {Rastogi}, {Reddy Janga},
  {Sabater}, {Sakurikar}, {Seifert}, {Sherbert}, {Sherwood-Taylor}, {Shih},
  {Sick}, {Silbiger}, {Singanamalla}, {Singer}, {Sladen}, {Sooley},
  {Sornarajah}, {Streicher}, {Teuben}, {Thomas}, {Tremblay}, {Turner},
  {Terr{\'o}n}, {van Kerkwijk}, {de la Vega}, {Watkins}, {Weaver}, {Whitmore},
  {Woillez}, {Zabalza}, \& {Astropy Contributors}}]{astropy:2018}
{Astropy Collaboration}, {Price-Whelan}, A.~M., {Sip{\H{o}}cz}, B.~M., {et~al.}
  2018, \aj, 156, 123

\bibitem[{{Astropy Collaboration} {et~al.}(2013){Astropy Collaboration},
  {Robitaille}, {Tollerud}, {Greenfield}, {Droettboom}, {Bray}, {Aldcroft},
  {Davis}, {Ginsburg}, {Price-Whelan}, {Kerzendorf}, {Conley}, {Crighton},
  {Barbary}, {Muna}, {Ferguson}, {Grollier}, {Parikh}, {Nair}, {Unther},
  {Deil}, {Woillez}, {Conseil}, {Kramer}, {Turner}, {Singer}, {Fox}, {Weaver},
  {Zabalza}, {Edwards}, {Azalee Bostroem}, {Burke}, {Casey}, {Crawford},
  {Dencheva}, {Ely}, {Jenness}, {Labrie}, {Lim}, {Pierfederici}, {Pontzen},
  {Ptak}, {Refsdal}, {Servillat}, \& {Streicher}}]{astropy:2013}
{Astropy Collaboration}, {Robitaille}, T.~P., {Tollerud}, E.~J., {et~al.} 2013,
  \aap, 558, A33

\bibitem[{{Beuzit} {et~al.}(2019){Beuzit}, {Vigan}, {Mouillet}, {Dohlen},
  {Gratton}, {Boccaletti}, {Sauvage}, {Schmid}, {Langlois}, {Petit},
  {Baruffolo}, {Feldt}, {Milli}, {Wahhaj}, {Abe}, {Anselmi}, {Antichi},
  {Barette}, {Baudrand}, {Baudoz}, {Bazzon}, {Bernardi}, {Blanchard}, {Brast},
  {Bruno}, {Buey}, {Carbillet}, {Carle}, {Cascone}, {Chapron}, {Charton},
  {Chauvin}, {Claudi}, {Costille}, {De Caprio}, {de Boer}, {Delboulb{\'e}},
  {Desidera}, {Dominik}, {Downing}, {Dupuis}, {Fabron}, {Fantinel}, {Farisato},
  {Feautrier}, {Fedrigo}, {Fusco}, {Gigan}, {Ginski}, {Girard}, {Giro},
  {Gisler}, {Gluck}, {Gry}, {Henning}, {Hubin}, {Hugot}, {Incorvaia}, {Jaquet},
  {Kasper}, {Lagadec}, {Lagrange}, {Le Coroller}, {Le Mignant}, {Le Ruyet},
  {Lessio}, {Lizon}, {Llored}, {Lundin}, {Madec}, {Magnard}, {Marteaud},
  {Martinez}, {Maurel}, {M{\'e}nard}, {Mesa}, {M{\"o}ller-Nilsson}, {Moulin},
  {Moutou}, {Orign{\'e}}, {Parisot}, {Pavlov}, {Perret}, {Pragt}, {Puget},
  {Rabou}, {Ramos}, {Reess}, {Rigal}, {Rochat}, {Roelfsema}, {Rousset}, {Roux},
  {Saisse}, {Salasnich}, {Santambrogio}, {Scuderi}, {Segransan}, {Sevin},
  {Siebenmorgen}, {Soenke}, {Stadler}, {Suarez}, {Tiph{\`e}ne}, {Turatto},
  {Udry}, {Vakili}, {Waters}, {Weber}, {Wildi}, {Zins}, \&
  {Zurlo}}]{Beuzit2019}
{Beuzit}, J.~L., {Vigan}, A., {Mouillet}, D., {et~al.} 2019, \aap, 631, A155

\bibitem[{{Buchner} {et~al.}(2014){Buchner}, {Georgakakis}, {Nandra}, {Hsu},
  {Rangel}, {Brightman}, {Merloni}, {Salvato}, {Donley}, \&
  {Kocevski}}]{Buchner2014}
{Buchner}, J., {Georgakakis}, A., {Nandra}, K., {et~al.} 2014, \aap, 564, A125

\bibitem[{{Crotts} {et~al.}(2024){Crotts}, {Matthews}, {Duch{\^e}ne},
  {Esposito}, {Dong}, {Hom}, {Oppenheimer}, {Rice}, {Wolff}, {Chen}, {Do
  {\'O}}, {Kalas}, {Lewis}, {Weinberger}, {Wilner}, {Ammons}, {Arriaga}, {De
  Rosa}, {Debes}, {Fitzgerald}, {Gonzales}, {Hines}, {Hinkley}, {Hughes},
  {Kolokolova}, {Lee}, {L{\'o}pez}, {Macintosh}, {Mazoyer}, {Metchev},
  {Millar-Blanchaer}, {Nielsen}, {Patience}, {Perrin}, {Pueyo},
  {Rantakyr{\"o}}, {Ren}, {Schneider}, {Soummer}, \& {Stark}}]{Crotts2024}
{Crotts}, K.~A., {Matthews}, B.~C., {Duch{\^e}ne}, G., {et~al.} 2024, \apj,
  961, 245

\bibitem[{{Currie} {et~al.}(2015){Currie}, {Lisse}, {Kuchner}, {Madhusudhan},
  {Kenyon}, {Thalmann}, {Carson}, \& {Debes}}]{Currie2015}
{Currie}, T., {Lisse}, C.~M., {Kuchner}, M., {et~al.} 2015, \apjl, 807, L7

\bibitem[{{Daley} {et~al.}(2019){Daley}, {Hughes}, {Carter}, {Flaherty},
  {Lambros}, {Pan}, {Schlichting}, {Chiang}, {Wyatt}, {Wilner}, {Andrews}, \&
  {Carpenter}}]{Daley2019}
{Daley}, C., {Hughes}, A.~M., {Carter}, E.~S., {et~al.} 2019, \apj, 875, 87

\bibitem[{{de Boer} {et~al.}(2020){de Boer}, {Langlois}, {van Holstein},
  {Girard}, {Mouillet}, {Vigan}, {Dohlen}, {Snik}, {Keller}, {Ginski}, {Stam},
  {Milli}, {Wahhaj}, {Kasper}, {Schmid}, {Rabou}, {Gluck}, {Hugot}, {Perret},
  {Martinez}, {Weber}, {Pragt}, {Sauvage}, {Boccaletti}, {Le Coroller},
  {Dominik}, {Henning}, {Lagadec}, {M{\'e}nard}, {Turatto}, {Udry}, {Chauvin},
  {Feldt}, \& {Beuzit}}]{deBoer2020}
{de Boer}, J., {Langlois}, M., {van Holstein}, R.~G., {et~al.} 2020, \aap, 633,
  A63

\bibitem[{{Dohlen} {et~al.}(2008){Dohlen}, {Saisse}, {Origne}, {Moreaux},
  {Fabron}, {Zamkotsian}, {Lanzoni}, \& {Lemarquis}}]{Dohlen2008}
{Dohlen}, K., {Saisse}, M., {Origne}, A., {et~al.} 2008, in Society of
  Photo-Optical Instrumentation Engineers (SPIE) Conference Series, Vol. 7018,
  Advanced Optical and Mechanical Technologies in Telescopes and
  Instrumentation, ed. E.~{Atad-Ettedgui} \& D.~{Lemke}, 701859

\bibitem[{{Dominik} {et~al.}(2021){Dominik}, {Min}, \& {Tazaki}}]{optool}
{Dominik}, C., {Min}, M., \& {Tazaki}, R. 2021, {OpTool: Command-line driven
  tool for creating complex dust opacities}, Astrophysics Source Code Library,
  record ascl:2104.010

\bibitem[{{Dorschner} {et~al.}(1995){Dorschner}, {Begemann}, {Henning},
  {Jaeger}, \& {Mutschke}}]{Dorschner1995}
{Dorschner}, J., {Begemann}, B., {Henning}, T., {Jaeger}, C., \& {Mutschke}, H.
  1995, \aap, 300, 503

\bibitem[{{Draper} {et~al.}(2016){Draper}, {Duch{\^e}ne}, {Millar-Blanchaer},
  {Matthews}, {Wang}, {Kalas}, {Graham}, {Padgett}, {Ammons}, {Bulger}, {Chen},
  {Chilcote}, {Doyon}, {Fitzgerald}, {Follette}, {Gerard}, {Greenbaum},
  {Hibon}, {Hinkley}, {Macintosh}, {Ingraham}, {Lafreni{\`e}re}, {Marchis},
  {Marois}, {Nielsen}, {Oppenheimer}, {Patel}, {Patience}, {Perrin}, {Pueyo},
  {Rajan}, {Rameau}, {Sivaramakrishnan}, {Vega}, {Ward-Duong}, \&
  {Wolff}}]{Draper2016}
{Draper}, Z.~H., {Duch{\^e}ne}, G., {Millar-Blanchaer}, M.~A., {et~al.} 2016,
  \apj, 826, 147

\bibitem[{{Duch{\^e}ne} {et~al.}(2020){Duch{\^e}ne}, {Rice}, {Hom}, {Zalesky},
  {Esposito}, {Millar-Blanchaer}, {Ren}, {Kalas}, {Fitzgerald}, {Arriaga},
  {Bruzzone}, {Bulger}, {Chen}, {Chiang}, {Cotten}, {Czekala}, {De Rosa},
  {Dong}, {Draper}, {Follette}, {Graham}, {Hung}, {Lopez}, {Macintosh},
  {Matthews}, {Mazoyer}, {Metchev}, {Patience}, {Perrin}, {Rameau}, {Song},
  {Stahl}, {Wang}, {Wolff}, {Zuckerman}, {Ammons}, {Bailey}, {Barman},
  {Chilcote}, {Doyon}, {Gerard}, {Goodsell}, {Greenbaum}, {Hibon}, {Ingraham},
  {Konopacky}, {Maire}, {Marchis}, {Marley}, {Marois}, {Nielsen},
  {Oppenheimer}, {Palmer}, {Poyneer}, {Pueyo}, {Rajan}, {Rantakyr{\"o}},
  {Ruffio}, {Savransky}, {Schneider}, {Sivaramakrishnan}, {Soummer}, {Thomas},
  \& {Ward-Duong}}]{Duchene2020}
{Duch{\^e}ne}, G., {Rice}, M., {Hom}, J., {et~al.} 2020, \aj, 159, 251

\bibitem[{{Engler} {et~al.}(2019){Engler}, {Boccaletti}, {Schmid}, {Milli},
  {Augereau}, {Mazoyer}, {Maire}, {Henning}, {Avenhaus}, \&
  {Baudoz}}]{Engler2019}
{Engler}, N., {Boccaletti}, A., {Schmid}, H.~M., {et~al.} 2019, \aap, 622, A192

\bibitem[{{Engler} {et~al.}(2023){Engler}, {Milli}, {Gratton}, {Ulmer-Moll},
  {Vigan}, {Lagrange}, {Kiefer}, {Rubini}, {Grandjean}, {Schmid}, {Messina},
  {Squicciarini}, {Olofsson}, {Th{\'e}bault}, {van Holstein}, {Janson},
  {M{\'e}nard}, {Marshall}, {Chauvin}, {Lendl}, {Bhowmik}, {Boccaletti},
  {Bonnefoy}, {del Burgo}, {Choquet}, {Desidera}, {Feldt}, {Fusco}, {Girard},
  {Gisler}, {Hagelberg}, {Langlois}, {Maire}, {Mesa}, {Meyer}, {Rabou},
  {Rodet}, {Schmidt}, \& {Zurlo}}]{Engler2023}
{Engler}, N., {Milli}, J., {Gratton}, R., {et~al.} 2023, \aap, 672, A1

\bibitem[{{Esposito} {et~al.}(2018){Esposito}, {Duch{\^e}ne}, {Kalas}, {Rice},
  {Choquet}, {Ren}, {Perrin}, {Chen}, {Arriaga}, {Chiang}, {Nielsen}, {Graham},
  {Wang}, {De Rosa}, {Follette}, {Ammons}, {Ansdell}, {Bailey}, {Barman},
  {Sebasti{\'a}n Bruzzone}, {Bulger}, {Chilcote}, {Cotten}, {Doyon},
  {Fitzgerald}, {Goodsell}, {Greenbaum}, {Hibon}, {Hung}, {Ingraham},
  {Konopacky}, {Larkin}, {Macintosh}, {Maire}, {Marchis}, {Marois}, {Mazoyer},
  {Metchev}, {Millar-Blanchaer}, {Oppenheimer}, {Palmer}, {Patience},
  {Poyneer}, {Pueyo}, {Rajan}, {Rameau}, {Rantakyr{\"o}}, {Ryan}, {Savransky},
  {Schneider}, {Sivaramakrishnan}, {Song}, {Soummer}, {Thomas}, {Wallace},
  {Ward-Duong}, {Wiktorowicz}, \& {Wolff}}]{Esposito2018}
{Esposito}, T.~M., {Duch{\^e}ne}, G., {Kalas}, P., {et~al.} 2018, \aj, 156, 47

\bibitem[{{Esposito} {et~al.}(2020){Esposito}, {Kalas}, {Fitzgerald},
  {Millar-Blanchaer}, {Duch{\^e}ne}, {Patience}, {Hom}, {Perrin}, {De Rosa},
  {Chiang}, {Czekala}, {Macintosh}, {Graham}, {Ansdell}, {Arriaga}, {Bruzzone},
  {Bulger}, {Chen}, {Cotten}, {Dong}, {Draper}, {Follette}, {Hung}, {Lopez},
  {Matthews}, {Mazoyer}, {Metchev}, {Rameau}, {Ren}, {Rice}, {Song}, {Stahl},
  {Wang}, {Wolff}, {Zuckerman}, {Ammons}, {Bailey}, {Barman}, {Chilcote},
  {Doyon}, {Gerard}, {Goodsell}, {Greenbaum}, {Hibon}, {Hinkley}, {Ingraham},
  {Konopacky}, {Maire}, {Marchis}, {Marley}, {Marois}, {Nielsen},
  {Oppenheimer}, {Palmer}, {Poyneer}, {Pueyo}, {Rajan}, {Rantakyr{\"o}},
  {Ruffio}, {Savransky}, {Schneider}, {Sivaramakrishnan}, {Soummer}, {Thomas},
  \& {Ward-Duong}}]{Esposito2020}
{Esposito}, T.~M., {Kalas}, P., {Fitzgerald}, M.~P., {et~al.} 2020, \aj, 160,
  24

\bibitem[{{Feroz} {et~al.}(2009){Feroz}, {Hobson}, \& {Bridges}}]{Feroz2009}
{Feroz}, F., {Hobson}, M.~P., \& {Bridges}, M. 2009, \mnras, 398, 1601

\bibitem[{{Flasseur} {et~al.}(2021){Flasseur}, {Th{\'e}}, {Denis},
  {Thi{\'e}baut}, \& {Langlois}}]{Flasseur2021}
{Flasseur}, O., {Th{\'e}}, S., {Denis}, L., {Thi{\'e}baut}, {\'E}., \&
  {Langlois}, M. 2021, \aap, 651, A62

\bibitem[{{Gaia Collaboration} {et~al.}(2021){Gaia Collaboration}, {Brown},
  {Vallenari}, {Prusti}, {de Bruijne}, {Babusiaux}, {Biermann}, {Creevey},
  {Evans}, {Eyer}, {Hutton}, {Jansen}, {Jordi}, {Klioner}, {Lammers},
  {Lindegren}, {Luri}, {Mignard}, {Panem}, {Pourbaix}, {Randich}, {Sartoretti},
  {Soubiran}, {Walton}, {Arenou}, {Bailer-Jones}, {Bastian}, {Cropper},
  {Drimmel}, {Katz}, {Lattanzi}, {van Leeuwen}, {Bakker}, {Cacciari},
  {Casta{\~n}eda}, {De Angeli}, {Ducourant}, {Fabricius}, {Fouesneau},
  {Fr{\'e}mat}, {Guerra}, {Guerrier}, {Guiraud}, {Jean-Antoine Piccolo},
  {Masana}, {Messineo}, {Mowlavi}, {Nicolas}, {Nienartowicz}, {Pailler},
  {Panuzzo}, {Riclet}, {Roux}, {Seabroke}, {Sordo}, {Tanga}, {Th{\'e}venin},
  {Gracia-Abril}, {Portell}, {Teyssier}, {Altmann}, {Andrae}, {Bellas-Velidis},
  {Benson}, {Berthier}, {Blomme}, {Brugaletta}, {Burgess}, {Busso}, {Carry},
  {Cellino}, {Cheek}, {Clementini}, {Damerdji}, {Davidson}, {Delchambre},
  {Dell'Oro}, {Fern{\'a}ndez-Hern{\'a}ndez}, {Galluccio}, {Garc{\'\i}a-Lario},
  {Garcia-Reinaldos}, {Gonz{\'a}lez-N{\'u}{\~n}ez}, {Gosset}, {Haigron},
  {Halbwachs}, {Hambly}, {Harrison}, {Hatzidimitriou}, {Heiter},
  {Hern{\'a}ndez}, {Hestroffer}, {Hodgkin}, {Holl}, {Jan{\ss}en}, {Jevardat de
  Fombelle}, {Jordan}, {Krone-Martins}, {Lanzafame}, {L{\"o}ffler}, {Lorca},
  {Manteiga}, {Marchal}, {Marrese}, {Moitinho}, {Mora}, {Muinonen}, {Osborne},
  {Pancino}, {Pauwels}, {Petit}, {Recio-Blanco}, {Richards}, {Riello},
  {Rimoldini}, {Robin}, {Roegiers}, {Rybizki}, {Sarro}, {Siopis}, {Smith},
  {Sozzetti}, {Ulla}, {Utrilla}, {van Leeuwen}, {van Reeven}, {Abbas}, {Abreu
  Aramburu}, {Accart}, {Aerts}, {Aguado}, {Ajaj}, {Altavilla}, {{\'A}lvarez},
  {{\'A}lvarez Cid-Fuentes}, {Alves}, {Anderson}, {Anglada Varela}, {Antoja},
  {Audard}, {Baines}, {Baker}, {Balaguer-N{\'u}{\~n}ez}, {Balbinot}, {Balog},
  {Barache}, {Barbato}, {Barros}, {Barstow}, {Bartolom{\'e}}, {Bassilana},
  {Bauchet}, {Baudesson-Stella}, {Becciani}, {Bellazzini}, {Bernet}, {Bertone},
  {Bianchi}, {Blanco-Cuaresma}, {Boch}, {Bombrun}, {Bossini}, {Bouquillon},
  {Bragaglia}, {Bramante}, {Breedt}, {Bressan}, {Brouillet}, {Bucciarelli},
  {Burlacu}, {Busonero}, {Butkevich}, {Buzzi}, {Caffau}, {Cancelliere},
  {C{\'a}novas}, {Cantat-Gaudin}, {Carballo}, {Carlucci}, {Carnerero},
  {Carrasco}, {Casamiquela}, {Castellani}, {Castro-Ginard}, {Castro Sampol},
  {Chaoul}, {Charlot}, {Chemin}, {Chiavassa}, {Cioni}, {Comoretto}, {Cooper},
  {Cornez}, {Cowell}, {Crifo}, {Crosta}, {Crowley}, {Dafonte}, {Dapergolas},
  {David}, {David}, {de Laverny}, {De Luise}, {De March}, {De Ridder}, {de
  Souza}, {de Teodoro}, {de Torres}, {del Peloso}, {del Pozo}, {Delbo},
  {Delgado}, {Delgado}, {Delisle}, {Di Matteo}, {Diakite}, {Diener},
  {Distefano}, {Dolding}, {Eappachen}, {Edvardsson}, {Enke}, {Esquej}, {Fabre},
  {Fabrizio}, {Faigler}, {Fedorets}, {Fernique}, {Fienga}, {Figueras},
  {Fouron}, {Fragkoudi}, {Fraile}, {Franke}, {Gai}, {Garabato},
  {Garcia-Gutierrez}, {Garc{\'\i}a-Torres}, {Garofalo}, {Gavras}, {Gerlach},
  {Geyer}, {Giacobbe}, {Gilmore}, {Girona}, {Giuffrida}, {Gomel}, {Gomez},
  {Gonzalez-Santamaria}, {Gonz{\'a}lez-Vidal}, {Granvik},
  {Guti{\'e}rrez-S{\'a}nchez}, {Guy}, {Hauser}, {Haywood}, {Helmi}, {Hidalgo},
  {Hilger}, {H{\l}adczuk}, {Hobbs}, {Holland}, {Huckle}, {Jasniewicz},
  {Jonker}, {Juaristi Campillo}, {Julbe}, {Karbevska}, {Kervella}, {Khanna},
  {Kochoska}, {Kontizas}, {Kordopatis}, {Korn}, {Kostrzewa-Rutkowska},
  {Kruszy{\'n}ska}, {Lambert}, {Lanza}, {Lasne}, {Le Campion}, {Le Fustec},
  {Lebreton}, {Lebzelter}, {Leccia}, {Leclerc}, {Lecoeur-Taibi}, {Liao},
  {Licata}, {Lindstr{\o}m}, {Lister}, {Livanou}, {Lobel}, {Madrero Pardo},
  {Managau}, {Mann}, {Marchant}, {Marconi}, {Marcos Santos}, {Marinoni},
  {Marocco}, {Marshall}, {Martin Polo}, {Mart{\'\i}n-Fleitas}, {Masip},
  {Massari}, {Mastrobuono-Battisti}, {Mazeh}, {McMillan}, {Messina},
  {Michalik}, {Millar}, {Mints}, {Molina}, {Molinaro}, {Moln{\'a}r},
  {Montegriffo}, {Mor}, {Morbidelli}, {Morel}, {Morris}, {Mulone}, {Munoz},
  {Muraveva}, {Murphy}, {Musella}, {Noval}, {Ord{\'e}novic}, {Orr{\`u}},
  {Osinde}, {Pagani}, {Pagano}, {Palaversa}, {Palicio}, {Panahi}, {Pawlak},
  {Pe{\~n}alosa Esteller}, {Penttil{\"a}}, {Piersimoni}, {Pineau}, {Plachy},
  {Plum}, {Poggio}, {Poretti}, {Poujoulet}, {Pr{\v{s}}a}, {Pulone}, {Racero},
  {Ragaini}, {Rainer}, {Raiteri}, {Rambaux}, {Ramos}, {Ramos-Lerate}, {Re
  Fiorentin}, {Regibo}, {Reyl{\'e}}, {Ripepi}, {Riva}, {Rixon}, {Robichon},
  {Robin}, {Roelens}, {Rohrbasser}, {Romero-G{\'o}mez}, {Rowell}, {Royer},
  {Rybicki}, {Sadowski}, {Sagrist{\`a} Sell{\'e}s}, {Sahlmann}, {Salgado},
  {Salguero}, {Samaras}, {Sanchez Gimenez}, {Sanna}, {Santove{\~n}a},
  {Sarasso}, {Schultheis}, {Sciacca}, {Segol}, {Segovia}, {S{\'e}gransan},
  {Semeux}, {Shahaf}, {Siddiqui}, {Siebert}, {Siltala}, {Slezak}, {Smart},
  {Solano}, {Solitro}, {Souami}, {Souchay}, {Spagna}, {Spoto}, {Steele},
  {Steidelm{\"u}ller}, {Stephenson}, {S{\"u}veges}, {Szabados}, {Szegedi-Elek},
  {Taris}, {Tauran}, {Taylor}, {Teixeira}, {Thuillot}, {Tonello}, {Torra},
  {Torra}, {Turon}, {Unger}, {Vaillant}, {van Dillen}, {Vanel}, {Vecchiato},
  {Viala}, {Vicente}, {Voutsinas}, {Weiler}, {Wevers}, {Wyrzykowski}, {Yoldas},
  {Yvard}, {Zhao}, {Zorec}, {Zucker}, {Zurbach}, \& {Zwitter}}]{Gaia2020}
{Gaia Collaboration}, {Brown}, A.~G.~A., {Vallenari}, A., {et~al.} 2021, \aap,
  649, A1

\bibitem[{{Gaia Collaboration} {et~al.}(2016){Gaia Collaboration}, {Prusti},
  {de Bruijne}, {Brown}, {Vallenari}, {Babusiaux}, {Bailer-Jones}, {Bastian},
  {Biermann}, {Evans}, {Eyer}, {Jansen}, {Jordi}, {Klioner}, {Lammers},
  {Lindegren}, {Luri}, {Mignard}, {Milligan}, {Panem}, {Poinsignon},
  {Pourbaix}, {Randich}, {Sarri}, {Sartoretti}, {Siddiqui}, {Soubiran},
  {Valette}, {van Leeuwen}, {Walton}, {Aerts}, {Arenou}, {Cropper}, {Drimmel},
  {H{\o}g}, {Katz}, {Lattanzi}, {O'Mullane}, {Grebel}, {Holland}, {Huc},
  {Passot}, {Bramante}, {Cacciari}, {Casta{\~n}eda}, {Chaoul}, {Cheek}, {De
  Angeli}, {Fabricius}, {Guerra}, {Hern{\'a}ndez}, {Jean-Antoine-Piccolo},
  {Masana}, {Messineo}, {Mowlavi}, {Nienartowicz}, {Ord{\'o}{\~n}ez-Blanco},
  {Panuzzo}, {Portell}, {Richards}, {Riello}, {Seabroke}, {Tanga},
  {Th{\'e}venin}, {Torra}, {Els}, {Gracia-Abril}, {Comoretto},
  {Garcia-Reinaldos}, {Lock}, {Mercier}, {Altmann}, {Andrae}, {Astraatmadja},
  {Bellas-Velidis}, {Benson}, {Berthier}, {Blomme}, {Busso}, {Carry},
  {Cellino}, {Clementini}, {Cowell}, {Creevey}, {Cuypers}, {Davidson}, {De
  Ridder}, {de Torres}, {Delchambre}, {Dell'Oro}, {Ducourant}, {Fr{\'e}mat},
  {Garc{\'\i}a-Torres}, {Gosset}, {Halbwachs}, {Hambly}, {Harrison}, {Hauser},
  {Hestroffer}, {Hodgkin}, {Huckle}, {Hutton}, {Jasniewicz}, {Jordan},
  {Kontizas}, {Korn}, {Lanzafame}, {Manteiga}, {Moitinho}, {Muinonen},
  {Osinde}, {Pancino}, {Pauwels}, {Petit}, {Recio-Blanco}, {Robin}, {Sarro},
  {Siopis}, {Smith}, {Smith}, {Sozzetti}, {Thuillot}, {van Reeven}, {Viala},
  {Abbas}, {Abreu Aramburu}, {Accart}, {Aguado}, {Allan}, {Allasia},
  {Altavilla}, {{\'A}lvarez}, {Alves}, {Anderson}, {Andrei}, {Anglada Varela},
  {Antiche}, {Antoja}, {Ant{\'o}n}, {Arcay}, {Atzei}, {Ayache}, {Bach},
  {Baker}, {Balaguer-N{\'u}{\~n}ez}, {Barache}, {Barata}, {Barbier}, {Barblan},
  {Baroni}, {Barrado y Navascu{\'e}s}, {Barros}, {Barstow}, {Becciani},
  {Bellazzini}, {Bellei}, {Bello Garc{\'\i}a}, {Belokurov}, {Bendjoya},
  {Berihuete}, {Bianchi}, {Bienaym{\'e}}, {Billebaud}, {Blagorodnova},
  {Blanco-Cuaresma}, {Boch}, {Bombrun}, {Borrachero}, {Bouquillon}, {Bourda},
  {Bouy}, {Bragaglia}, {Breddels}, {Brouillet}, {Br{\"u}semeister},
  {Bucciarelli}, {Budnik}, {Burgess}, {Burgon}, {Burlacu}, {Busonero}, {Buzzi},
  {Caffau}, {Cambras}, {Campbell}, {Cancelliere}, {Cantat-Gaudin}, {Carlucci},
  {Carrasco}, {Castellani}, {Charlot}, {Charnas}, {Charvet}, {Chassat},
  {Chiavassa}, {Clotet}, {Cocozza}, {Collins}, {Collins}, {Costigan}, {Crifo},
  {Cross}, {Crosta}, {Crowley}, {Dafonte}, {Damerdji}, {Dapergolas}, {David},
  {David}, {De Cat}, {de Felice}, {de Laverny}, {De Luise}, {De March}, {de
  Martino}, {de Souza}, {Debosscher}, {del Pozo}, {Delbo}, {Delgado},
  {Delgado}, {di Marco}, {Di Matteo}, {Diakite}, {Distefano}, {Dolding}, {Dos
  Anjos}, {Drazinos}, {Dur{\'a}n}, {Dzigan}, {Ecale}, {Edvardsson}, {Enke},
  {Erdmann}, {Escolar}, {Espina}, {Evans}, {Eynard Bontemps}, {Fabre},
  {Fabrizio}, {Faigler}, {Falc{\~a}o}, {Farr{\`a}s Casas}, {Faye}, {Federici},
  {Fedorets}, {Fern{\'a}ndez-Hern{\'a}ndez}, {Fernique}, {Fienga}, {Figueras},
  {Filippi}, {Findeisen}, {Fonti}, {Fouesneau}, {Fraile}, {Fraser}, {Fuchs},
  {Furnell}, {Gai}, {Galleti}, {Galluccio}, {Garabato}, {Garc{\'\i}a-Sedano},
  {Gar{\'e}}, {Garofalo}, {Garralda}, {Gavras}, {Gerssen}, {Geyer}, {Gilmore},
  {Girona}, {Giuffrida}, {Gomes}, {Gonz{\'a}lez-Marcos},
  {Gonz{\'a}lez-N{\'u}{\~n}ez}, {Gonz{\'a}lez-Vidal}, {Granvik}, {Guerrier},
  {Guillout}, {Guiraud}, {G{\'u}rpide}, {Guti{\'e}rrez-S{\'a}nchez}, {Guy},
  {Haigron}, {Hatzidimitriou}, {Haywood}, {Heiter}, {Helmi}, {Hobbs},
  {Hofmann}, {Holl}, {Holland}, {Hunt}, {Hypki}, {Icardi}, {Irwin}, {Jevardat
  de Fombelle}, {Jofr{\'e}}, {Jonker}, {Jorissen}, {Julbe}, {Karampelas},
  {Kochoska}, {Kohley}, {Kolenberg}, {Kontizas}, {Koposov}, {Kordopatis},
  {Koubsky}, {Kowalczyk}, {Krone-Martins}, {Kudryashova}, {Kull}, {Bachchan},
  {Lacoste-Seris}, {Lanza}, {Lavigne}, {Le Poncin-Lafitte}, {Lebreton},
  {Lebzelter}, {Leccia}, {Leclerc}, {Lecoeur-Taibi}, {Lemaitre}, {Lenhardt},
  {Leroux}, {Liao}, {Licata}, {Lindstr{\o}m}, {Lister}, {Livanou}, {Lobel},
  {L{\"o}ffler}, {L{\'o}pez}, {Lopez-Lozano}, {Lorenz}, {Loureiro},
  {MacDonald}, {Magalh{\~a}es Fernandes}, {Managau}, {Mann}, {Mantelet},
  {Marchal}, {Marchant}, {Marconi}, {Marie}, {Marinoni}, {Marrese},
  {Marschalk{\'o}}, {Marshall}, {Mart{\'\i}n-Fleitas}, {Martino}, {Mary},
  {Matijevi{\v{c}}}, {Mazeh}, {McMillan}, {Messina}, {Mestre}, {Michalik},
  {Millar}, {Miranda}, {Molina}, {Molinaro}, {Molinaro}, {Moln{\'a}r},
  {Moniez}, {Montegriffo}, {Monteiro}, {Mor}, {Mora}, {Morbidelli}, {Morel},
  {Morgenthaler}, {Morley}, {Morris}, {Mulone}, {Muraveva}, {Musella},
  {Narbonne}, {Nelemans}, {Nicastro}, {Noval}, {Ord{\'e}novic},
  {Ordieres-Mer{\'e}}, {Osborne}, {Pagani}, {Pagano}, {Pailler}, {Palacin},
  {Palaversa}, {Parsons}, {Paulsen}, {Pecoraro}, {Pedrosa}, {Pentik{\"a}inen},
  {Pereira}, {Pichon}, {Piersimoni}, {Pineau}, {Plachy}, {Plum}, {Poujoulet},
  {Pr{\v{s}}a}, {Pulone}, {Ragaini}, {Rago}, {Rambaux}, {Ramos-Lerate},
  {Ranalli}, {Rauw}, {Read}, {Regibo}, {Renk}, {Reyl{\'e}}, {Ribeiro},
  {Rimoldini}, {Ripepi}, {Riva}, {Rixon}, {Roelens}, {Romero-G{\'o}mez},
  {Rowell}, {Royer}, {Rudolph}, {Ruiz-Dern}, {Sadowski}, {Sagrist{\`a}
  Sell{\'e}s}, {Sahlmann}, {Salgado}, {Salguero}, {Sarasso}, {Savietto},
  {Schnorhk}, {Schultheis}, {Sciacca}, {Segol}, {Segovia}, {Segransan},
  {Serpell}, {Shih}, {Smareglia}, {Smart}, {Smith}, {Solano}, {Solitro},
  {Sordo}, {Soria Nieto}, {Souchay}, {Spagna}, {Spoto}, {Stampa}, {Steele},
  {Steidelm{\"u}ller}, {Stephenson}, {Stoev}, {Suess}, {S{\"u}veges}, {Surdej},
  {Szabados}, {Szegedi-Elek}, {Tapiador}, {Taris}, {Tauran}, {Taylor},
  {Teixeira}, {Terrett}, {Tingley}, {Trager}, {Turon}, {Ulla}, {Utrilla},
  {Valentini}, {van Elteren}, {Van Hemelryck}, {van Leeuwen}, {Varadi},
  {Vecchiato}, {Veljanoski}, {Via}, {Vicente}, {Vogt}, {Voss}, {Votruba},
  {Voutsinas}, {Walmsley}, {Weiler}, {Weingrill}, {Werner}, {Wevers},
  {Whitehead}, {Wyrzykowski}, {Yoldas}, {{\v{Z}}erjal}, {Zucker}, {Zurbach},
  {Zwitter}, {Alecu}, {Allen}, {Allende Prieto}, {Amorim},
  {Anglada-Escud{\'e}}, {Arsenijevic}, {Azaz}, {Balm}, {Beck}, {Bernstein},
  {Bigot}, {Bijaoui}, {Blasco}, {Bonfigli}, {Bono}, {Boudreault}, {Bressan},
  {Brown}, {Brunet}, {Bunclark}, {Buonanno}, {Butkevich}, {Carret}, {Carrion},
  {Chemin}, {Ch{\'e}reau}, {Corcione}, {Darmigny}, {de Boer}, {de Teodoro}, {de
  Zeeuw}, {Delle Luche}, {Domingues}, {Dubath}, {Fodor}, {Fr{\'e}zouls},
  {Fries}, {Fustes}, {Fyfe}, {Gallardo}, {Gallegos}, {Gardiol}, {Gebran},
  {Gomboc}, {G{\'o}mez}, {Grux}, {Gueguen}, {Heyrovsky}, {Hoar}, {Iannicola},
  {Isasi Parache}, {Janotto}, {Joliet}, {Jonckheere}, {Keil}, {Kim},
  {Klagyivik}, {Klar}, {Knude}, {Kochukhov}, {Kolka}, {Kos}, {Kutka}, {Lainey},
  {LeBouquin}, {Liu}, {Loreggia}, {Makarov}, {Marseille}, {Martayan},
  {Martinez-Rubi}, {Massart}, {Meynadier}, {Mignot}, {Munari}, {Nguyen},
  {Nordlander}, {Ocvirk}, {O'Flaherty}, {Olias Sanz}, {Ortiz}, {Osorio},
  {Oszkiewicz}, {Ouzounis}, {Palmer}, {Park}, {Pasquato}, {Peltzer}, {Peralta},
  {P{\'e}turaud}, {Pieniluoma}, {Pigozzi}, {Poels}, {Prat}, {Prod'homme},
  {Raison}, {Rebordao}, {Risquez}, {Rocca-Volmerange}, {Rosen}, {Ruiz-Fuertes},
  {Russo}, {Sembay}, {Serraller Vizcaino}, {Short}, {Siebert}, {Silva},
  {Sinachopoulos}, {Slezak}, {Soffel}, {Sosnowska}, {Strai{\v{z}}ys}, {ter
  Linden}, {Terrell}, {Theil}, {Tiede}, {Troisi}, {Tsalmantza}, {Tur},
  {Vaccari}, {Vachier}, {Valles}, {Van Hamme}, {Veltz}, {Virtanen}, {Wallut},
  {Wichmann}, {Wilkinson}, {Ziaeepour}, \& {Zschocke}}]{Gaia2016}
{Gaia Collaboration}, {Prusti}, T., {de Bruijne}, J.~H.~J., {et~al.} 2016,
  \aap, 595, A1

\bibitem[{{Gibbs} {et~al.}(2019){Gibbs}, {Wagner}, {Apai}, {Mo{\'o}r},
  {Currie}, {Bonnefoy}, {Langlois}, \& {Lisse}}]{Gibbs2019}
{Gibbs}, A., {Wagner}, K., {Apai}, D., {et~al.} 2019, \aj, 157, 39

\bibitem[{{Gledhill} {et~al.}(1991){Gledhill}, {Scarrott}, \&
  {Wolstencroft}}]{Gledhill1991}
{Gledhill}, T.~M., {Scarrott}, S.~M., \& {Wolstencroft}, R.~D. 1991, \mnras,
  252, 50P

\bibitem[{{Graham} {et~al.}(2007){Graham}, {Kalas}, \& {Matthews}}]{Graham2007}
{Graham}, J.~R., {Kalas}, P.~G., \& {Matthews}, B.~C. 2007, \apj, 654, 595

\bibitem[{Harris {et~al.}(2020)Harris, Millman, van~der Walt, Gommers,
  Virtanen, Cournapeau, Wieser, Taylor, Berg, Smith, Kern, Picus, Hoyer, van
  Kerkwijk, Brett, Haldane, del R{\'{i}}o, Wiebe, Peterson,
  G{\'{e}}rard-Marchant, Sheppard, Reddy, Weckesser, Abbasi, Gohlke, \&
  Oliphant}]{numpy}
Harris, C.~R., Millman, K.~J., van~der Walt, S.~J., {et~al.} 2020, Nature, 585,
  357

\bibitem[{{Hom} {et~al.}(2024){Hom}, {Patience}, {Chen}, {Duch{\^e}ne},
  {Mazoyer}, {Millar-Blanchaer}, {Esposito}, {Kalas}, {Crotts}, {Gonzales},
  {Kolokolova}, {Lewis}, {Matthews}, {Rice}, {Weinberger}, {Wilner}, {Wolff},
  {Bruzzone}, {Choquet}, {Debes}, {De Rosa}, {Donaldson}, {Draper},
  {Fitzgerald}, {Hines}, {Hinkley}, {Hughes}, {L{\'o}pez}, {Marchis},
  {Metchev}, {Moro-Martin}, {Nesvold}, {Nielsen}, {Oppenheimer}, {Padgett},
  {Perrin}, {Pueyo}, {Rantakyr{\"o}}, {Ren}, {Schneider}, {Soummer}, {Song}, \&
  {Stark}}]{Hom2024}
{Hom}, J., {Patience}, J., {Chen}, C.~H., {et~al.} 2024, \mnras, 528, 6959

\bibitem[{Hunter(2007)}]{matplotlib}
Hunter, J.~D. 2007, Computing in Science \& Engineering, 9, 90

\bibitem[{{Juillard} {et~al.}(2023){Juillard}, {Christiaens}, \&
  {Absil}}]{Juillard2023}
{Juillard}, S., {Christiaens}, V., \& {Absil}, O. 2023, \aap, 679, A52

\bibitem[{{Kennedy} {et~al.}(2018){Kennedy}, {Marino}, {Matr{\`a}},
  {Pani{\'c}}, {Wilner}, {Wyatt}, \& {Yelverton}}]{Kennedy2018}
{Kennedy}, G.~M., {Marino}, S., {Matr{\`a}}, L., {et~al.} 2018, \mnras, 475,
  4924

\bibitem[{{Kral} {et~al.}(2020){Kral}, {Matr{\`a}}, {Kennedy}, {Marino}, \&
  {Wyatt}}]{Kral2020}
{Kral}, Q., {Matr{\`a}}, L., {Kennedy}, G.~M., {Marino}, S., \& {Wyatt}, M.~C.
  2020, \mnras, 497, 2811

\bibitem[{{Krivov}(2010)}]{Krivov2010}
{Krivov}, A. 2010, ArXiv e-prints [\eprint[arXiv]{1003.5229}]

\bibitem[{{Krivov} \& {Wyatt}(2021)}]{Krivov2021}
{Krivov}, A.~V. \& {Wyatt}, M.~C. 2021, \mnras, 500, 718

\bibitem[{Lam {et~al.}(2015)Lam, Pitrou, \& Seibert}]{numba}
Lam, S.~K., Pitrou, A., \& Seibert, S. 2015, in Proceedings of the Second
  Workshop on the LLVM Compiler Infrastructure in HPC, LLVM '15 (New York, NY,
  USA: Association for Computing Machinery)

\bibitem[{{Lawson} {et~al.}(2021){Lawson}, {Currie}, {Wisniewski}, {Tamura},
  {Augereau}, {Brandt}, {Guyon}, {Kasdin}, {Groff}, {Lozi}, {Deo}, {Vievard},
  {Chilcote}, {Jovanovic}, {Martinache}, {Skaf}, {Henning}, {Knapp}, {Kwon},
  {McElwain}, {Pyo}, {Sitko}, {Uyama}, \& {Wagner}}]{Lawson2021}
{Lawson}, K., {Currie}, T., {Wisniewski}, J.~P., {et~al.} 2021, \aj, 162, 293

\bibitem[{{Lee} \& {Chiang}(2016)}]{Lee2016}
{Lee}, E.~J. \& {Chiang}, E. 2016, \apj, 827, 125

\bibitem[{{Marino} {et~al.}(2016){Marino}, {Matr{\`a}}, {Stark}, {Wyatt},
  {Casassus}, {Kennedy}, {Rodriguez}, {Zuckerman}, {Perez}, {Dent}, {Kuchner},
  {Hughes}, {Schneider}, {Steele}, {Roberge}, {Donaldson}, \&
  {Nesvold}}]{Marino2016}
{Marino}, S., {Matr{\`a}}, L., {Stark}, C., {et~al.} 2016, \mnras, 460, 2933

\bibitem[{{Matthews} {et~al.}(2017){Matthews}, {Hinkley}, {Vigan}, {Kennedy},
  {Rizzuto}, {Stapelfeldt}, {Mawet}, {Booth}, {Chen}, \&
  {Jang-Condell}}]{Matthews2017}
{Matthews}, E., {Hinkley}, S., {Vigan}, A., {et~al.} 2017, \apjl, 843, L12

\bibitem[{{Mie}(1908)}]{Mie1908}
{Mie}, G. 1908, Annalen der Physik, 330, 377

\bibitem[{{Millar-Blanchaer} {et~al.}(2016){Millar-Blanchaer}, {Wang}, {Kalas},
  {Graham}, {Duch{\^e}ne}, {Nielsen}, {Perrin}, {Moon}, {Padgett}, {Metchev},
  {Ammons}, {Bailey}, {Barman}, {Bruzzone}, {Bulger}, {Chen}, {Chilcote},
  {Cotten}, {De Rosa}, {Doyon}, {Draper}, {Esposito}, {Fitzgerald}, {Follette},
  {Gerard}, {Greenbaum}, {Hibon}, {Hinkley}, {Hung}, {Ingraham},
  {Johnson-Groh}, {Konopacky}, {Larkin}, {Macintosh}, {Maire}, {Marchis},
  {Marley}, {Marois}, {Matthews}, {Oppenheimer}, {Palmer}, {Patience},
  {Poyneer}, {Pueyo}, {Rajan}, {Rameau}, {Rantakyr{\"o}}, {Savransky},
  {Schneider}, {Sivaramakrishnan}, {Song}, {Soummer}, {Thomas}, {Vega},
  {Wallace}, {Ward-Duong}, {Wiktorowicz}, \& {Wolff}}]{Millar-Blanchaer2016}
{Millar-Blanchaer}, M.~A., {Wang}, J.~J., {Kalas}, P., {et~al.} 2016, \aj, 152,
  128

\bibitem[{{Milli} {et~al.}(2024){Milli}, {Choquet}, {Tazaki}, {M{\'e}nard},
  {Augereau}, {Olofsson}, {Th{\'e}bault}, {Poch}, {Levasseur-Regourd}, {Lasue},
  {Renard}, {Hadamcik}, {Baruteau}, {Schmid}, {Engler}, {van Holstein},
  {Zubko}, {Lagrange}, {Marino}, {Pinte}, {Dominik}, {Boccaletti}, {Langlois},
  {Zurlo}, {Desgrange}, {Gluck}, {Mouillet}, {Costille}, \&
  {Sauvage}}]{Milli2024}
{Milli}, J., {Choquet}, E., {Tazaki}, R., {et~al.} 2024, \aap, 683, A22

\bibitem[{{Milli} {et~al.}(2012){Milli}, {Mouillet}, {Lagrange}, {Boccaletti},
  {Mawet}, {Chauvin}, \& {Bonnefoy}}]{Milli2012}
{Milli}, J., {Mouillet}, D., {Lagrange}, A.-M., {et~al.} 2012, \aap, 545, A111

\bibitem[{{Min} {et~al.}(2005){Min}, {Hovenier}, \& {de Koter}}]{Min2005}
{Min}, M., {Hovenier}, J.~W., \& {de Koter}, A. 2005, \aap, 432, 909

\bibitem[{{Min} {et~al.}(2016){Min}, {Rab}, {Woitke}, {Dominik}, \&
  {M{\'e}nard}}]{Min2016}
{Min}, M., {Rab}, C., {Woitke}, P., {Dominik}, C., \& {M{\'e}nard}, F. 2016,
  \aap, 585, A13

\bibitem[{{Mu{\~n}oz} {et~al.}(2021){Mu{\~n}oz}, {Frattin}, {Jardiel},
  {G{\'o}mez-Mart{\'\i}n}, {Moreno}, {Ramos}, {Guirado}, {Peiteado},
  {Caballero}, {Milli}, \& {M{\'e}nard}}]{Munoz2021}
{Mu{\~n}oz}, O., {Frattin}, E., {Jardiel}, T., {et~al.} 2021, \apjs, 256, 17

\bibitem[{{Olofsson} {et~al.}(2020){Olofsson}, {Milli}, {Bayo}, {Henning}, \&
  {Engler}}]{Olofsson2020}
{Olofsson}, J., {Milli}, J., {Bayo}, A., {Henning}, T., \& {Engler}, N. 2020,
  \aap, 640, A12

\bibitem[{{Olofsson} {et~al.}(2016){Olofsson}, {Samland}, {Avenhaus},
  {Caceres}, {Henning}, {Mo{\'o}r}, {Milli}, {Canovas}, {Quanz}, {Schreiber},
  {Augereau}, {Bayo}, {Bazzon}, {Beuzit}, {Boccaletti}, {Buenzli}, {Casassus},
  {Chauvin}, {Dominik}, {Desidera}, {Feldt}, {Gratton}, {Janson}, {Lagrange},
  {Langlois}, {Lannier}, {Maire}, {Mesa}, {Pinte}, {Rouan}, {Salter},
  {Thalmann}, \& {Vigan}}]{Olofsson2016}
{Olofsson}, J., {Samland}, M., {Avenhaus}, H., {et~al.} 2016, \aap, 591, A108

\bibitem[{{Olofsson} {et~al.}(2023){Olofsson}, {Th{\'e}bault}, {Bayo}, {Milli},
  {van Holstein}, {Henning}, {Medina-Olea}, {Godoy}, \&
  {Mauc{\'o}}}]{Olofsson2023}
{Olofsson}, J., {Th{\'e}bault}, P., {Bayo}, A., {et~al.} 2023, \aap, 674, A84

\bibitem[{{Olofsson} {et~al.}(2022{\natexlab{a}}){Olofsson}, {Th{\'e}bault},
  {Kennedy}, \& {Bayo}}]{Olofsson2022b}
{Olofsson}, J., {Th{\'e}bault}, P., {Kennedy}, G.~M., \& {Bayo}, A.
  2022{\natexlab{a}}, \aap, 664, A122

\bibitem[{{Olofsson} {et~al.}(2022{\natexlab{b}}){Olofsson}, {Th{\'e}bault},
  {Kral}, {Bayo}, {Boccaletti}, {Godoy}, {Henning}, {van Holstein},
  {Mauc{\'o}}, {Milli}, {Montesinos}, {Rein}, \& {Sefilian}}]{Olofsson2022a}
{Olofsson}, J., {Th{\'e}bault}, P., {Kral}, Q., {et~al.} 2022{\natexlab{b}},
  \mnras, 513, 713

\bibitem[{{Pairet} {et~al.}(2021){Pairet}, {Cantalloube}, \&
  {Jacques}}]{Pairet2021}
{Pairet}, B., {Cantalloube}, F., \& {Jacques}, L. 2021, \mnras, 503, 3724

\bibitem[{{Ren} {et~al.}(2019){Ren}, {Choquet}, {Perrin}, {Duch{\^e}ne},
  {Debes}, {Pueyo}, {Rice}, {Chen}, {Schneider}, {Esposito}, {Poteet}, {Wang},
  {Ammons}, {Ansdell}, {Arriaga}, {Bailey}, {Barman}, {Sebasti{\'a}n Bruzzone},
  {Bulger}, {Chilcote}, {Cotten}, {De Rosa}, {Doyon}, {Fitzgerald}, {Follette},
  {Goodsell}, {Gerard}, {Graham}, {Greenbaum}, {Hagan}, {Hibon}, {Hines},
  {Hung}, {Ingraham}, {Kalas}, {Konopacky}, {Larkin}, {Macintosh}, {Maire},
  {Marchis}, {Marois}, {Mazoyer}, {M{\'e}nard}, {Metchev}, {Millar-Blanchaer},
  {Mittal}, {Moerchen}, {Nielsen}, {N'Diaye}, {Oppenheimer}, {Palmer},
  {Patience}, {Pinte}, {Poyneer}, {Rajan}, {Rameau}, {Rantakyr{\"o}}, {Ruffio},
  {Ryan}, {Savransky}, {Schneider}, {Sivaramakrishnan}, {Song}, {Soummer},
  {Stark}, {Thomas}, {Vigan}, {Wallace}, {Ward-Duong}, {Wiktorowicz}, {Wolff},
  {Ygouf}, \& {Norman}}]{Ren2019}
{Ren}, B., {Choquet}, {\'E}., {Perrin}, M.~D., {et~al.} 2019, \apj, 882, 64

\bibitem[{{Ren} {et~al.}(2021){Ren}, {Choquet}, {Perrin}, {Mawet}, {Chen},
  {Milli}, {Debes}, {Rebollido}, {Stark}, {Hagan}, {Hines}, {Millar-Blanchaer},
  {Pueyo}, {Roberge}, {Schneider}, {Serabyn}, {Soummer}, \& {Wolff}}]{Ren2021}
{Ren}, B., {Choquet}, {\'E}., {Perrin}, M.~D., {et~al.} 2021, \apj, 914, 95

\bibitem[{{Ren} {et~al.}(2020){Ren}, {Pueyo}, {Chen}, {Choquet}, {Debes},
  {Duch{\^e}ne}, {M{\'e}nard}, \& {Perrin}}]{Ren2020}
{Ren}, B., {Pueyo}, L., {Chen}, C., {et~al.} 2020, \apj, 892, 74

\bibitem[{{Tamura} {et~al.}(2006){Tamura}, {Fukagawa}, {Kimura}, {Yamamoto},
  {Suto}, \& {Abe}}]{Tamura2006}
{Tamura}, M., {Fukagawa}, M., {Kimura}, H., {et~al.} 2006, \apj, 641, 1172

\bibitem[{{Tazaki} \& {Dominik}(2022)}]{Tazaki2022}
{Tazaki}, R. \& {Dominik}, C. 2022, \aap, 663, A57

\bibitem[{{Tazaki} {et~al.}(2023){Tazaki}, {Ginski}, \& {Dominik}}]{Tazaki2023}
{Tazaki}, R., {Ginski}, C., \& {Dominik}, C. 2023, \apjl, 944, L43

\bibitem[{{Thalmann} {et~al.}(2013){Thalmann}, {Janson}, {Buenzli}, {Brandt},
  {Wisniewski}, {Dominik}, {Carson}, {McElwain}, {Currie}, {Knapp},
  {Moro-Mart{\'{\i}}n}, {Usuda}, {Abe}, {Brandner}, {Egner}, {Feldt}, {Golota},
  {Goto}, {Guyon}, {Hashimoto}, {Hayano}, {Hayashi}, {Hayashi}, {Henning},
  {Hodapp}, {Ishii}, {Iye}, {Kandori}, {Kudo}, {Kusakabe}, {Kuzuhara}, {Kwon},
  {Matsuo}, {Mayama}, {Miyama}, {Morino}, {Nishimura}, {Pyo}, {Serabyn},
  {Suto}, {Suzuki}, {Takami}, {Takato}, {Terada}, {Tomono}, {Turner},
  {Watanabe}, {Yamada}, {Takami}, \& {Tamura}}]{Thalmann2013}
{Thalmann}, C., {Janson}, M., {Buenzli}, E., {et~al.} 2013, \apjl, 763, L29

\bibitem[{{Th{\'e}bault}(2012)}]{Thebault2012}
{Th{\'e}bault}, P. 2012, \aap, 537, A65

\bibitem[{{Thebault} \& {Kral}(2019)}]{Thebault2019}
{Thebault}, P. \& {Kral}, Q. 2019, \aap, 626, A24

\bibitem[{{Thebault} {et~al.}(2014){Thebault}, {Kral}, \&
  {Augereau}}]{Thebault2014}
{Thebault}, P., {Kral}, Q., \& {Augereau}, J.-C. 2014, \aap, 561, A16

\bibitem[{{Thebault} {et~al.}(2023){Thebault}, {Olofsson}, \&
  {Kral}}]{Thebault2023}
{Thebault}, P., {Olofsson}, J., \& {Kral}, Q. 2023, \aap, 674, A51

\bibitem[{{Th{\'e}bault} \& {Wu}(2008)}]{Thebault2008}
{Th{\'e}bault}, P. \& {Wu}, Y. 2008, \aap, 481, 713

\bibitem[{{van Holstein} {et~al.}(2020){van Holstein}, {Girard}, {de Boer},
  {Snik}, {Milli}, {Stam}, {Ginski}, {Mouillet}, {Wahhaj}, {Schmid}, {Keller},
  {Langlois}, {Dohlen}, {Vigan}, {Pohl}, {Carbillet}, {Fantinel}, {Maurel},
  {Orign{\'e}}, {Petit}, {Ramos}, {Rigal}, {Sevin}, {Boccaletti}, {Le
  Coroller}, {Dominik}, {Henning}, {Lagadec}, {M{\'e}nard}, {Turatto}, {Udry},
  {Chauvin}, {Feldt}, \& {Beuzit}}]{vanHolstein2020}
{van Holstein}, R.~G., {Girard}, J.~H., {de Boer}, J., {et~al.} 2020, \aap,
  633, A64

\bibitem[{Virtanen {et~al.}(2020)Virtanen, Gommers, Oliphant, Haberland, Reddy,
  Cournapeau, Burovski, Peterson, Weckesser, Bright, {van der Walt}, Brett,
  Wilson, Millman, Mayorov, Nelson, Jones, Kern, Larson, Carey, Polat, Feng,
  Moore, {VanderPlas}, Laxalde, Perktold, Cimrman, Henriksen, Quintero, Harris,
  Archibald, Ribeiro, Pedregosa, {van Mulbregt}, \& {SciPy 1.0
  Contributors}}]{scipy}
Virtanen, P., Gommers, R., Oliphant, T.~E., {et~al.} 2020, Nature Methods, 17,
  261

\bibitem[{{Vizgan} {et~al.}(2022){Vizgan}, {Hughes}, {Carter}, {Flaherty},
  {Pan}, {Chiang}, {Schlichting}, {Wilner}, {Andrews}, {Carpenter}, {Mo{\'o}r},
  \& {MacGregor}}]{Vizgan2022}
{Vizgan}, D., {Hughes}, A.~M., {Carter}, E.~S., {et~al.} 2022, \apj, 935, 131

\bibitem[{{Wahhaj} {et~al.}(2021){Wahhaj}, {Milli}, {Romero}, {Cieza}, {Zurlo},
  {Vigan}, {Pe{\~n}a}, {Valdes}, {Cantalloube}, {Girard}, \&
  {Pantoja}}]{Wahhaj2021}
{Wahhaj}, Z., {Milli}, J., {Romero}, C., {et~al.} 2021, \aap, 648, A26

\bibitem[{{Warren} \& {Brandt}(2008)}]{Warren2008}
{Warren}, S.~G. \& {Brandt}, R.~E. 2008, Journal of Geophysical Research
  (Atmospheres), 113, D14220

\bibitem[{{Wyatt}(2008)}]{Wyatt2008}
{Wyatt}, M.~C. 2008, \araa, 46, 339

\bibitem[{{Xie} {et~al.}(2022){Xie}, {Choquet}, {Vigan}, {Cantalloube},
  {Benisty}, {Boccaletti}, {Bonnefoy}, {Desgrange}, {Garufi}, {Girard},
  {Hagelberg}, {Janson}, {Kenworthy}, {Lagrange}, {Langlois}, {Menard}, \&
  {Zurlo}}]{Xie2022}
{Xie}, C., {Choquet}, E., {Vigan}, A., {et~al.} 2022, \aap, 666, A32

\bibitem[{{Zubko} {et~al.}(1996){Zubko}, {Mennella}, {Colangeli}, \&
  {Bussoletti}}]{Zubko1996}
{Zubko}, V.~G., {Mennella}, V., {Colangeli}, L., \& {Bussoletti}, E. 1996,
  \mnras, 282, 1321

\end{thebibliography}

\appendix

\section{Observations and data processing}\label{app:data}

\subsection{Observing log}

Table\,\ref{tab:obslog} summarizes all the observations obtained in Programmes 105.20GP.001 and 109.237K.001.

\begin{table*}
	\centering
    \caption{Log of the IRDIS BB\_H pupil-tracking DPI observations.}
	\label{tab:obslog}
	\begin{tabular}{lcccccccccc}
		\hline\hline
        Star & Type & Date & DIT & $N_\mathrm{f}$ & <Airmass> & <Seeing>    & $<\tau>$ & $\Delta \mathrm{PA}$ & OB grade & Used?\\
             &      &      & [s] &                &           & [$\arcsec$] & [ms]     & [$^{\circ}$] & & \\
		\hline
\rowcolor{Gray} HD\,191089 & SCI & 2021-07-20 & 32 & 64 & 1.10 & 0.85 & 3.7 & 10.45 & A \& B & \cmark \\
\rowcolor{Gray} $\hookrightarrow$ HD\,191131 & CAL &   & 32 & 24 & 1.10 & 0.86 & 3.7 & 10.64 &  &   \\
HD\,191089 & DPI & 2021-07-21 & 32 & 96 & 1.45 & 0.96 & 3.8 & 5.11 & A & \cmark \\
\rowcolor{Gray} HD\,191089 & SCI & 2021-07-21 & 32 & 69 & 1.07 & 0.74 & 6.5 & 10.54 & Problem & \xmark \\
\rowcolor{Gray} $\hookrightarrow$ HD\,191131 & CAL &   & 32 & 24 & 1.06 & 0.74 & 5.0 & 14.04 &  & \xmark \\
\rowcolor{Gray} HD\,191089 & SCI & 2021-09-04 & 32 & 64 & 1.09 & 0.53 & 4.5 & 11.71 & B & \cmark \\
\rowcolor{Gray} $\hookrightarrow$ HD\,191131 & CAL &   & 32 & 24 & 1.09 & 0.58 & 5.2 & 11.78 &  &   \\
\rowcolor{Gray} HD\,191089 & SCI & 2022-05-04 & 32 & 64 & 1.05 & 0.77 & 5.3 & 18.76 & A & \cmark \\
\rowcolor{Gray} $\hookrightarrow$ HD\,191131 & CAL &   & 32 & 24 & 1.04 & 0.81 & 5.4 & 14.60 &  & \cmark \\
HD\,191089 & DPI & 2022-05-17 & 32 & 16 & 1.35 & 1.03 & 2.8 & 0.88 & C & \xmark \\
\hline
HD\,157587 & DPI & 2021-07-16 & 64 & 52 & 1.18 & 0.67 & 3.0 & 2.53 & C & \cmark \\
HD\,157587 & DPI & 2021-09-29 & 64 & 8 & 1.16 & 0.99 & 2.8 & 0.40 & X & \xmark \\
\rowcolor{Gray} HD\,157587 & SCI & 2022-05-01 & 64 & 24 & 1.01 & 0.62 & 2.9 & 354.54 & C & \xmark \\
\rowcolor{Gray} $\hookrightarrow$ HD\,158018 & CAL &   & 64 & 12 & 1.01 & 0.57 & 2.9 & 327.77 &  & \xmark \\
\rowcolor{Gray} HD\,157587 & SCI & 2022-05-10 & 64 & 8 & 1.01 & 1.02 & 4.6 & 17.67 & Incomplete & \xmark \\
\rowcolor{Gray} $\hookrightarrow$ HD\,158018 & CAL &   & 64 & 4 & 1.01 & 1.01 & 4.0 & 5.12 &  & \xmark \\
HD\,157587 & DPI & 2022-05-11 & 64 & 52 & 1.17 & 1.22 & 2.6 & 2.73 & C & \cmark \\
HD\,157587 & DPI & 2022-05-14 & 64 & 52 & 1.10 & 0.77 & 2.9 & 7.20 & A & \cmark \\
\rowcolor{Gray} HD\,157587 & SCI & 2022-08-01 & 64 & 32 & 1.03 & 0.56 & 5.5 & 53.23 & A & \cmark \\
\rowcolor{Gray} $\hookrightarrow$ HD\,158018 & CAL &   & 64 & 12 & 1.02 & 0.53 & 6.0 & 48.90 &  & \cmark \\
\hline
\rowcolor{Gray} HD\,115600 & SCI & 2021-07-11 & 64 & 11 & 1.25 & 0.52 & 4.2 & 13.91 & Problem & \xmark \\
HD\,115600 & DPI & 2021-07-16 & 64 & 52 & 1.58 & 0.61 & 4.2 & 18.08 & A & \cmark \\
\rowcolor{Gray} HD\,115600 & SCI & 2022-02-10 & 64 & 8 & 1.26 & 0.74 & 6.9 & 4.13 & Problem & \xmark \\
\rowcolor{Gray} HD\,115600 & SCI & 2022-02-14 & 64 & 32 & 1.25 & 0.45 & 6.8 & 27.54 & A & \cmark \\
\rowcolor{Gray} $\hookrightarrow$ HD\,117255 & CAL &   & 64 & 12 & 1.25 & 0.49 & 8.2 & 26.17 &  & \cmark \\
\rowcolor{Gray} HD\,115600 & SCI & 2022-04-11 & 64 & 32 & 1.22 & 1.08 & 1.8 & 29.03 & C & \cmark \\
\rowcolor{Gray} $\hookrightarrow$ HD\,117255 & CAL &   & 64 & 12 & 1.22 & 1.07 & 2.4 & 27.25 &  &   \\
HD\,115600 & DPI & 2022-04-17 & 64 & 64 & 1.66 & 0.60 & 10.2 & 21.11 & A & \cmark \\
\rowcolor{Gray} HD\,115600 & SCI & 2022-05-01 & 64 & 32 & 1.24 & 0.73 & 4.0 & 30.34 & A & \cmark \\
\rowcolor{Gray} $\hookrightarrow$ HD\,117255 & CAL &   & 64 & 12 & 1.24 & 0.75 & 4.4 & 25.93 &  &   \\
\rowcolor{Gray} HD\,115600 & SCI & 2022-05-13 & 64 & 8 & 1.28 & 1.03 & 3.9 & 3.99 & Incomplete & \xmark \\
\hline
HD\,129590 & DPI & 2021-07-16 & 64 & 52 & 1.45 & 0.66 & 4.4 & 12.38 & B & \cmark \\
\rowcolor{Gray} HD\,129590 & SCI & 2022-04-01 & 64 & 32 & 1.06 & 0.66 & 4.8 & 64.05 & A & \cmark \\
\rowcolor{Gray} $\hookrightarrow$ HD\,129280 & CAL &   & 64 & 11 & 1.06 & 0.68 & 5.6 & 53.76 &  & \cmark \\
HD\,129590 & DPI & 2022-04-04 & 64 & 52 & 1.08 & 0.92 & 2.6 & 36.79 & B & \cmark \\
\rowcolor{Gray} HD\,129590 & SCI & 2022-04-28 & 64 & 32 & 1.07 & 0.78 & 3.3 & 42.48 & A & \cmark \\
\rowcolor{Gray} $\hookrightarrow$ HD\,129280 & CAL &   & 64 & 12 & 1.07 & 0.76 & 3.7 & 33.97 &  &   \\
\rowcolor{Gray} HD\,129590 & SCI & 2022-05-14 & 64 & 34 & 1.04 & 0.62 & 4.5 & 64.01 & A \& ? & \cmark \\
\rowcolor{Gray} $\hookrightarrow$ HD\,129280 & CAL &   & 64 & 12 & 1.04 & 0.65 & 4.5 & 52.38 &  &   \\
\hline

	\end{tabular}
\tablefoot{We report the type of observations (SCI, CAL, or DPI), the observing date, the detector integration time (DIT), the number of total frames ($N_\mathrm{f}$), the average airmass, seeing, coherence time ($\tau$), the range of parallactic angle during the sequence, the grades of the Observing Blocks, and whether the observations were used in this study. For clarity, star-hopping sequences are highlighted in light gray and the calibrator is marked with an arrow ($\hookrightarrow$).}
\end{table*}

\subsection{Gallery of final products}

\begin{figure*}
  \centering
  \includegraphics[width=0.49\hsize]{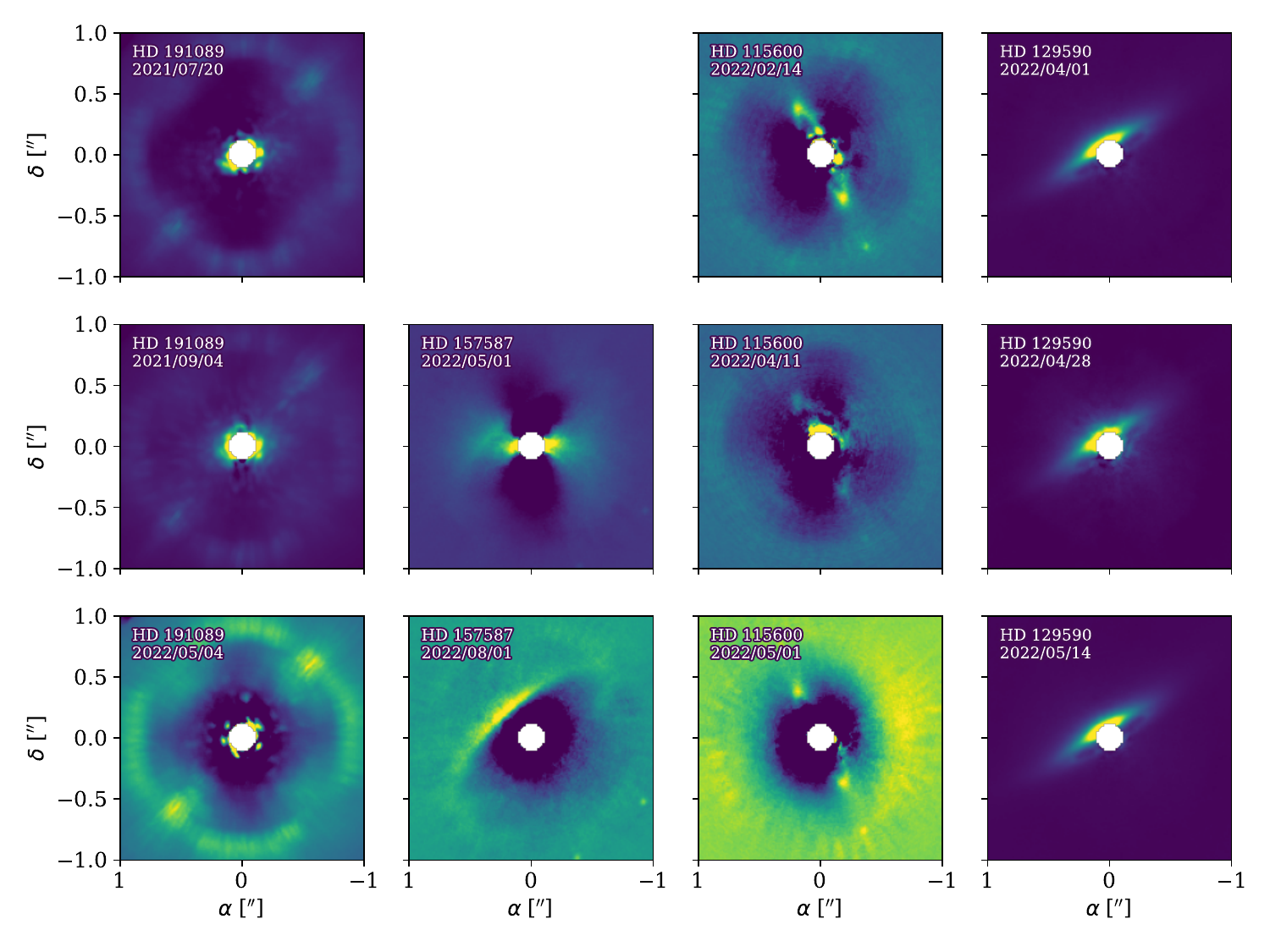}
  \includegraphics[width=0.49\hsize]{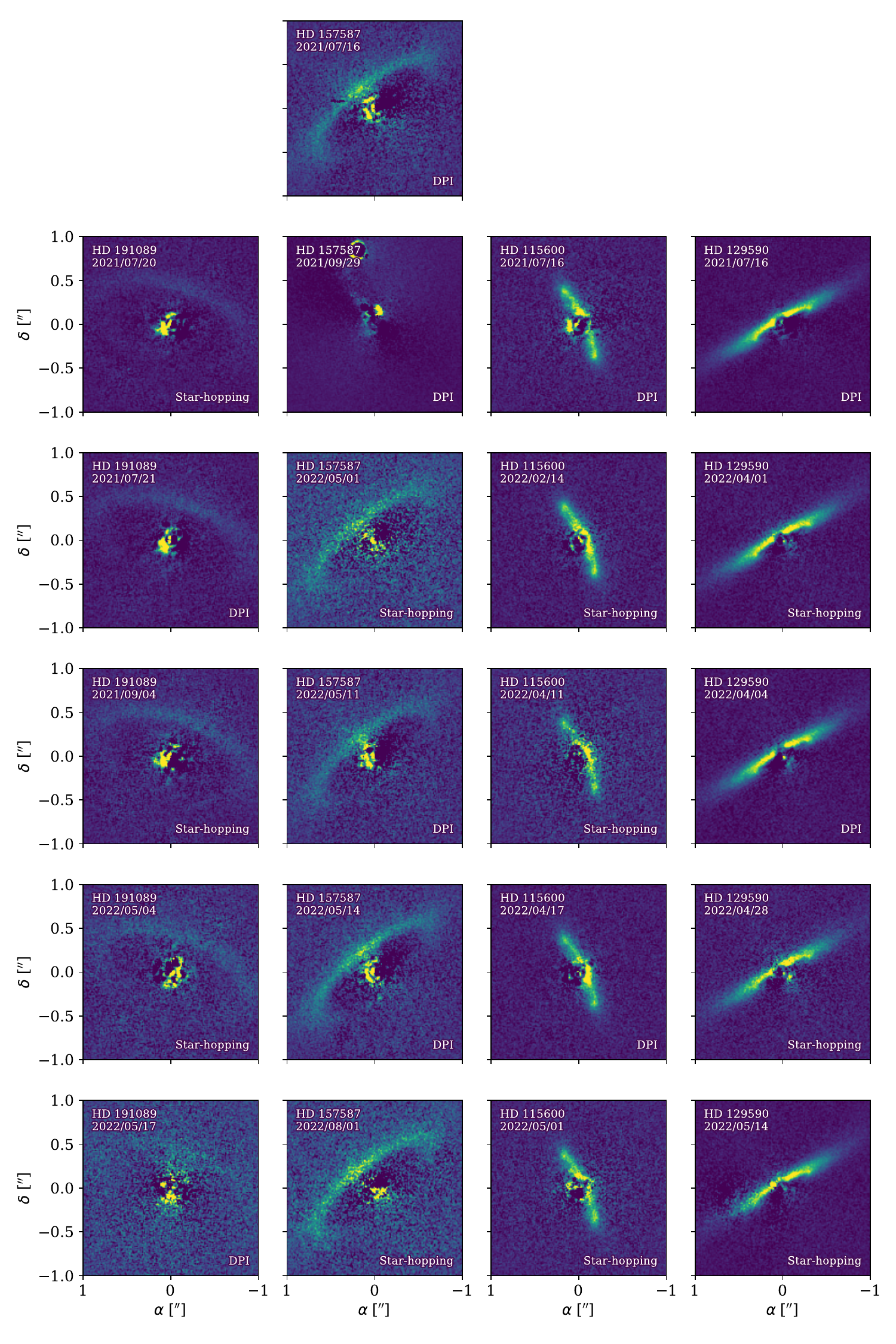}
    \caption{Gallery of the final data products for all the epochs listed in Table\,\ref{tab:obslog}. The four columns on the left are for total intensity observations, the four on the right are for polarized intensity.}
  \label{fig:all_pdi}
\end{figure*}

Figure\,\ref{fig:all_pdi} shows a gallery of all the reduced observations obtained within Programme IDs 105.20GP.001 and 109.237K.001. The left half shows observations in total intensity and the right half shows the polarimetric $Q_\phi$ images.

\subsection{Assessment of self-subtraction effects}

\begin{figure*}
  \centering
  \includegraphics[width=\hsize]{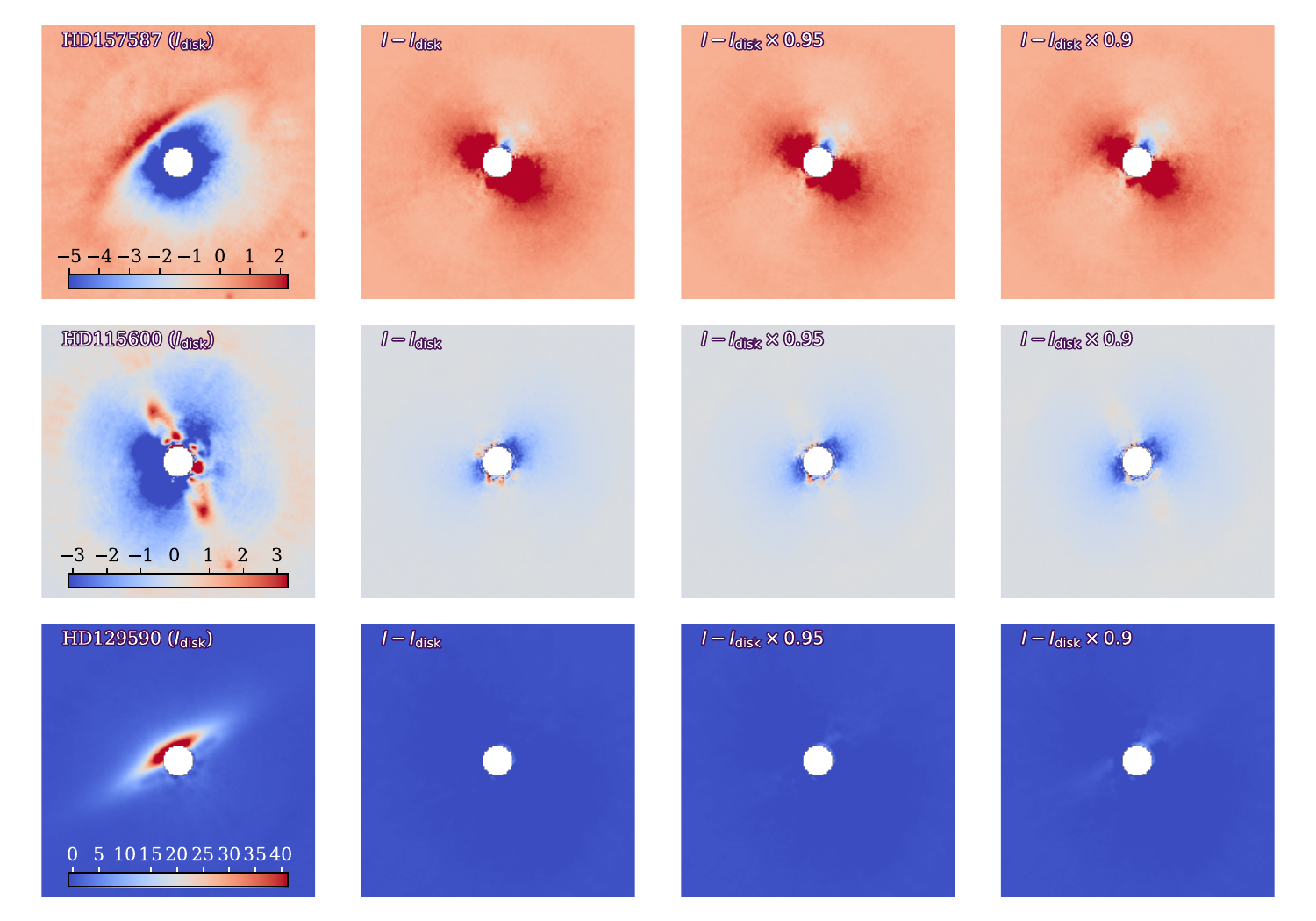}
    \caption{From left to right: total intensity image computed using DI-sNMF, that will serve as input for the "forward modeling"-like exercise ($I_\mathrm{disk}$). The next three panels show the results when first subtracting $I_\mathrm{disk}$, $I_\mathrm{disk} \times 0.95$, and $I_\mathrm{disk} \times 0.9$ to the original datacube and performing DI-sNMF. The color scale is the same for all panels of one row. From top to bottom we show the results for HD\,157857, HD\,115600, and HD\,129590, respectively.}
  \label{fig:selfsub}
\end{figure*}

Self-subtraction remains a significant challenge for many post-processing algorithms and can prevent us from properly measuring the total intensity phase function of the disks, as well as their morphology (e.g., \citealp{Milli2012}). Since the degree of polarization depends on the total intensity images, we here want to estimate if they are severely impacted by self-subtraction effects. Instead of relying on a model-dependent approach, our goal is to check whether our final total intensity images are an accurate representation of the disks. We therefore duplicated the original SPHERE data cube and for each frame, we subtracted the final total intensity image to it (rotated by the corresponding parallactic angle). We then run the DI-sNMF pipeline on this newly created cube, using the same reference star to build the components, and compute a residual map. This is equivalent to performing forward modeling of the disk, an approach that has been routinely used in the past decade. This exercise is then repeated twice more, with a disk image that is $5\%$ and $10\%$ fainter to estimate how stringent the constraints are. The results are presented in Figure\,\ref{fig:selfsub}. While there is still some residual signal close to the coronagraph, the disks are overall very well removed. For all three stars we can see some signal along the trace of the disks on the rightmost panels ($10\%$ flux decrease). For HD\,129590 and HD\,115600 this is also the case with a $5\%$ decrease in flux, while for HD\,157857 it is more difficult to assert whether we start seeing some signal coming from the disk. In all cases, when subtracting  the unaltered $I_\mathrm{disk}$ image to the datacube, there is no clear trace of any scattered light. This means that the final total intensity images (left panels) are indeed reliable representation of the surface brightness of the disks and that self-subtraction effects are well mitigated with this approach (at the $<5-10\%$ level).

\section{Geometric modeling of the polarimetric data}\label{sec:ddit}

\begin{figure*}
  \centering
  \includegraphics[width=\hsize]{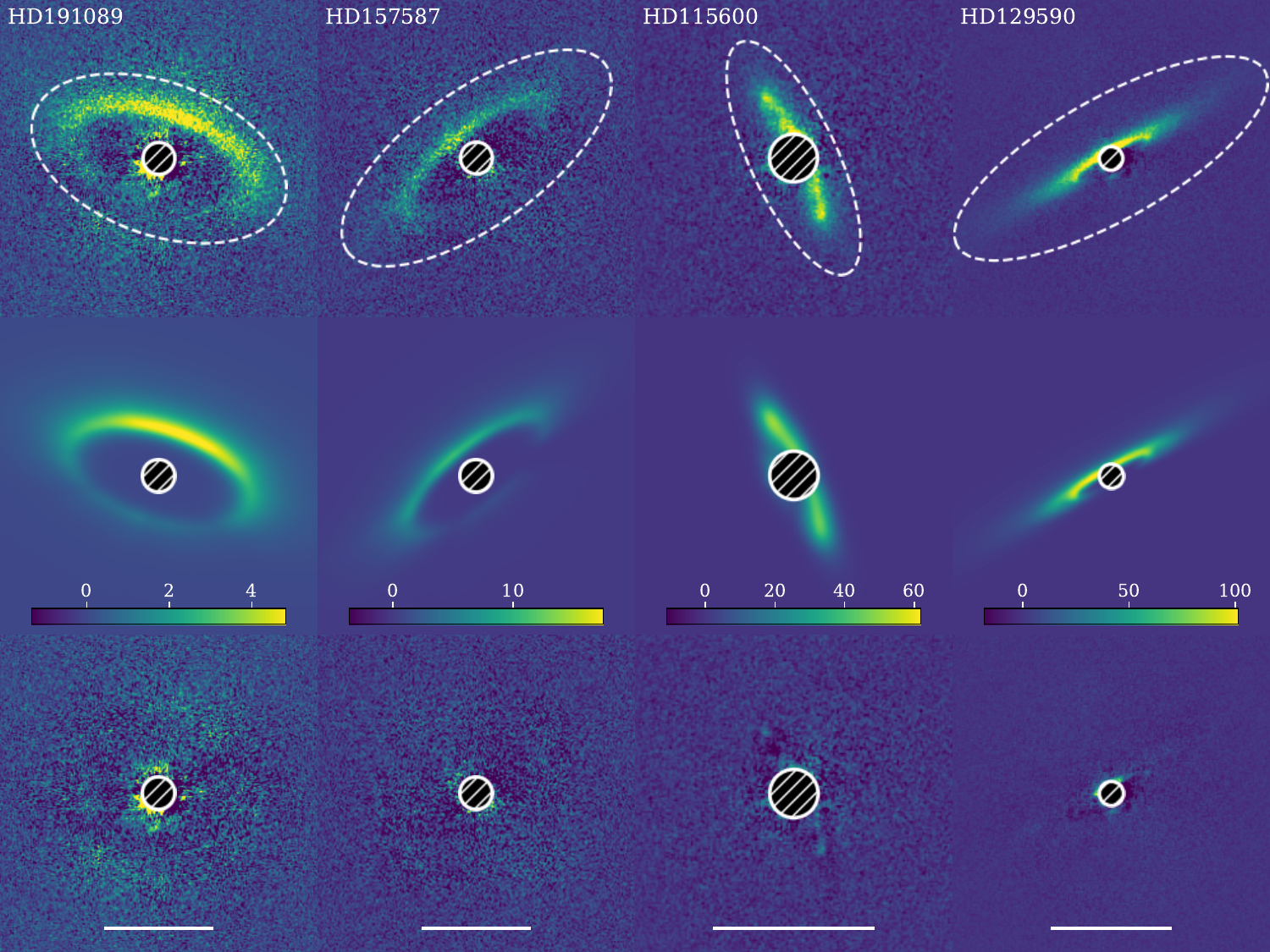}
    \caption{Modeling results for the four debris disks discussed in this study. The $Q_\phi$ image (the ellipse represents where the goodness of fit is estimated), best fit model, and residuals are shown from top to bottom. From left to right, the stars are HD\,191089, HD\,157587, HD\,115600, and HD\,129590 (same as in \citealp{Olofsson2022a} for the latter two stars). The scaling is linear and the same for each column (colorbar shown in the middle panel). The horizontal bar in the bottom panel represents $1\arcsec$.}
  \label{fig:qphi_model}
\end{figure*}

To determine the position angle and inclination of the disks studied here we modeled the $Q_\phi$ observations similarly to what was described in \citet{Olofsson2022a}. As a matter of fact, for HD\,129590 and HD\,115600, we used the results presented in the aforementioned paper. As discussed in \citet{Olofsson2023}, polarimetric observations are best suited to derive the morphological parameters of the disk compared to total intensity observations. Figure\,\ref{fig:qphi_model} shows the observations, best fit results, and residuals, and Table\,\ref{tab:ddit} shows the best fit parameters. The interested reader is referred to \citet{Olofsson2022a} for further detail on the modeling approach, but in short, the volumetric dust density distribution $N_\mathrm{dens}$ of the disk follows

\begin{equation}
    N_\mathrm{dens}(r,z) \propto \left[ \left(\frac{r}{a_0}\right)^{-2 \alpha_\mathrm{in}} + \left(\frac{r}{a_0}\right)^{-2 \alpha_\mathrm{out}}  \right]^{-1/2} \times \mathrm{exp}\left[-\left(\frac{|z|}{\mathrm{tan}(\psi) r}\right)^{\gamma} \right],
\end{equation}

where $r$ is the stellocentric distance, $z$ the height above the midplane, $a_0$ a reference radius, $\alpha_\mathrm{in}$ and $\alpha_\mathrm{out}$ two indices for the dust radial distribution. The vertical structure of the disk is parametrized by an opening angle $\psi$ and an exponent $\gamma$ to control the fall-off along the $z$ direction.

\begin{table*}
	\centering
    \caption{Results of the geometric modeling of the observations, ordered by increasing distance from earth.}
	\label{tab:ddit}
	\begin{tabular}{lccccccccc}
		\hline\hline
        Star & $d_\star$ & $a_0$ & $i$ & $\phi$ & $\alpha_\mathrm{in}$ & $\alpha_\mathrm{out}$ & $\psi$ & $\gamma$ & $\chi_\mathrm{r}^2$\\
             & [pc]      & [$\arcsec$] & [$^\circ$] & [$^\circ$] &  & & [$10^{-3}$ rad] & & \\
		\hline
HD\,191089 & $50.11 \pm 0.05$ & $0.88 \pm 0.02$ & $61.3 \pm 0.3$ & $-108.0 \pm 0.3$ & $8.7 \pm 0.7$ & $-3.4 \pm 0.2$ & $46 \pm 19$ & $5.21 \pm 2.74$ & $0.23$ \\
HD\,157587 & $99.87 \pm 0.23$ & $0.79 \pm 0.03$ & $70.1 \pm 0.3$ & $-50.7 \pm 0.3$ & $9.7 \pm 0.3$ & $-2.9 \pm 0.1$ & $15 \pm 11$ & $5.79 \pm 2.93$ & $0.25$ \\
HD\,115600 & $109.04 \pm 0.25$ & $0.45 \pm 0.02$ & $76.6 \pm 0.3$ & $-156.0 \pm 0.2$ & $2.2 \pm 0.2$ & $-6.0 \pm 0.2$ & $132 \pm  4$ & $8.32 \pm 1.52$ & $1.04$ \\
HD\,129590 & $136.32 \pm 0.44$ & $0.35 \pm 0.02$ & $82.0 \pm 0.2$ & $-60.6 \pm 0.3$ & $32.4 \pm 2.5$ & $-1.8 \pm 0.1$ & $ 1 \pm  1$ & $9.70 \pm 0.23$ & $1.27$ \\
		\hline
	\end{tabular}
\tablefoot{The columns show the stellar names, the distance, the reference radius $a_0$, the inclination $i$, position angle $\phi$, inner and outer slopes of the density distribution ($\alpha_\mathrm{in}$ and $\alpha_\mathrm{out}$, respectively), opening angle $\psi$, the exponential fall-off $\gamma$, and the reduced $\chi^2$.}
\end{table*}

\begin{figure*}
  \centering
  \includegraphics[width=\hsize]{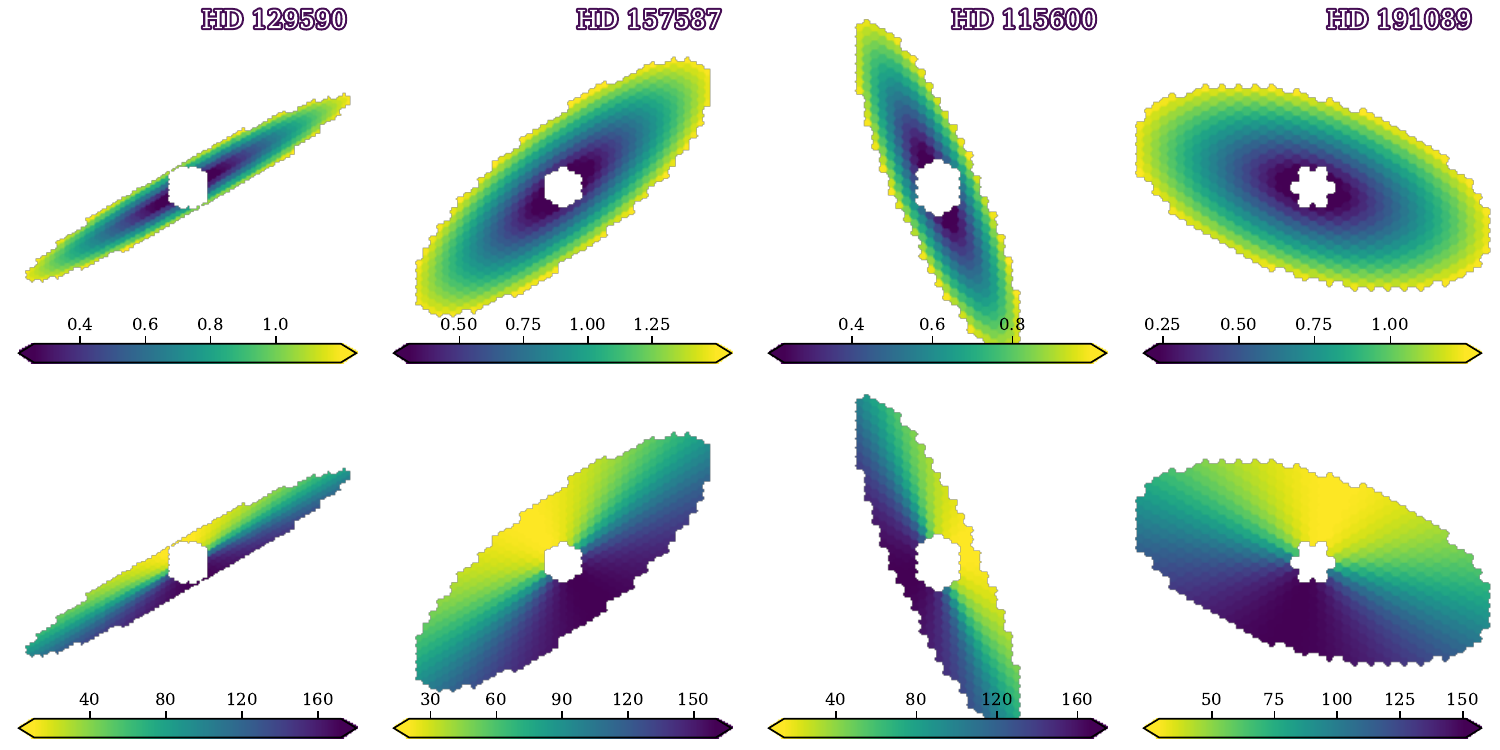}
    \caption{Stellocentric distance in arcsec and scattering angle in degrees (top and bottom, respectively) in the midplane, binned to the hexagonal grid.}
  \label{fig:scatdist}
\end{figure*}

\end{document}